\documentclass{tlp}
\usepackage{latexsym}
\usepackage{amssymb}
\usepackage{aopmath}
\usepackage{epsf}
\usepackage{pslatex}

\title[On termination of meta-programs]
{On termination of meta-programs} 
\author[Alexander Serebrenik and Danny De Schreye]
         {ALEXANDER SEREBRENIK\thanks{This work has been done during the author's stay at Department of Computer Science, K.U. Leuven, Belgium}\\
	{\'E}cole Polytechnique (STIX) \\
        91128 Palaiseau Cedex, France\\
	\email{Alexander.Serebrenik@polytechnique.fr}
         \and DANNY DE SCHREYE\\
         Department of Computer Science, K.U. Leuven\\
Celestijnenlaan 200A, B-3001, Heverlee,
Belgium\\
         \email{Danny.DeSchreye@cs.kuleuven.ac.be}}

\newtheorem{example}{Example}
\newtheorem{definition}{Definition}

\submitted{16 July 2003}
\revised{17 December 2003} 
\accepted{18 December 2003}
\begin{document}
\newcommand{\eat}[1]{}

\maketitle

\begin{abstract}
The term {\em meta-programming\/} refers to the ability of writing programs
that have other programs as data and exploit their semantics.

The aim of this paper is presenting a methodology 
allowing us to perform a correct termination analysis for a broad 
class of practical meta-interpreters, including negation and 
performing different tasks during the execution.
It is based on combining the power of general orderings,
used in proving termination of term-rewrite systems and programs, 
and on the well-known acceptability condition, used in  proving 
termination of logic programs.

The methodology establishes a relationship between the ordering needed to
prove termination of the interpreted program and the ordering
needed to prove 
termination of the meta-interpreter together with this interpreted program.
If such a relationship is established, termination of one of those
implies termination of the other one, i.e., the meta-interpreter
preserves termination.

Among the meta-interpreters that are analysed correctly are
a proof trees constructing meta-interpreter, 
different kinds of tracers and reasoners. 

To appear without appendix in Theory and Practice of Logic Programming.

\end{abstract}

\section{Introduction}
\label{meta:section:introduction}
The choice of logic programming as a basis for meta-programming offers a 
number of practical and theoretical advantages. One of them is the possibility
of tackling critical foundation problems of meta-programming within a framework
with a strong theoretical basis. Another is the surprising ease of programming.
These advantages motivated intensive research on meta-programming inside
the logic programming community~\cite{Apt:Turini,Hill:Gallagher,Levi:Ramundo,%
Martens:DeSchreye,Pedreschi:Ruggieri}. Meta-programming in logic languages
is also a powerful technique for many different application areas such
as aspect-oriented programming~\cite{DeVolder:DHondt,Brichau:Mens:DeVolder}
and constraints solving~\cite{Lamma:Milano:Mello}.

Moreover, termination analysis is one of the most intensive research 
areas in logic programming as well (see e.g.~\cite{Bossi:Cocco:Etalle:Rossi:modular,Bruynooghe:Codish:Genaim:Vanhoof,Decorte:DeSchreye:Vandecasteele,Dershowitz:Lindenstrauss:Sagiv:Serebrenik,Genaim:Codish:Gallagher:Lagoon,Lee:gpce,Mesnard:Ruggieri,Verbaeten:Sagonas:DeSchreye}).

Traditionally, termination analysis of logic programs has been done either
by the ``transformational'' approach or by the ``direct'' one. A 
transformational approach first transforms the logic program into an 
``equivalent'' term-rewrite system (or, in some cases, into an equivalent 
functional program). Here, equivalence means that, at the very least, the 
termination of the term-rewrite system should imply the termination of the 
logic program, for some predefined collection of queries\footnote{The 
approach of Arts~\cite{Arts:PhD} is exceptional in the sense that 
the termination of the logic program is concluded from a weaker property
of {\em single-redex normalisation\/} of the term-rewrite system.}. Direct 
approaches do not include such a transformation, but prove the termination 
directly on the basis of the logic program. These approaches are usually
based on {\em level mappings}, functions that map atoms to natural numbers,
and {\em norms} that map terms to natural numbers.
De Schreye {\em et al.} proved in \cite{DeSchreye:Verschaetse:Bruynooghe}
that termination is equivalent to {\em acceptability}, i.e. to 
existence of a level mapping that
decreases from the call to the head of a clause to the appropriately 
instantiated call to the recursive body subgoal.
In~\cite{DeSchreye:Serebrenik:Kowalski} we 
have developed an approach that provides the best of both worlds: a
means to incorporate into ``direct'' acceptability-based
approaches the generality of general term-orderings.

The aim of this paper is to present a methodology that
allows us to perform a correct termination analysis for a broad 
class of meta-interpreters. 
This methodology is based on the ``combined'' approach
to termination analysis mentioned above. It makes possible the
reuse of termination proofs obtained for the interpreted program as a base
for the termination proof of the meta-program. As 
Example~\ref{meta:example:nogood:linear:lm} will
illustrate, with the level mappings based techniques the reuse 
would be impossible,
even if the simplest ``vanilla'' meta-interpreter, presented in the
following example, was considered. 

\begin{example}
Our research has been motivated by the famous ``vanilla'' meta-interpreter
$M_0$, undoubtedly belonging to logic programming classics.
\label{example:vanilla}
\begin{eqnarray*}
&& \mbox{\sl solve}(\mbox{\sl true})\mbox{.}\\
&& \mbox{\sl solve}((\mbox{\sl Atom},\mbox{\sl Atoms})) \leftarrow \mbox{\sl solve}(\mbox{\sl Atom}), \mbox{\sl solve}(\mbox{\sl Atoms})\mbox{.}\\
&& \mbox{\sl solve}(\mbox{\sl Head}) \leftarrow \mbox{\sl clause}(\mbox{\sl Head},\mbox{\sl Body}), \mbox{\sl solve}(\mbox{\sl Body})\mbox{.}
\end{eqnarray*}
Atoms of the form $\mbox{\sl clause}(\mbox{\sl Head},\mbox{\sl Body})$
represent the interpreted program. 
Termination of the ``vanilla'' meta-interpreter
has been studied by Pedreschi and Ruggieri. They have 
proved that termination of the query $Q$ with respect to a program $P$
implies termination of the query $\mbox{\sl solve}(Q)$  with 
respect to $M_0$ and $P$ (Corollary 40,~\cite{Pedreschi:Ruggieri}). We are 
going to see that the two statements are equivalent, i.e.,
the query $Q$ terminates with respect to a program $P$ if and only if
the query $\mbox{\sl solve}(Q)$ LD-terminates with 
respect to $M_0$ and $P$.
$\hfill\Box$\end{example}

Even though the termination of an interpreted program might easily be
proved with level mappings, the termination proof of the meta-interpreter
extended by this program with respect to the set of the corresponding queries
might be much more complex. As the following example demonstrates, 
in some cases no linear level mapping can prove termination of the 
meta-program, despite the fact that termination of an interpreted program 
can be shown with linear level mappings. Recall, that a level mapping 
$\mid\cdot\mid$ and a norm $\|\cdot\|$ are called {\em linear} if 
\begin{eqnarray*}
&& \mid p(t_1, \ldots, t_n)\mid = c^p + \Sigma^n_{i=1} a^p_i \|t_i\|,\\
&& \|f(t_1, \ldots, t_n)\| = c^f + \Sigma^n_{i=1} a^f_i \|t_i\|,
\end{eqnarray*}
and for all $p,f$ and for all $i$, the coefficients are non-negative
integers.

\begin{example}
\label{meta:example:nogood:linear:lm}
Let $P$ be the following program:
\eat{%
\begin{eqnarray*}
&& l(X) \leftarrow p(X), r(X)\mbox{.}\\
&& p(X) \leftarrow q(X,Y), p(Y)\mbox{.}\\
&& r(f(X)) \leftarrow s(Y), r(X)\mbox{.}\\
&& q(f(Z), Z)\mbox{.}\\
&& p(0)\mbox{.}\\
&& r(0)\mbox{.}\\
&& s(0)\mbox{.}
\end{eqnarray*}
}\[
\begin{array}{lll}
l(X) \leftarrow p(X), r(X)\mbox{.} & p(X) \leftarrow q(X,Y), p(Y)\mbox{.} &
r(f(X)) \leftarrow s(Y), r(X)\mbox{.}\\
q(f(Z), Z)\mbox{.} & p(0)\mbox{.} & r(0)\mbox{.}\\
s(0)\mbox{.} &&
\end{array}\]

This program clearly terminates for $l(t)$ for every ground
term $t$. To show termination one may, for example, use a term-size norm
$\|\cdot\|$, defined on a term $t$ as a number of nodes in the 
tree-representation of $t$, and a level mapping based on $\|\cdot\|$:
$\mid l(t)\mid\;=\;\|t\|$, $\mid p(t)\mid\;=\;\|t\|$,
$\mid r(t)\mid\;=\;\|t\|$, $\mid s(t)\mid\;=\;0$ and
$\mid q(t_1,t_2)\mid\;=\;0$. One can show that the program above
satisfies the acceptability condition of De Schreye {\em et 
al.}~\cite{DeSchreye:Verschaetse:Bruynooghe} with respect to this
level mapping and, hence, queries of the form $l(t)$ with a ground
argument $t$ terminate with respect to the program.

Corollary 40,~\cite{Pedreschi:Ruggieri} implies that
$\mbox{\sl solve}(l(t))$ terminates for every ground $t$
as well. However, if a linear level mapping $\mid\cdot\mid$ and a
linear norm $\|\cdot\|$ are used in the traditional way to prove termination
of the meta-program, 
the following constraints are obtained among others:
\begin{eqnarray}
&& \mid\mbox{\sl solve}(p(X))\mid\;>\;\mid\mbox{\sl solve}((q(X,Y), p(Y)))\mid \label{meta:nolin:1} \\
&& \mid\mbox{\sl solve}(r(f(X)))\mid\;>\;\mid\mbox{\sl solve}((s(Y), r(X)))\mid \\
&& \mid\mbox{\sl solve}((s(Y), r(X)))\mid\;>\;\mid\mbox{\sl solve}(r(X))\mid,\label{meta:nolin:3} 
\end{eqnarray}
where $>$ denotes the traditional ordering on natural numbers.
In general, the last constraint should take into consideration
the intermediate body atom $s(Y)$ as well, but one can prove that it cannot
affect $r(X)$. Observe that, unlike the interpreted program, the comma of
``$(s(Y), r(X))$'' in the meta-program is a functor to be considered during
the termination analysis. 

One can show that there is no linear level mapping 
that can satisfy (\ref{meta:nolin:1})--(\ref{meta:nolin:3}).
Indeed, constraints (\ref{meta:nolin:1})--(\ref{meta:nolin:3})
can be reduced to the following (without loss of generality,
$c^{\mbox{\sl solve}} = 0$ and $a^{\mbox{\sl solve}}_1 = 1$). For
functors having only one argument the subscript is dropped.
\begin{eqnarray*}
&& c^p + a^p \|X\| > c^{,} + a^{,}_1 c^{q} + a^{,}_1 a^{q}_1 \|X\| +
a^{,}_1 a^{q}_2 \|Y\| +  a^{,}_2 c^{p} + a^{,}_2 a^{p} \|Y\|\\
&& c^r + a^r c^f + a^r a^f \|X\| > c^{,} + a^{,}_1 c^{s} + a^{,}_1 a^{s} \|Y\| + a^{,}_2 c^{r} + a^{,}_2 a^{r}\|X\|\\
&& c^{,} + a^{,}_1 c^{s} + a^{,}_1 a^{s} \|Y\| + a^{,}_2 c^{r} + a^{,}_2 a^{r} \|X\| > c^{r} + a^{r} \|X\|\\
\end{eqnarray*}
Further reduction gives, among others, the following inequalities:
\begin{eqnarray}
&& c^p > c^{,} + a^{,}_1 c^{q} +  a^{,}_2 c^{p}\label{meta:nolin:4}\\
&& c^r + a^r c^f > c^{,} + a^{,}_1 c^{s} + a^{,}_2 c^{r}\label{meta:nolin:5}\\
&& c^{,} + a^{,}_1 c^{s} + a^{,}_2 c^{r} > c^r\label{meta:nolin:6}\\
&& a^{,}_2 a^{r}\geq a^{r}\label{meta:nolin:7}
\end{eqnarray}
Condition (\ref{meta:nolin:4}) implies that $a^{,}_2 = 0$. Thus, by 
(\ref{meta:nolin:7}) $a^{r} = 0$ holds as well.
However, (\ref{meta:nolin:5}) and 
(\ref{meta:nolin:6}) imply that $a^{r} c^{f} > 0$, which provides
the desired contradiction.
$\hfill\Box$\end{example}

Note that by the well-known result of \cite{DeSchreye:Verschaetse:Bruynooghe}
termination of the meta-program implies existence of a 
non-linear level mapping that would prove termination.
However, such a level mapping might be difficult or even impossible
to generate automatically.

One can consider a number of possible solutions to this problem. First,
we may restrict attention to a specific class of interpreted programs,
such that if their termination can be proved via linear norms and level 
mappings, so should be the termination of the meta-program obtained from it
and $M_0$. As the following example illustrates even for this restricted
class of programs no obvious relation can be established between a 
level mapping required to prove termination of the interpreted program
and a level mapping required to prove termination of the meta-program.
Observe, that results on modular termination proofs for logic 
programs~\cite{Apt:Pedreschi,Bossi:Cocco:Etalle:Rossi:modular,Pedreschi:Ruggieri:modular,Verbaeten:Sagonas:DeSchreye} further discussed in Section~\ref{section:meta:conclusion}
are not applicable here, since the level-mapping required to prove
termination of calls to {\sl clause} is trivial.

\begin{example}
\label{example:linear:norel}
Let {\sl P} be the following program:
\begin{eqnarray*}
&& \mbox{\sl p}([X,Y|T]) \leftarrow  \mbox{\sl p}([Y|T]),\mbox{\sl p}(T). 
\end{eqnarray*}
Termination of the set of queries $\{\mbox{\sl p}(t)\mid\;t\;\mbox{\rm is a list of finite length}\}$ can easily be proved, for example by using a level 
mapping $\mid\!\!\mbox{\sl p}(X)\!\!\mid\; = \|X\|_l$, where $\|\cdot\|_l$
is a list-length norm defined as $\|[h|t]\|_l = 1 + \|t\|_l$ for lists and
as $\|t\|_l = 0$ for terms other than lists. However, 
when this program is considered together with $M_0$ this level mapping and this
norm cannot be extended in a way allowing us to prove termination, even though 
there exists a linear level mapping and a linear norm that provide a 
termination proof.
In our case, the following linear level mapping is sufficient
to prove termination: $\mid\!\!\mbox{\sl solve}(A)\!\!\mid\; = \|A\|$,
$\|(A, B)\| = 1+\|A\|+\|B\|$, $\|p(X)\| = 1+\|X\|$, $\|[H|T]\| = 1+3\|T\|$.
$\hfill\Box$\end{example}

Thus, even though modern termination analysis techniques, such as
the con\-straint-based approach 
of~\cite{Decorte:DeSchreye:Vandecasteele}, are able to derive the level mapping
required, they cannot reuse any information from a termination proof of the 
interpreted program to do so, and the process has to be restarted from scratch.
Moreover, the constraints set up for such examples are fairly complex 
($n$ body atoms are interpreted as a $,/2$-term of depth $n$ and reasoning on 
them requires products of (at least) $n$ parameters). Other approaches 
based on level mappings work on a basis of fixed 
norms~\cite{Dershowitz:Lindenstrauss:Sagiv:Serebrenik,Codish:Taboch}, like list-length and 
term-size, and therefore fail to prove termination of the example.
Hence, we follow a different way and instead of considering level 
mappings and norms, we 
move to the general orderings based framework originally 
presented in \cite{DeSchreye:Serebrenik:Kowalski}.

In order for meta-interpreters to be useful in applications they should
be able to cope with a richer language than the one of the
``vanilla'' meta-interpreter, including, for example, negation. Moreover, 
typical applications of meta-interpreters, such as debuggers, also require 
the production of some additional output or the performance of 
some additional tasks 
during the execution, such as constructing proof trees or cutting 
``unlikely'' branches for an uncertainty reasoner with cutoff. These 
extensions can and usually will influence termination properties of 
the meta-interpreter. To this end
we first identify popular classes of meta-interpreters, 
including the important 
{\em extended meta-inter\-preters}~\cite{Martens:DeSchreye}. Next,
we use the orderings-based framework to find conditions implying that
termination is not violated or not improved. By combining these 
conditions one obtains the requirements for termination preservation. 

The rest of this paper is organised as follows. After some preliminary
remarks, we present the  general orderings based framework 
of \cite{DeSchreye:Serebrenik:Kowalski}. Next we
introduce basic definitions and discuss the
methodology developed, as it applies 
to the ``vanilla'' meta-interpreter $M_0$. 
Afterwards, we show how the same methodology can be applied to more advanced 
meta-interpreters.

\section{Preliminaries}
We follow the standard notation for terms and atoms. A {\em query} is a 
finite sequence of atoms. Given an atom $A$, $\mbox{\sl rel}(A)$ denotes
the predicate occurring in $A$. $\mbox{\sl Term}_P$ and $\mbox{\sl Atom}_P$ 
denote, respectively, the 
sets of all terms and atoms that can be constructed from
the language underlying $P$. 
The extended Herbrand Universe $U^E_P$ (the 
extended Herbrand base $B^E_P$) is a quotient set of $\mbox{\sl Term}_P$ 
($\mbox{\sl Atom}_P$) modulo the variant relation. Recall that  the 
{\em quotient set} of a set $X$  with respect to an equivalence relation
$\sim$ is the set consisting of all equivalence classes induced on $X$ 
by $\sim$.

We refer to an SLD-tree constructed using the left-to-right selection rule of
Prolog as an LD-tree. We will say that a query $Q$ {\it LD-terminates} for
a program $P$, if the LD-tree for $\{Q\}\cup P$ is finite.

The following definition is similar to Definition 6.30~\cite{Apt:Book}.
\begin{definition}
\label{def:depend}
Let $P$ be a program and $p$, $q$ be predicates occurring in it.
\begin{itemize}
\item We say that {\em $p$ refers to $q$ in $P$\/} if there is a clause in $P$
that uses $p$ in its head and $q$ in its body.
\item We say that {\em $p$ depends on $q$ in $P$\/} and write $p\sqsupseteq q$,
if $(p,q)$ is in the transitive closure of the relation {\em refers t
o}.
\item We say that {\em $p$ and $q$  are mutually recursive\/} and write 
$p\simeq q$, if $p\sqsupseteq q$ and $q\sqsupseteq p$.
\end{itemize}
\end{definition}
The only difference between this definition and the original definition of 
Apt is that we define $\sqsupseteq$ as a transitive closure 
of the {\em refers to} relation and not as a reflexive transitive closure of 
it. Thus, we can say that the predicate $p$ is recursive if and only 
if $p\simeq p$, 
while if the definition of Apt is followed, $p\simeq p$ holds for all $p$.
We also abbreviate $p\sqsupseteq q$, $q\not \sqsupseteq p$ by $p\sqsupset q$.

Results for termination of meta-interpreters presented in this paper are
based on the notion of order-acceptability with respect to a set of queries, 
studied in~\cite{DeSchreye:Serebrenik:Kowalski}. 
This notion of order-acceptability 
generalises the notion of acceptability with respect to 
a set~\cite{Decorte:DeSchreye:98} in two ways: 1) it generalises it to general
orderings, 2) it generalises it to mutual recursion, using
the standard notion of mutual recursion~\cite{Apt:Book}---the original 
definition of acceptability required decrease only for calls to the
predicate that appears in the head of the clause. This restriction limited
the approach to programs with direct recursion only.

We start by reviewing some properties of orderings.
A {\em quasi-ordering} over a set $S$ is a reflexive and transitive
relation $\geq$ defined on elements of $S$.  We define the associated
equivalence relation $\leq\geq$ as $s \leq\geq t$ if and only if 
$s\geq t$ and $t\geq s$, and the associated {\em ordering\/} $>$ 
as $s > t$ if and only if $s\geq t$ but
not $t\geq s$. If neither $s\geq t$, nor $t\geq s$ we write
$s\|_\geq t$. If $s\geq t$ or $t\geq s$ hold for all $s$ and $t$, the
quasi-ordering is called {\em total}, otherwise it is called {\em partial}.
Sometimes, in order to distinguish between 
different quasi-orderings and associated relations, we also use 
$\succeq$, $\succ$, $\preceq\succeq$ and $\|_\succeq$.
An ordered set $S$ is said to be {\em well-founded}
if there are no infinite descending sequences $s_1 > s_2 > \ldots$ of elements
of $S$. If the set $S$ is clear from the context we will say that the ordering,
defined on it, is well-founded. 

Before introducing the order-acceptability we need the following notion.
The {\em call set}, $\mbox{\sl Call}(P,S)$, 
is the set of all atoms $A$, such that a variant of $A$ is a selected atom 
in some branch of the LD-tree for $P\cup \{Q\}$, for some $Q\in S$. 
Techniques
for inferring the supersets of the call set were suggested 
in~\cite{Janssens:Bruynooghe,Janssens:Bruynooghe:Englebert}. 
For the sake of simplicity we write 
$\mbox{\sl Call}(P,Q)$ instead of $\mbox{\sl Call}(P,\{Q\})$.

\begin{definition}
\label{def:taset}
Let $S$ be a set of atomic queries and $P$ a definite 
program. $P$ is {\em order-acceptable with respect to\ $S$} if there exists a 
well-founded ordering $>$ over $\mbox{\sl Call}(P,Q)$, such that
\begin{itemize}
\item for any $A\in\mbox{\sl Call}(P,S)$ 
\item for any clause $A'\leftarrow B_1,\ldots,B_n$ in $P$, such that 
$\mbox{\rm mgu}(A,A') = \theta$ exists,
\item for any atom $B_i$, such that $\mbox{\sl rel}(B_i)\simeq \mbox{\sl rel}(A)
$
\item for any computed answer substitution $\sigma$ for 
$\leftarrow (B_1, \ldots, B_{i-1})\theta$:
\[A > B_i\theta\sigma\]
\end{itemize}
\end{definition}

In~\cite{DeSchreye:Serebrenik:Kowalski} we prove the following theorem.
\begin{theorem}
\label{taset:term}
Let $P$ be a program. $P$ is order-acceptable with respect to\ a set of atomic
queries $S$ if and only if $P$ is LD-terminating for all queries in $S$.
\end{theorem}
We discovered that order-acceptability is a powerful notion, allowing us 
a wide variety of programs, such as {\sl normalisation}~\cite{Decorte:DeSchreye:Vandecasteele}, {\sl derivative}~\cite{DM79:cacm}, 
{\sl bid}~\cite{Maria:Benchmarks}, and credit evaluation expert 
system~\cite{Sterling:Shapiro} to mention a few. In this paper we will see 
that order-acceptability plays a key role in analysing termination behaviour
of meta-programs. We also remark that the declarative version of our
termination proof method, so called {\em rigid acceptability} also
presented in~\cite{DeSchreye:Serebrenik:Kowalski}, cannot be used to analyse
termination behaviour of meta-programs as rigid acceptability implies
termination but it is no longer equivalent to it.

\section{Basic definitions}
In this section, we present a number of basic definitions. We start
by defining the kind of program we call a meta-program. Then we
introduce two semantic notions that relate computed answers of the
interpreted program to computed answers 
of the meta-program. Finally, we conclude by discussing
an appropriate notion of termination for meta-interpreters.

We 
have seen already in Example~\ref{example:vanilla} that the input program is 
represented as a set of atoms of the predicate $\mbox{\sl clause}$. We call this representation a {\em clause-encoding} and define it formally as follows: 

\begin{definition} 
\label{definition:clause:encoding}
Let $P$ be a program. The {\em clause-encoding} $\mbox{\sl ce}(P)$
is a collection of facts of a new predicate $\mbox{\sl clause}$, 
such that $\mbox{\sl clause}(H,B)\in \mbox{\sl ce}(P)$ if and only if
$H\leftarrow B$ is a clause in $P$.
\end{definition}

\begin{example}
\label{example:clause:encoding}
Let $P$ be the following program: 
\[\begin{array}{lll}
p(X) \leftarrow q(X)\mbox{.} & q(b)\mbox{.} & s \leftarrow r, t.
\end{array}\]
Then, the following program is $\mbox{\sl ce}(P)$:
\[\begin{array}{lll}
\mbox{\sl clause}(p(X), q(X))\mbox{.} &
\mbox{\sl clause}(q(b), \mbox{\sl true})\mbox{.} &
\mbox{\sl clause}(s, (r, t))\mbox{.}
\end{array}\]
$\hfill\Box$\end{example}

An alternative representation of an interpreted program and the related
meta-interpreter will be discussed in Section~\ref{section:ground:representation} (Example~\ref{example:idemo}).

A {\em meta-interpreter} for a language is an interpreter for the language
written in the language itself. We follow~\cite{Sterling:Shapiro} by
using a predicate {\sl solve} for the meta-interpreter predicate. Sometimes
a name {\sl demo} is used~\cite{Pedreschi:Ruggieri}. One of the first 
meta-interpreters was introduced by Kowalski in~\cite{Kowalski:Book}.

\begin{definition}
The program $P$ is called a {\em meta-program\/} if it can be represented
as $M\cup I$, such that:
\begin{itemize}
\item $I$ is a clause-encoding of some program $P'$.
\item $M$ defines a predicate {\sl solve} that does not appear in $P'$.
\end{itemize}
$M$ is called the {\em meta-interpreter}. $P'$ is called the
{\em interpreted program}.
\end{definition}

We also assume that neither $,/2$ nor $\mbox{\sl clause}/2$
appears in the language underlying the interpreted program.
Observe that if this assumption is violated, a clear distinction between the 
meta-interpreter and the interpreted program is no longer possible.
Note that this restriction implies that we cannot study higher-order 
meta-programs (i.e., programs with two or more meta-layers).
Distinguishing between the meta-interpreter and the interpreted program
is essential, for example, to ease the recognition of a program
as a meta-program. Observe that meta-interpreters like ``vanilla''
mix an interpreted language and the meta-language because of the call to
$\mbox{\sl clause}/2$ in the definition of {\sl solve}. An alternative class
of meta-interpreters keeping these languages strictly separate is
considered in Section~\ref{section:ground:representation} (Example~\ref{example:idemo}).

Now we are going to define the notions of {\it soundness} and
{\it completeness} for meta-inter\-preters, that relate
computed answers of the interpreted program to the computed
answers of the meta-program. It should be noted that we
define these notions for $\mbox{\sl solve}$ with arity $n\geq 1$.
Second, third, etc. arguments are often required to support 
added functionality (see $M_2$ in Example~\ref{example:simple:meta-interpreters} below).
Often these arguments will store some information about the interpreted 
program obtained during the execution of the meta-interpreter. For example,
$M_2$ in Example~\ref{example:simple:meta-interpreters} returns 
the maximal depth of the LD-tree of the interpreted program and the 
interpreted query.

\eat{
\begin{definition}
The meta-interpreter $M$ defining $\mbox{\sl solve}$ with arity $n$
is called {\em sound} if for every program $P$ and 
every query $Q_0\in B^E_P$: 
\begin{itemize}
\item if $n=1$, then if $\mbox{\sl solve}(t_0)$ is a computed
answer for $\{\mbox{\sl solve}(Q_0)\}\cup M\cup \mbox{\sl ce}(P)$ 
then $t_0$ is a correct answer for $\{Q_0\}\cup P$.
\item if $n>1$, then for every $s_1, \ldots, s_{n-1}\in U^E_M$ if 
$\mbox{\sl solve}(t_0, t_1,\ldots, t_{n-1})$ is a computed
answer for $\{\mbox{\sl solve}(Q_0, s_1,\ldots,s_{n-1})\}\cup M\cup \mbox{\sl ce}(P)$ 
then $t_0$ is a correct answer $t$ for $\{Q_0\}\cup P$.
\end{itemize}
\end{definition}
}
\begin{definition}
The meta-interpreter $M$ defining $\mbox{\sl solve}$ with arity $n$
is called {\em sound} if for every program $P$ and 
every query $Q_0\in B^E_P$ for every $s_1, \ldots, s_{n-1}\in U^E_M$ if 
$\mbox{\sl solve}(t_0, t_1,\ldots,$ $t_{n-1})$ is a computed
answer for $\{\mbox{\sl solve}(Q_0, s_1,\ldots,s_{n-1})\}\cup M\cup \mbox{\sl ce}(P)$ then $t_0$ is a correct answer for $\{Q_0\}\cup P$.
\end{definition}

The definition of soundness, as well as further definitions, requires
some property to hold for {\em all} programs. These
definitions do not depend on the considered class of programs. However,
constructing meta-interpreters that will satisfy the properties required
for all Prolog programs can be difficult. Thus, we start by restricting
the class of programs considered to definite logic programs. 

To simplify the presentation, 
we denote $\mbox{\it Vars}_n$ a (countably
infinite) set of linear sequences of length $n$ of free variables.
Recall that a sequence of free variables is called {\em linear} 
if all the variables are different.

\begin{definition}
\label{def:complete}
The meta-interpreter $M$ defining $\mbox{\sl solve}$ with arity $n$
is called {\em complete} if for every program $P$, every query 
$Q_0\in B^E_P$ and every computed answer $t_0$ for $\{Q_0\}\cup P$
holds:
\begin{itemize} 
\item if $n=1$, then there exists $t\in U^E_{M\cup \mbox{\sl ce}(P)}$
such that $\mbox{\sl solve}(t)$ is a computed
answer for $\{\mbox{\sl solve}(Q_0)\}\cup M\cup \mbox{\sl ce}(P)$ 
and $t_0$ is an instance of $t$
\item if $n>1$, then there exist $(v_1, \ldots, v_{n-1})\in \mbox{\sl Vars}_{n-1}$ 
and
$t,t_1,\ldots,t_{n-1}\in U^E_{M\cup \mbox{\sl ce}(P)}$ such that
$\mbox{\sl solve}(t, t_1,\ldots, t_{n-1})$ is a computed
answer for 
$\{\mbox{\sl solve}(Q_0,v_1,\ldots,v_{n-1})\}\cup (M\cup \mbox{\sl ce}(P))$,
 and $t_0$ is an instance of $t$.
\end{itemize}
\end{definition}

\begin{example}
\label{example:simple:meta-interpreters}
The following meta-interpreter $M_1$ is sound, but is not complete:
$\mbox{\sl solve}(A) \leftarrow \mbox{\sl fail}$.
The meta-interpreter $\mbox{\sl solve}(A,X)\mbox{.}$
is complete, but it is not sound.
The ``vanilla'' meta-interpreter $M_0$ 
(Example~\ref{example:vanilla}) is both sound and complete,
as shown in~\cite{Levi:Ramundo}.

The following meta-interpreter $M_2$ is also sound
and complete. Given a program $P$ and a query $Q$, computing
$\mbox{\sl solve}(Q,v)$, where $v$ is a free variable,
with respect to $M_2\cup \mbox{\sl ce}(P)$ not only mimics the
execution of $Q$ with respect to $P$ but also calculates the
maximal depth of the LD-tree.

\begin{eqnarray*}
&& \mbox{\sl solve}(\mbox{\sl true}, 0)\mbox{.}\\
&& \mbox{\sl solve}((A,B), K)\leftarrow \mbox{\sl solve}(A, M), \mbox{\sl solve}(B, N), \mbox{\sl max}(M, N, K)\mbox{.}\\
&& \mbox{\sl solve}(A, s(N))\leftarrow \mbox{\sl clause}(A, B), \mbox{\sl solve}(B, N)\mbox{.}\\ 
&& \\
&& \mbox{\sl max}(0, 0, 0)\mbox{.}\\
&& \mbox{\sl max}(s(X), 0, s(X))\mbox{.}\\
&& \mbox{\sl max}(0, s(X), s(X))\mbox{.}\\
&& \mbox{\sl max}(s(X), s(Y), s(Z))\leftarrow \mbox{\sl max}(X, Y, Z).
\end{eqnarray*}
It is intuitively clear why this meta-interpreter is sound. In the latter
part of the paper we investigate an important class of meta-interpreters,
including $M_2$, and prove that all meta-interpreters in this class
are sound (Lemma~\ref{lemma:de:sound}).

To see that $M_2$ is complete recall that the second argument of 
$\mbox{\sl solve}(\mbox{\sl Query})$
that is required by Definition~\ref{def:complete} should be a
variable. Indeed, in this case unifications of calls to {\em solve}
with the corresponding clauses will depend only on the first argument
of {\em solve} and calls to {\em max} cannot affect meta-variables, fail or
introduce infinite execution. Formally, completeness of this 
meta-interpreter follows from Lemma~\ref{lemma:de:complete} and from the
fact that for any interpreted program $P$ and any interpreted query $Q$, 
$M_2$ extended by $P$ terminates for all calls to {\sl max} obtained with 
respect to $\mbox{\sl solve}(Q, v)$, where $v$ is a free variable.
$\hfill\Box$\end{example}

Recall that our aim is to study termination of
meta-interpreters, that is termination of queries of the form 
$\mbox{\sl solve}(Q_0, H_1, \ldots,$ $H_n)$, where $Q_0$
is a query with respect to the interpreted program.  Thus, the
crucial issue is to define an appropriate {\em notion of termination\/}
for meta-interpreters. 
It should be observed that requiring termination of
$\mbox{\sl solve}(Q_0, H_1, \ldots,$ $H_n)$ for all possible queries $Q_0$ 
and all possible interpreted programs $P$ is undesirable. In fact,
there are no interesting meta-interpreters satisfying this property.
Therefore, instead of termination we consider termination preservation.
For many applications, such as debuggers, this is the desired behaviour of a
meta-interpreter. However, there are many
meta-interpreters that may change termination behaviour 
of the interpreted program, either by improving or by violating it. 

\begin{definition} (non-violating LD-termination)
\label{definition:nv}
\begin{enumerate}
\item Let $M$ be a meta-interpreter defining $\mbox{\sl solve}$ with arity $1$.
$M$ is called {\em non-violating LD-termination} 
if for every program $P$ and every query $Q_0\in B^E_P$ 
 if the LD-tree of $\{Q_0\}\cup P$ is finite, then the LD-tree of 
$\{\mbox{\sl solve}(Q_0)\}\cup (M\cup \mbox{\sl ce}(P))$ is finite as well.
\item
Let $M$ be a meta-interpreter defining $\mbox{\sl solve}$ with arity $n+1$,
$n > 0$. $M$ is called {\em non-violating LD-termination with respect to 
$S\subseteq (U^E_M)^n$} if for every program $P$ and every query $Q_0\in B^E_P$ 
if the LD-tree of $\{Q_0\}\cup P$ is finite, then
for every sequence $(H_1, \ldots, H_n)\in S$, the LD-tree of 
$\{\mbox{\sl solve}(Q_0, H_1, \ldots, H_n)\}\cup (M\cup \mbox{\sl ce}(P))$
is finite as well.
\end{enumerate}
\end{definition}

Observe that $S$ formalises the intuitive notion of a set of sequences of 
``arguments other than the meta-argument''. It should be noted that the two 
cases can not be collapsed to one, i.e., $n > 0$ is essential. Indeed,
assume that $n = 0$ in the second case. Then, $(U^E_M)^n$ is empty and, 
therefore, $S$ is empty as well. Hence, there exists no sequence
$(H_1, \ldots, H_n)\in S$ and universally quantified term in the 
``then''-clause is always true. In other words, $M$ would be called
non-violating LD-termination if for every program $P$ and every query $Q_0$
the LD-tree of $\{Q_0\}\cup P$ is finite, which is useless as a definition.
Therefore, the case $n = 0$ should be considered separately.

It should be noted that traditionally the feature introduced
in Definition~\ref{definition:nv} is called
{\em improving termination\/}. However, this term is not
quite accurate, since by improving we do not mean that 
the meta-program terminates {\em more often\/} than the original one, but
that it terminates {\em at least as often} as the original one. Thus,
we chose to use more clear terminology.

It also follows from the definition of non-violation that every 
meta-interpreter defining $\mbox{\sl solve}$ with arity greater than 1
does not violate termination with respect to the empty set. 

\begin{example}
Recall the meta-interpreters shown in Example~\ref{example:simple:meta-interpreters}. $M_1$ does not violate termination, and $M_2$ does not violate 
termination with respect to $(U^E_{M_2})^1$, that is with respect to 
$U^E_{M_2}$.
$\hfill\Box$\end{example}

The dual notion is of termination non-improving.
\begin{definition} (non-improving LD-termination)
\begin{enumerate}
\item Let $M$ be a meta-interpreter defining $\mbox{\sl solve}$ with arity $1$.
$M$ is called {\em  non-improving LD-termination} 
if for every program $P$ and every 
$\mbox{\sl solve}(Q_0)\in B^E_{M\cup \mbox{\sl ce}(P)}$, 
finiteness of the LD-tree of 
$\{\mbox{\sl solve}(Q_0)\}\cup (M\cup \mbox{\sl ce}(P))$
implies finiteness of the LD-tree of $\{Q_0\}\cup P$.
\item 
Let $M$ be a meta-interpreter defining $\mbox{\sl solve}$ with arity $n+1$,
$n > 0$. $M$ is called {\em  non-improving LD-termination with respect to 
$S\subseteq (U^E_M)^n$} if for every program $P$ and every query 
$\mbox{\sl solve}(Q_0, H_1, \ldots, H_n)\in B^E_{M\cup \mbox{\sl ce}(P)}$,
such that $(H_1, \ldots, H_n)\in S$, finiteness of the LD-tree of 
$\{\mbox{\sl solve}(Q_0, H_1, \ldots, H_n)\}\cup (M\cup \mbox{\sl ce}(P))$
implies finiteness of the LD-tree of $\{Q_0\}\cup P$.
\end{enumerate}
\end{definition}

\begin{example}
The meta-interpreter $M_2$ does not improve termination with respect to 
$\mbox{\sl Vars}$, where $\mbox{\sl Vars}$ is a countably infinite set 
of variables. 
$\hfill\Box$\end{example}

Finally, we define {\em termination preservation}.
\begin{definition}
Let $M$ be a meta-interpreter defining $\mbox{\sl solve}$ 
with arity $n+1$. We say that $M$ is {\em preserving termination} 
({\em preserving termination
with respect to $S\subseteq (U^E_M)^n$}, if $n>0$), if it is non-violating 
LD-termination (non-violating LD-termination with respect to $S$) and 
non-improving LD-termination (non-improving LD-termination with respect to $S$).
\end{definition}

The ``vanilla'' meta-interpreter $M_0$ preserves termination and the 
meta-interpreter 
$M_2$ preserves termination with respect to $\mbox{\sl Vars}$, that is
if it is used to measure the depth of LD-refutation of a given query, and not
to bound it. In the next sections we prove these statements.

\section{Termination of the ``vanilla'' meta-interpreter}
Termination of the ``vanilla'' meta-interpreter, presented in 
Example~\ref{example:vanilla}, has been studied by Pedreschi and Ruggieri.
They have proved that ``vanilla'' does not violate termination (Corollary 
40,~\cite{Pedreschi:Ruggieri}). However, we can claim more---this meta-interpreter
preserves termination.

We base our proof on soundness and completeness of ``vanilla'',
proved in~\cite{Levi:Ramundo}. Observe that, in general, soundness and 
completeness are not sufficient for the call set to be preserved. 
Indeed, consider the following example, motivated by the ideas of 
unfolding~\cite{Bossi:Cocco}. 

\begin{example} 
\label{example:Bossi}
The following meta-interpreter $M_3$ eliminates calls to undefined predicates.
\begin{eqnarray*}
&& \mbox{\sl solve}(\mbox{\sl true})\mbox{.}\\
&& \mbox{\sl solve}((A,B)) \leftarrow \mbox{\sl solve}(A), \mbox{\sl solve}(B)\mbox{.}\\
&& \mbox{\sl solve}(A) \leftarrow \mbox{\sl clause}(A,B), \mbox{\sl check}(B), \mbox{\sl solve}(B)\mbox{.}\\
&& \\
&& \mbox{\sl check}((A,B)) \leftarrow \mbox{\sl check}(A),\mbox{\sl check}(B)\mbox{.}\\
&& \mbox{\sl check}(A) \leftarrow \mbox{\sl clause}(A,\_)\mbox{.}\\
&& \mbox{\sl check}(\mbox{\sl true})\mbox{.}
\end{eqnarray*}

This meta-interpreter is sound and complete, i.e., preserves 
computed answers. However, it does not preserve termination. Indeed, let $P$ 
be the following program:
\begin{eqnarray*}
&& p \leftarrow q, r\mbox{.}\\
&& q \leftarrow q\mbox{.}
\end{eqnarray*}
and let $p$ be the query. Then, $p$ with respect to $P$ does not terminate, 
while
$\mbox{\sl solve}(p)$ with respect to $M_3\cup \mbox{\sl ce}(P)$ 
terminates (finitely fails). Thus, this meta-interpreter does not 
preserve LD-termination.
Observe that unfolding may only improve
termination~\cite{Bossi:Cocco}. Thus, this meta-interpreter is non-violating
LD-termination. 
$\hfill\Box$\end{example}

Thus, the claim that the ``vanilla'' 
meta-inter\-preter preserves the calls set should be proven separately.
\begin{lemma}
\label{lemma:vanilla:calls:preserved}
Let $P$ be an interpreted program, $M_0$ be the ``vanilla''
meta-interpreter and $Q\in B^E_{P}$, then:
\begin{itemize}
\item for every call $A\in \mbox{\sl Call}(P, Q)$, there exists 
$\mbox{\sl solve}(A')\in \mbox{\sl Call}(M_0\cup \mbox{\sl ce}(P), 
\mbox{\sl solve}(Q))$ such that $A$ and $A'$ are variants;
\item for every call $\mbox{\sl solve}(A)\in \mbox{\sl Call}(M_0\cup \mbox{\sl ce}(P), \mbox{\sl solve}(Q))$, such that $A\in B^E_{P}$, there exists 
$A'\in \mbox{\sl Call}(P, Q)$, such that $A$ and $A'$ are variants.
\end{itemize}
\end{lemma}
\begin{proof}
The proof can be found in ~\ref{appendix:vanilla:preserves:calls}.
\end{proof}

This lemma extends Theorem 9~\cite{Pedreschi:Ruggieri}, by claiming not only \
that every call of the meta-program and of the meta-query ``mimics''
the original execution, but also that every call of the original program and
query is ``mimicked'' by the meta-program.

Now we can complete the analysis of the ``vanilla'' meta-interpreter,
namely, prove that it does not improve termination. The main idea 
is to construct a quasi-ordering relation $\succeq$ for atoms of the
interpreted program based on the quasi-ordering $\geq$ 
such that the meta-program is 
order-acceptable via the quasi-ordering relation $\geq$. 
To complete the proof, we have to show
that the interpreted program is order-acceptable via $\succeq$.

\begin{theorem}
\label{meta:interpreted}
Let $P$ be a definite program, $S$ a set of atomic queries, 
and  $M_0$ the ``vanilla'' meta-interpreter, such that 
$M_0\cup \mbox{\sl ce}(P)$ is LD-terminating for all queries in
$\{\mbox{\sl solve}(Q) \mid Q\in S\}$. Then,
$P$ is LD-terminating for all queries in $S$.
\end{theorem}
\begin{proof} 
By Theorem~\ref{taset:term} 
$M_0\cup \mbox{\sl ce}(P)$ is order-acceptable with respect to a set
$\{\mbox{\sl solve}(Q) \mid Q\in S\}$. We are going to prove  
order-acceptability of $P$ with respect to $S$. By Theorem~\ref{taset:term}
termination will be implied.

Since $M_0\cup \mbox{\sl ce}(P)$ is order-acceptable with respect to 
$\{\mbox{\sl solve}(Q) \mid Q\in S\}$ there exists 
a well-founded quasi-ordering 
$\geq$, satisfying requirements of Definition~\ref{def:taset}. Let $\succeq$ 
be a new quasi-ordering on $B^E_P$ defined as $A \succ B$ if 
$\mbox{\sl solve}(A) > \mbox{\sl solve}(B)$ and $A\preceq\succeq A$ for all
$A$.

The ordering is defined on $\{A \mid A\in B^E_P\;\wedge\;\exists Q\in S\;\mbox{\rm such that}\;\mbox{\sl solve}(A)\in \mbox{\sl Call}(M_0 \cup \mbox{\sl ce}(P), \mbox{\sl solve}(Q))\}$. 
By Lemma~\ref{lemma:vanilla:calls:preserved} this set coincides with 
$\mbox{\sl Call}(P, S)$. The ordering $\succ$ is well-defined and
well-founded. These properties follow immediately from the corresponding
properties of $>$. 
\eat{
Indeed:
\begin{itemize}
\item Assume that there exists an atom $A$, such that $A\succ A$.
Then, by definition of $\succ$, $\mbox{\sl solve}(A) > \mbox{\sl solve}(A)$ 
holds, contradicting irreflexivity of $>$.
\item Assume that there exist atoms $A$ and $B$, such that $A\succ B$ and
$B\succ A$. Then, by definition of $\succ$, $\mbox{\sl solve}(A) > 
\mbox{\sl solve}(B)$ and $\mbox{\sl solve}(B) > \mbox{\sl solve}(A)$
hold, contradicting asymmetry of $>$.
\item Let $A, B$ and $C$, atoms such that $A\succ B$ and
$B\succ C$. Then, by definition of $\succ$, $\mbox{\sl solve}(A) > 
\mbox{\sl solve}(B)$ and $\mbox{\sl solve}(B) > \mbox{\sl solve}(C)$
hold. By transitivity of $>$, $\mbox{\sl solve}(A) > \mbox{\sl solve}(C)$.
Thus, by definition of $\succ$, $A\succ C$.
\item Let $A_1,A_2,\ldots$ be an infinitely descending chain with
respect to $\succ$, i.e., let $A_n \succ A_{n+1}$ hold for all $n$.
Then, $\mbox{\sl solve}(A_n) > \mbox{\sl solve}(A_{n+1})$ holds for all $n$
and $\mbox{\sl solve}(A_1),$ $\mbox{\sl solve}(A_2),\ldots$ forms an infinitely
descending chain with respect to $>$, contradicting well-foundedness of $>$.
\end{itemize}
}

Next, we prove that $P$ is order-acceptable with respect to $S$ 
via $\succeq$. Let $Q\in S$, $A\in \mbox{\sl Call}(P,Q)$ and 
let $A'\leftarrow B_1,\ldots,B_n$ be a clause in $P$, such that 
$\mbox{\sl mgu}(A,A') = \theta$ exists. Let $B_i$ be such that 
$\mbox{\sl rel}(B_i)\simeq \mbox{\sl rel}(A)$ and let $\sigma$ be 
a computed answer substitution for $(B_1,\ldots,B_{i-1})\theta$.
We have to show that $A\succ B_i\theta\sigma$.

By Lemma~\ref{lemma:vanilla:calls:preserved} $\mbox{\sl solve}(A)\in \mbox{\sl Call}(M_0 \cup \mbox{\sl ce}(P), \mbox{\sl solve}(Q))$. The only clause that can be used in the
resolution with it is $\mbox{\sl solve}(\mbox{\sl Head}) \leftarrow 
\mbox{\sl clause}(\mbox{\sl Head},\mbox{\sl Body}), \mbox{\sl solve}(\mbox{\sl Body})$.
Observe that $\mbox{\sl mgu}(\mbox{\sl solve}(A), \mbox{\sl solve}(\mbox{\sl Head}))$
affects neither $A$ nor $\mbox{\sl Body}$. Let $\tau$ be
a computed answer 
substitution that unifies $A$ and $A'$, $\mbox{\sl Body}$ and $(B_1,\ldots,B_n)$. By the choice of $\tau$, 
$\tau$ maps $\mbox{\sl Body}$ to $(B_1,\ldots,B_n)\theta$. 
Then, $\mbox{\sl solve}(\mbox{\sl Body})\tau = \mbox{\sl solve}((B_1,\ldots,B_n))\theta$ and
$\mbox{\sl solve}(A) > \mbox{\sl solve}((B_1,\ldots,$ $B_n)\theta)$
by the order-acceptability of $M_0\cup \mbox{\sl ce}(P)$ via $\geq$.

If $n=1$ then $\mbox{\sl solve}(A) > \mbox{\sl solve}(B_1\theta)$.
By definition of $\succ$, $A\succ B_1\theta$, completing the proof.

Assume that $n>1$. In this case,
$\mbox{\sl solve}((B_1,\ldots,B_n)\theta)$ is another call in the call set.
The clause
$\mbox{\sl solve}((\mbox{\sl Atom},\mbox{\sl Atoms})) \leftarrow \mbox{\sl solve}(\mbox{\sl Atom}), \mbox{\sl solve}(\mbox{\sl Atoms})$
is the only one that can be used at the resolution step. 
Let $\delta$ be the most general unifier of
$\mbox{\sl solve}((B_1,\ldots,B_n)\theta)$ with the head of the clause above.
The substitution $\delta$
does not affect the variables appearing in $(B1,\ldots,B_n)\theta$,
since $\mbox{\sl Atom}$ and $\mbox{\sl Atoms}$ are variables. Thus,
$B_j\theta\delta = B_j\theta$ for all $j$ and $\delta$ is omitted in the
lines to come.

Order-acceptability of the meta-program implies that
$\mbox{\sl solve}((B_1,\ldots,B_n)\theta) > \mbox{\sl solve}(B_1\theta)$
and for any computed answer substitution $\sigma_1$ for 
$\mbox{\sl solve}(B_1\theta)$,
$\mbox{\sl solve}((B_1,\ldots,B_n)\theta) > 
\mbox{\sl solve}((B_2,\ldots,B_n)\theta\sigma_1)$.
Proceeding in the same way, we obtain
\begin{eqnarray*}
&& \mbox{\sl solve}((B_2,\ldots,B_n)\theta\sigma_1) > 
\mbox{\sl solve}((B_3,\ldots,B_n)\theta\sigma_1\sigma_2)\\
&& \vdots\\
&& \mbox{\sl solve}((B_{i-1},\ldots,B_n)\theta\sigma_1\ldots\sigma_{i-2}) > 
\mbox{\sl solve}((B_{i},\ldots,B_n)\theta\sigma_1\ldots\sigma_{i-1}),
\end{eqnarray*}
where $\sigma_j$ is a computed answer substitution for
$\mbox{\sl solve}(B_j\theta\sigma_1\ldots\sigma_{j-1})$.

Moreover, 
$\mbox{\sl solve}((B_{i},\ldots,B_n)\theta\sigma_1\ldots\sigma_{i-1}) >
\mbox{\sl solve}(B_{i}\theta\sigma_1\ldots\sigma_{i-1})$.
Transitivity of the $>$-ordering implies
$\mbox{\sl solve}(A) > \mbox{\sl solve}(B_{i}\theta\sigma_1\ldots\sigma_{i-1})$.
Computed answer substitutions are preserved by $M_0$~\cite{Levi:Ramundo}.
Thus, 
for all $j$, $\sigma_j$ is also a computed answer substitution for 
$B_j\theta\sigma_1\ldots\sigma_{j-1}$. Therefore, $\sigma_j$'s can be
chosen such that $\sigma = \sigma_1\ldots\sigma_{i-1}$. In other words,
$\mbox{\sl solve}(A) > \mbox{\sl solve}(B_i\theta\sigma)$.
Thus, 
$A \succ B_i\theta\sigma$ by definition of $\succ$, completing the proof.
\end{proof}

The other direction of the theorem has been proved by Pedreschi and
Ruggieri~\cite{Pedreschi:Ruggieri}. It allows us to
state the following corollary.
\begin{corollary}
\label{corollary:pres:term}
The ``vanilla'' meta-interpreter $M_0$ preserves LD-termination.
\end{corollary}

The proof of Theorem~\ref{meta:interpreted} presented above suggests the
following methodology for proving that a particular 
meta-interpreter does not improve 
LD-termination. First, define an ordering on the set of calls to 
the meta-interpreter, such that the ordering
reflects its behaviour. Then, establish the relationship between a new ordering
and the one that reflects order-acceptability with respect to a set of the interpreted
program. Prove, using this relationship, that the newly defined ordering is 
well-defined, well-founded and reflects order-acceptability of the
meta-program with respect to a corresponding set of calls. In order for 
the proofs to be
correct, one may need to assume (or to prove as a prerequisite) that the
meta-interpreter is sound and that the set of calls of the interpreted
program and of the meta-program correspond to each other.
The opposite direction, i.e., that the meta-interpreter does not violate 
termination, can be proved using a similar methodology. Therefore, in the
following section we will define an ordering for advanced meta-interpreters
based on the existing ordering for $M_0$ and vice versa.

We illustrate the methodology proposed by considering $M_0$ and
Example~\ref{meta:example:nogood:linear:lm}. 
Recall that we have seen in 
Section~\ref{meta:section:introduction}
that level-mapping based approaches experience serious difficulties
with analysing termination of this example.
\begin{example}
Let $P$ be the program discussed in 
Example~\ref{meta:example:nogood:linear:lm}.
Acceptability (and thus, termination) 
of $\{\mbox{\sl solve}(l(t))\mid t\;\mbox{\rm is a ground term}\}$ with respect
to the corresponding meta-program can be established via the 
ordering $>$ that satisfies for all ground terms $t_1$ and $t_2$:
\begin{eqnarray*}
&& \mbox{\sl solve}(l(t_1)) > \mbox{\sl solve}((p(t_1),r(t_1)))\\
&& \mbox{\sl solve}((p(t_1),r(t_1))) > \mbox{\sl solve}(p(t_1))\\
&& \mbox{\sl solve}((p(t_1),r(t_1))) > \mbox{\sl solve}(r(t_1))\\
&& \mbox{\sl solve}(p(t_1)) > \mbox{\sl solve}((q(t_1, t_2),p(t_2)))\\
&& \mbox{\sl solve}((q(t_1, t_2),p(t_2))) > \mbox{\sl solve}(q(t_1, t_2))\\
&& \mbox{\sl solve}((q(t_1, t_2),p(t_2))) > \mbox{\sl solve}(p(t_2))\\
&& \mbox{\sl solve}(r(f(t_1))) > \mbox{\sl solve}((s(t_2),r(t_1)))\\
&& \mbox{\sl solve}((s(t_1),r(t_2))) > \mbox{\sl solve}(s(t_1))\\
&& \mbox{\sl solve}((s(t_1),r(t_2))) > \mbox{\sl solve}(r(t_2))
\end{eqnarray*}
These inequalities imply:
\begin{eqnarray}
&& \mbox{\sl solve}(p(t_1)) > \mbox{\sl solve}(p(t_2))\;\mbox{\rm for terms such that $\mbox{\sl solve}(q(t_1, t_2))$ holds} \label{meta:example:nogood:1} \\
&& \mbox{\sl solve}(r(f(t_1))) > \mbox{\sl solve}(r(t_1))\label{meta:example:nogood:2}
\end{eqnarray}
Recall the construction we applied in the proof of 
Theorem~\ref{meta:interpreted}. A new ordering has been proposed for 
the atoms of $B^E_P$:
$A \succ B$ if $\mbox{\sl solve}(A) > \mbox{\sl solve}(B)$.
Hence, (\ref{meta:example:nogood:1}) and (\ref{meta:example:nogood:2})
lead, in our case, to the following definition of $\succ$:
$p(t_1) \succ p(t_2)$, for terms such that $q(t_1, t_2)$ holds,
and $r(f(t_1))\succ r(t_1)$.
One can easily see that $\succ$ is indeed well-founded and
that $P$ is order-acceptable with respect to 
$\{l(t)\mid t\;\mbox{\rm is a ground term}\}$ via this ordering.
$\hfill\Box$\end{example}

\eat{
\section{Advanced meta-interpreters}

In this section we start by identifying an important class of 
meta-interpreters, called {\em double extended meta-interpreters},
that are able to perform additional tasks such as constructing proof trees
or keeping track on execution depth. This class includes many 
meta-interpreters appearing in the literature. 

Next we discuss meta-interpreters for normal programs 
and obtain results similar to the definite case. Finally, we conclude this 
section by considering alternative way of representing clauses, so called
{\em ground representation}~\cite{Hill:Gallagher}, and study a corresponding 
meta-interpreter. Unlike the former meta-interpreters amalgamating the 
language of the interpreted program
and the meta language, the latter uses the encoding of the object level by
meta level terms.
}

\section{Double extended meta-interpreters}
\label{section:double:extended}
Typical applications of
meta-interpreters require the production of some
additional output or the performance of some additional tasks during the 
execution,
such as constructing proof trees (essential for debugging and explanation
applications) or cutting ``unlikely'' branches (required for 
uncertainty reasoners with cutoff). As we are going to see in 
the examples to come,
these extensions can and usually will 
influence termination properties of the meta-interpreter. 

In this section, we still consider definite meta-interpreters, but their 
clauses, which still follow the general outline of $M_0$, are
enriched with extra subgoals, providing additional functionality. 
This class of meta-interpreters expands the class of extended meta-interpreters
studied by~\cite{Martens:DeSchreye}. It includes many useful 
meta-interpreters, such as a meta-interpreter which constructs 
proof trees~\cite{Sterling:Shapiro} and can
be used as a basis for explanation facilities in expert system, as well as
meta-interpreters which allow reasoning about theories and provability~\cite{Brogi:Mancarella:Pedreschi:Turini,Martens:DeSchreye:inbook} or meta-interpreters
which implement reasoning
with uncertainty~\cite{Sterling:Shapiro}. Moreover, this class 
also describes a depth tracking tracer for Prolog, a reasoner with threshold 
cutoff~\cite{Sterling:Shapiro} 
and a pure four port box execution model tracer~\cite{Bowles:Wilk}.
The methodology presented so far
is expanded to analyse double extended meta-interpreters, and 
conditions implying termination non-violation and non-improvement
are established. 

\begin{definition}
\label{definition:de}
A definite program of the following form 
\begin{eqnarray*}
&& \mbox{\sl solve}(\mbox{\sl true}, t_{11}, \ldots, t_{1n})\leftarrow C_{11},\ldots, C_{1m_1}\mbox{.}\\
&& \mbox{\sl solve}((A,B), t_{21}, \ldots, t_{2n})\leftarrow \\
&& \hspace{1.0cm}D_{11},\ldots,D_{1k_1}, \mbox{\sl solve}(A, t_{31}, \ldots, t_{3n}), \\
&& \hspace{1.0cm}D_{21},\ldots,D_{2k_2}, \mbox{\sl solve}(B, t_{41}, \ldots, t_{4n})\\
&& \hspace{1.0cm} C_{21},\ldots, C_{2m_2}\mbox{.}\\
&& \mbox{\sl solve}(A, t_{51}, \ldots, t_{5n})\leftarrow \\
&& \hspace{1.0cm}D_{31},\ldots,D_{3k_3},\mbox{\sl clause}(A,B,s_{1}, \ldots, s_{k}), \\
&& \hspace{1.0cm}D_{41},\ldots,D_{4k_4},\mbox{\sl solve}(B, t_{61}, \ldots, t_{6n})\\
&& \hspace{1.0cm} C_{31},\ldots, C_{3m_3}\mbox{.}
\end{eqnarray*}
together with clauses defining any other predicate occurring in the 
$C_{kl}$ and $D_{pq}$ (none of which contain $\mbox{\sl solve}$ or 
$\mbox{\sl clause}$) is called a {\em double extended meta-interpreter}.
\end{definition}

The name of this class of the meta-interpreters stems from the fact that
they further generalise the class of extended meta-interpreters~\cite{Martens:DeSchreye:inbook}. Extended meta-interpreters are double 
extended meta-interpreters, such that for all $p$ and $q$, $D_{pq}$ is 
{\sl true}. Note that despite the similarity between the definition 
and Example~\ref{example:Bossi},  
Example~\ref{example:Bossi} is not a double extended meta-interpreter
due to the call to predicate {\sl clause} in the definition of {\sl check}.
Thus, the results established in this section are not applicable
to it.

\begin{example}
\label{example:4port:box}
The following program~\cite{Bowles:Wilk} shows the pure Prolog tracer for the 
four port box execution model of Byrd~\cite{Byrd} (in the original paper
{\sl interp} was used instead of {\sl solve} to denote the meta-predicate; we
renamed the predicate for the sake of uniformity).
\begin{eqnarray}
&& \mbox{\sl solve}(\mbox{\sl true}). \nonumber \\
&& \mbox{\sl solve}((G1,G2)) \leftarrow \mbox{\sl solve}(G1), \mbox{\sl solve}(G2). \nonumber \\
&& \mbox{\sl solve}(G)\leftarrow \mbox{\sl before}(G), \mbox{\sl clause}(G,B), \mbox{\sl solve}(B), \mbox{\sl after}(G). \label{example:debugger:1}\\
&& \nonumber  \\
&& \mbox{\sl before}(G)\leftarrow \mbox{\sl write}('\mbox{\sl call} '), \mbox{\sl write}(G), \mbox{\sl nl}.\label{example:debugger:non-fail} \\
&& \mbox{\sl before}(G)\leftarrow \mbox{\sl write}('\mbox{\sl fail} '), \mbox{\sl write}(G), \mbox{\sl nl}, \mbox{\sl fail}. \label{example:debugger:2} \\
&& \nonumber \\
&& \mbox{\sl after}(G)\leftarrow \mbox{\sl write}('\mbox{\sl succeed} '), \mbox{\sl write}(G), \mbox{\sl nl}.\nonumber \\
&& \mbox{\sl after}(G)\leftarrow \mbox{\sl write}('\mbox{\sl redo} '), \mbox{\sl write}(G), \mbox{\sl nl}, \mbox{\sl fail}.\nonumber 
\end{eqnarray}
This program is a double extended meta-interpreter, since it clearly has
the form prescribed by Definition~\ref{definition:de} and
neither $\mbox{\sl before}$ nor $\mbox{\sl after}$ is depending on
$\mbox{\sl solve}$ or $\mbox{\sl clause}$. 
$\hfill\Box$\end{example}

The next example of a double extended meta-interpreter 
is motivated by program 17.8 of~\cite{Sterling:Shapiro}.
Intuitively, a {\em proof tree} is a convenient way to represent the proof.
The root of the proof tree for an atomic query is the query itself. If a clause
$H\leftarrow B_1,\ldots, B_n$ has been used to resolve an atomic query $A$
via an mgu $\theta$, there is a directed edge (represented by $\leftarrow$ in 
the example to come) from a node representing $A\theta$ to a node
corresponding to a query $(B_1,\ldots,B_n)\theta$. The proof tree for 
a conjunctive query is a collection of proof trees for the individual 
conjuncts.

\begin{example}
\label{example:proof:tree}
The following meta-interpreter constructs a proof tree, while solving
a query. Proof trees are often used both for debugging~\cite{Naish:3valued} 
and explanation~\cite{Hammond,%
Arora:Ramakrishnan:Roth:Seshadri:Srivastava} purposes.
\begin{eqnarray*}
&& \mbox{\sl solve}(\mbox{\sl true}, \mbox{\sl true})\mbox{.}\\
&& \mbox{\sl solve}((A,B), (\mbox{\it ProofA},\mbox{\it ProofB}))\leftarrow
\mbox{\sl solve}(A,\mbox{\it ProofA}),\mbox{\sl solve}(B,\mbox{\it ProofB})\mbox{.}\\
&& \mbox{\sl solve}(A,(A\leftarrow \mbox{\it Proof}))\leftarrow \mbox{\sl clause}(A,B),\mbox{\sl solve}(B, \mbox{\it Proof})\mbox{.}
\end{eqnarray*}
This meta-interpreter is a double extended meta-interpreter 
(for all $k,l,p$ and $q$,
$C_{kl} = \mbox{\sl true}$ and $D_{pq} = \mbox{\sl true}$). 
$\hfill\Box$\end{example}

The study of double extended meta-interpreters might require different
clause encoding than the encoding 
considered in the previous section. For example,
reasoners with uncertainty might require that a {\em certainty factor} is
integrated in the encoding. Therefore, we adjust the definition of the clause
encoding as follows:
\begin{definition}(cf.\ Definition~\ref{definition:clause:encoding})
\label{definition:clause:encoding:1}
Let $D$ be a double extended meta-interpreter, 
let $P$ be a program. The {\em clause-encoding} $\mbox{\sl ce}^D(P)$
is a collection of facts of a new predicate $\mbox{\sl clause}$, 
such that for every clause $H\leftarrow B$ in $P$ there exists a
unique atom $\mbox{\sl clause}(H,B,s_{1}, \ldots, s_{k})\in 
\mbox{\sl ce}^D(P)$ and for every atom 
$\mbox{\sl clause}(H,B,s_{1}, \ldots, s_{k})\in \mbox{\sl ce}^D(P)$
there exists a clause $H\leftarrow B$ in $P$.
\end{definition}

In the remainder of the section, we discuss non-violation of LD-termination
and non-improving of LD-termination. 

\subsection{Non-violating LD-termination}
\label{section:non-violation}
We start with a discussion of non-violation of LD-termination. Before proving
this formally, we reconsider Example~\ref{example:proof:tree} and apply to it
the designed methodology. The treatment is done on the intuitive level.
A more formal discussion is postponed until Example~\ref{example:nv}.

\begin{example}
\label{example:proof:tree:nv}
Example~\ref{example:proof:tree}, continued. In order to prove that the
meta-inter\-preter does not violate LD-termination, we have to show that
for any definite program $P$, and for any query $Q$ terminating with respect
to $P$, $\mbox{\sl solve}(Q, u)$ terminates with respect to
$D\cup\mbox{\sl ce}^D(P)$ for any term $u$, where $D$ is the double extended 
meta-interpreter that constructs proof-trees.

Given that $Q$ terminates with respect to $P$, $\mbox{\sl solve}(Q)$
terminates with respect to $M_0\cup \mbox{\sl ce}(P)$ and 
there exists a quasi-ordering
$\geq$ such that $M_0\cup\mbox{\sl ce}(P)$ is order-acceptable with respect to 
$\mbox{\sl solve}(Q)$ via $\geq$.
Let $\succeq$ be a quasi-ordering on $B^E_{D\cup\mbox{\sl ce}^D(P)}$ defined
as follows for any atom $a$ and any terms $s,t,u_1$ and $u_2$:
\begin{itemize}
\item $\mbox{\sl solve}(s, u_1)\succ \mbox{\sl solve}(t, u_2)$ if $\mbox{\sl solve}(s) > \mbox{\sl solve}(t)$
\item $\mbox{\sl solve}(s, u_1)\succ a$ if $\mbox{\sl rel}(a)\neq \mbox{\sl solve}$
\item $a\preceq\succeq a$
\end{itemize}
\eat{
It follows from the corresponding property of $\geq$ and from 
the fact that no predicate is mutually recursive with $\mbox{\sl solve}$
that the relation $\succeq$ is indeed a well-founded quasi-ordering.
}

Next, we have to show that $D\cup\mbox{\sl ce}^D(P)$ is order-acceptable with 
respect to $\mbox{\sl solve}(Q, u)$ via $\succeq$. This claim is intuitively
clear since $\succeq$ is defined to ignore the extra argument of {\sl solve},
and if this argument is dropped from the clauses of $D$, $M_0$ is obtained.
$\hfill\Box$\end{example}

In order to formalise the intuition presented in 
Example~\ref{example:proof:tree:nv} we need to prove a number of
auxiliary statements.
First of all, observe that double extended meta-interpreters are
sound. 

\begin{lemma}
\label{lemma:de:sound}
Let $D$ be a double extended meta-interpreter. Then, $D$ is sound.
\end{lemma}
\begin{proof}
In order to prove the lemma, we use the $s$-semantics
approach presented in~\cite{Bossi:Gabbrielli:Levi:Martelli}. The semantics
is recalled and the formal 
proof of the lemma is presented in~\ref{appendix:non-violation}.
\end{proof}

As the following example demonstrates, unlike the ``vanilla'' 
meta-interpreter
$M_0$, double extended meta-interpreters do not necessarily preserve
the set of calls. However, one can show that there is a certain 
correspondence between the calls obtained with respect to a 
double extended meta-interpreter $D$ and the ``vanilla'' meta-interpreter
$M_0$. Recall that Lemma~\ref{lemma:vanilla:calls:preserved} established
that $M_0$ preserves the calls set.

\begin{example}
\label{example:instance:needed}
Let $D$ be the following double extended meta-interpreter and let $P$ be
the program of Example~\ref{example:clause:encoding}.
\begin{eqnarray*}
&& \mbox{\sl solve}(\mbox{\sl true})\mbox{.}\\
&& \mbox{\sl solve}((A,B))\leftarrow \mbox{\sl solve}(A), \mbox{\sl solve}(B)\mbox{.}\\
&& \mbox{\sl solve}(A)\leftarrow \mbox{\sl clause}(A,B), B = q(f(Z)), \mbox{\sl solve}(B)\mbox{.}
\end{eqnarray*}
The set of calls of $D\cup\mbox{\sl ce}^D(P)$ and $\mbox{\sl solve}(p(X))$ 
is $$\{\mbox{\sl solve}(p(X)), \mbox{\sl solve}(q(f(Z)),
\mbox{\sl clause}(p(X), q(X))\}\mbox{.}$$ The set of
calls for $M_0\cup\mbox{\sl ce}(P)$ and $\mbox{\sl solve}(p(X))$ is 
$$\{\mbox{\sl solve}(p(X)),\mbox{\sl solve}(q(X)),
\mbox{\sl solve}(\mbox{\sl true}), \mbox{\sl clause}(p(X), q(X)),
\mbox{\sl clause}(q(b), \mbox{\sl true})\}\mbox{.}$$
 There is no call in
$\mbox{\sl Call}(M_0\cup\mbox{\sl ce}(P),\mbox{\sl solve}(p(X)))$ that is a
{\em variant} of $\mbox{\sl solve}(q(f(Z))$. However, there is a call
$\mbox{\sl solve}(q(X))$ in 
$\mbox{\sl Call}(M_0\cup\mbox{\sl ce}(P),\mbox{\sl solve}(p(X)))$ 
such that $\mbox{\sl solve}(q(f(Z))$ is its instance.
$\hfill\Box$\end{example}

In Example~\ref{example:instance:needed} a correspondence was established
between the sets of calls obtained with respect to $D$ and those obtained 
with respect to $M_0$. Lemma~\ref{lemma:de:call:set}
proves that such a correspondence can always be established for
double extended meta-interpreters. Observe that, in general, not every sound 
meta-interpreter has this property (Example~\ref{example:Bossi}). 

\begin{lemma}
\label{lemma:de:call:set}
Let $D$ be a double extended meta-interpreter. Let $P$ be an interpreted 
program, $Q_0$ be an interpreted query and let $u_1,\ldots, u_n$ be a 
sequence of terms. Then, for every call $\mbox{\sl solve}(Q,t_1,$ 
$\ldots,t_n)$
in $\mbox{\sl Call}(D\cup\mbox{\sl ce}^D(P),\mbox{\sl solve}(Q_0,u_1,\ldots, u_n))$ there exists a call $\mbox{\sl solve}(G)$ in $\mbox{\sl Call}(M_0\cup\mbox{\sl ce}(P),Q_0)$ such that $Q$ is an instance of $G$.
\end{lemma}
\begin{proof}
The proof is similar to the proof of Lemma~\ref{lemma:vanilla:calls:preserved}
and can be found in ~\ref{appendix:non-violation}.
\end{proof}

\eat{
\begin{lemma}
\label{lemma:de:call:set}
Let $D$ be a double extended meta-interpreter. Let $P$ be an interpreted 
program, $Q_0$ be an interpreted query and let $u_1,\ldots, u_n$ be a sequence
of terms. Then, the following holds:
\begin{eqnarray*}
&& \forall\;\mbox{\sl solve}(Q,t_1,\ldots,t_n)\in \mbox{\sl Call}(D\cup\mbox{\sl ce}^D(P),\mbox{\sl solve}(Q_0,u_1,\ldots, u_n))\\
&&\hspace{1.0cm}\exists\;\mbox{\sl solve}(G)\in \mbox{\sl Call}(M_0\cup\mbox{\sl ce}(P),Q_0)\;\mbox{\rm such that}\;Q\;\mbox{\rm is an instance of}\;G
\end{eqnarray*}
\end{lemma}
\begin{proof}
We prove by induction on the derivation length the following statement:
for any derivation 
$\xi^D = \mbox{\sl solve}(Q_0,u_1,\ldots,u_n) \Rightarrow 
\ldots \mbox{\sl solve}(Q_k,t^k_1,\ldots,t^k_n)$, there exists a derivation
$\xi^{M_0} = \mbox{\sl solve}(G_0)\Rightarrow \ldots\Rightarrow 
\mbox{\sl solve}(G_l)$, such that $G_0 = Q_0$ and 
$Q_k$ is an instance of $G_l$.

If the length of $\xi^D$ is $1$, $Q_0$ is $\mbox{\sl true}$,
$t^1_1,\ldots,t^1_n$ are unifiable with $t_{11},\ldots,t_{1n}$,
and $C_{11},\ldots,C_{1m_1}$ are {\sl true} as well ($t_{11},\ldots,t_{1n}$
and $C_{11},\ldots,C_{1m_1}$ are as defined by 
Definition~\ref{definition:de}). Observe that
$\mbox{\sl solve}(Q_k,t^k_1,\ldots,t^k_n)$ is identical to
$\mbox{\sl solve}(Q_0, u_1,\ldots,u_n)$.
Then $G_0$ is
$\mbox{\sl true}$ and $\xi^{M_0}$ can be defined as 
$\mbox{\sl solve}(G_0)$, $G_l = G_0$. Clearly, $G_0 = Q_0$, and
$Q_k$ is an instance of $G_l$.

Assuming that the claim holds for the derivations of length $k\leq r$,
we are going to prove it for derivations of length $r+1$. Let
$\mbox{\sl solve}(Q_m,t^m_1,\ldots,t^m_n)$ be a selected atom of one of the
queries in $\xi^{D}$,
such that there exists a clause $(H\leftarrow B_1\ldots,B_s)\in D$,
$\mbox{\sl solve}(Q_m,t^m_1,\ldots,t^m_n)$ can be unified with $H$
(via an mgu $\theta$),
and the next call $\mbox{\sl solve}(Q_{r+1},t^{r+1}_1,\ldots,t^{r+1}_n)$ is an 
instance of $B_j$ for some $j$. We distinguish between the following cases:
\begin{itemize}
\item none of $B_1,\ldots,B_{j-1}$ is an atom of {\sl solve}.
Then, let $\sigma$ be a computed answer substitution for 
$(B_1,\ldots,B_{j-1})\theta$. Thus, if the variable denoting the 
first argument of $B_j$ is $A$, $Q_{r+1}$ is $A\theta\sigma$.

Since $m\leq r$ the inductive assumption can be used.
There exists a derivation $\mbox{\sl solve}(G_0)\Rightarrow \ldots \Rightarrow \mbox{\sl solve}(G_l)$ such that $Q_m$ is an instance of $G_l$, i.e., $Q_m = G_l\delta$ for some substitution $\delta$. By definition of $\theta$, $\theta$ unifies
$Q_m$ with the first argument of $H$. Thus, $\delta\theta$ unifies $G_l$
with the first argument, say $B$, of the head of the corresponding clause in 
$M_0$. Let $\theta'$ be an mgu of $B$ and $G_l$. The first argument of the 
call to the first recursive subgoal is then $G_{l+1} = A\theta'$. 
By definition of an mgu there exists a substitution $\rho$, such that 
$\theta'\rho = \gamma\theta$. Therefore, $Q_{r+1} = A\theta\sigma = A\theta'\rho = G_{l+1}\rho$.
In other words, $\mbox{\sl solve}(G_0)\Rightarrow \ldots \Rightarrow \mbox{\sl solve}(G_l) \Rightarrow \mbox{\sl solve}(G_{l+1})$
satisfies conditions of the lemma.
\item $\mbox{\sl rel}(B_p) = {\sl solve}$ for some $1\leq p\leq j-1$. 
Length of the derivation of $\mbox{\sl solve}(Q_m,t^m_1,\ldots,t^m_n)$
is less than $r+1$. Thus, inductive
assumption can be used. Let $\mbox{\sl solve}(G_0)\Rightarrow \ldots \Rightarrow \mbox{\sl solve}(G_l)$ be a derivation, such that $G_0 = Q_0$ and 
$Q_m$ is an instance of 
$G_l$, i.e., $Q_m = G_l\delta$ for some substitution $\delta$.
It should be observed that the call to $B_p$ is $B_p\theta\sigma^D$,
where $\sigma^D$ is a computed answer substitution for
$(B_1,\ldots,B_{p-1})\theta$. Let $\theta'$ be an mgu of 
$\mbox{\sl solve}(G_l)$ and the head of the corresponding clause in $M_0$.
Then, $\theta = \theta'\theta''$ for some substitution $\theta''$. 
Since $D$ is sound by Lemma~\ref{lemma:de:sound}, 
the call corresponding to $(B_1,\ldots,B_{p-1})\theta$
in $M_0$ computes $\sigma^{M_0}$, such that 
$\theta\sigma^D = \theta'\sigma^{M_0}\rho$ for some substitution $\rho$. 
Let $B$ be the first argument of $B_j$. 
The next call produced by $M_0$ is thus, 
$\mbox{\sl solve}(B\theta'\sigma^{M_0})$. We denote this as 
$\mbox{\sl solve}(G_{l+1})$.

Then, given the computed answer substitution $\mu$ for
$(B_{p+1},\ldots,B_{j-1})\theta\sigma^D$ the
following holds $Q_{r+1} = B\theta\sigma^D\mu = B\theta'\sigma^{M_0}\rho\mu =
G_{l+1}\rho\mu$. In other words
$\mbox{\sl solve}(G_0)\Rightarrow \ldots \Rightarrow \mbox{\sl solve}(G_l)\Rightarrow \mbox{\sl solve}(G_{l+1})$ satisfies 
conditions of the lemma.
\end{itemize}
\end{proof}
}

Now we are ready to prove that double extended meta-interpreters do not 
violate termination. 
\eat{ 
Our proof 
makes use of the well-known result of~\cite{Apt:Book}, presented in
Lemma~\ref{apt:cor:323}.

\begin{definition}{\bf (Definition 3.20) of~\cite{Apt:Book}}
Consider an SLD-derivation of step $Q\Rightarrow Q_1$ via a clause $c$
and the most general unifier $\theta$. We say that the SLD-derivation step 
$Q'\Rightarrow Q'_1$ via a clause $c$
and the most general unifier $\theta'$ is a {\em lift} of
$Q\Rightarrow Q_1$ if
\begin{itemize}
\item $Q$ is an instance of $Q'$,
\item in $Q$ and $Q'$ atoms in the same positions are selected,
\item $Q_1$ is an instance of $Q'_1$.
\end{itemize}
\end{definition}

The notion of a lift can be extended to SLD-derivations.

\begin{lemma}{\bf (Corollary 3.23) of~\cite{Apt:Book}}
\label{apt:cor:323}
For every successful SLD-derivation $\xi$ of $P\cup \{Q\eta\}$ with computed
answer substitution $\theta$ there exists a successful SLD-derivation $\xi'$
of $P\cup \{Q\}$ with computed answer substitution $\theta'$ such that
\begin{itemize}
\item $\xi'$ is a lift of $\xi$ of the same length as $\xi$,
\item $Q\theta'$ is more general than $Q\eta\theta$.
\end{itemize}
\end{lemma}
} 
Before presenting the result formally we 
need to introduce the following auxiliary definition.
\begin{definition}
Let $P$ be a definite program and let $Q$ be an atomic query,
such that $P$ is order-acceptable with respect to $\{Q\}$ via a 
quasi-ordering $\geq$. The quasi-ordering $\geq$ is called {\em minimal} 
if there exists no quasi-ordering $\geq_1$ such that $P$ is 
order-acceptable with respect to $\{Q\}$ via $\geq_1$ and 
$\geq_1\subset \geq$.
\end{definition}

First of all, we need to show the existence of a minimal ordering.
\begin{lemma}
Let $P$ be a definite program and let $Q$ be an atomic query. If 
$P$ is order-acceptable with respect to $\{Q\}$, there exists a 
minimal quasi-ordering $\geq$ such that $P$ is order-acceptable 
with respect to $\{Q\}$ via $\geq$.
\end{lemma}
\begin{proof}
Let ${\cal O}$ be the set of all quasi-orderings such that $P$ 
is order-acceptable with respect to $\{Q\}$ via them. Since $P$ 
is order-acceptable with respect to $\{Q\}$, ${\cal O}$ is not empty.
Hence we define a new quasi-ordering on $\mbox{\sl Call}(P,Q)$ as following:
\begin{eqnarray*}
A\preceq\succeq B &\mbox{\rm if}& A\;\mbox{\rm is identical to}\;B\\
A\succ B &\mbox{\rm if}& A > B\;\mbox{\rm for all $\geq\in {\cal O}$}
\end{eqnarray*}
It is straightforward to check that $\succeq$ is again a quasi-ordering.
We are going to show that  $P$ 
is order-acceptable with respect to $\{Q\}$ via $\succeq$, i.e.,
that $\succeq \in {\cal O}$. Indeed, let $A$ be in $\mbox{\sl Call}(P,Q)$,
$A'\leftarrow B_1,\ldots,B_n$ be a clause such that 
$\mbox{\sl mgu}(A,A') = \theta$ exists, $B_i$ a body subgoal such that
$\mbox{\sl rel}(A) = \mbox{\sl rel}(B_i)$ and $\sigma$ be a computed
answer substitution for $\leftarrow (B_1,\ldots,B_{i-1})\theta$. Then,
order-acceptability of $P$ with respect to $\{Q\}$ via the quasi-orderings
in ${\cal O}$ implies that for any $\geq\in {\cal O}$, $A > B_i\theta\sigma$.
By definition of $\succ$, it holds that $A\succ B_i\theta\sigma$. Hence,
$P$ is order-acceptable with respect to $\{Q\}$ via $\succeq$.

The construction of $\succeq$ above also implies immediately that this
quasi-ordering is minimal. 
\end{proof} 

Intuitively, minimal quasi-orderings contain decreases that are
essential for proving order-acceptability, and only these decreases.
To prove the statement formally we need the following notions introduced
by Verschaetse in~\cite{Verschaetse:thesis}.

\begin{definition}
\label{definition:direct:descendant}
\begin{itemize}
\item 
Let $Q_0, Q_1, Q_2,\ldots,\theta_1, \theta_2,\ldots$ be a derivation with 
selected atoms $A_0, A_1, $ $A_2,\ldots$ and applied renamed clauses
$H^i\leftarrow B^i_1,\ldots, B^i_{n_i}$ ($i = 1,2,\ldots$). We say that $A_k$ 
is a {\em direct descendant\/} of $A_i$, if $k > i$  and $A_k$ is the atom
$B^{i+1}_j\theta_{i+1}\ldots\theta_k$, ($1\leq j\leq n_{i+1}$).
\item
Let $Q_0, Q_1, Q_2,\ldots,\theta_1, \theta_2,\ldots$ be a derivation with 
selected atoms $A_0, A_1, $ $A_2,\ldots$. A {\em subsequence\/} of derivation 
steps, 
$Q_{i(0)}, Q_{i(1)}, \ldots, \theta_{i(0)+1}, \ldots$ with selected atoms 
$A_{i(0)}, A_{i(1)},$ $A_{i(2)},\ldots$ is 
{\em directed}, if for each $k$ ($k > 1$), $A_{i(k)}$ is a direct descendant
of $A_{i(k-1)}$ in the given derivation.
A derivation $Q_0, Q_1, Q_2,\ldots,\theta_1, \theta_2,\ldots$ is {\em directed\/} if it is its own directed subsequence.
\end{itemize}
\end{definition}

Verschaetse~\cite{Verschaetse:thesis} also proved the following lemma:
\begin{lemma}
\label{Verschaetse:lemma}
Let $P$ be a definite program and $A$ be an atomic query. If $P$ and $A$ have an 
infinite derivation, then they have an infinite directed derivation.
\end{lemma}

Now we are ready to prove the statement of Lemma~\ref{lemma:minimal:ds}
formally.
\begin{lemma}
\label{lemma:minimal:ds}
Let $P$ be a program and $Q$ be a query. Let $\geq$ be a minimal quasi-ordering
such that  $P$ is order-acceptable with respect to $\{Q\}$ via it. Then,
for all $A$ and $B$ in $\mbox{\sl Call}(P,Q)$, if $A > B$ then
there exists a directed derivation $Q_0, \ldots, Q_n$ with 
selected atoms $A_0, \ldots, A_n$, such that
$A_0 = A$, $A_n = B$ and for all $0\leq i < n$, 
$A_i > A_{i+1}$.
\end{lemma}
\begin{proof}
For the sake of contradiction assume that there exist $A$ and $B$ such that
$A > B$ and no directed derivation exists as required. We are going define a new 
quasi-ordering $\succeq$ on $\mbox{\sl Call}(P,Q)$ that will contradict the 
minimality of $\geq$.

Let $K\succ M$ be defined as a transitive closure of the following 
relation:
``$K > M$ and $M$ is a direct descendant of $K$''.
Let $K\preceq\succeq M$ if and only if $K$ is identical to $M$.

By definition of the notion of a direct descendant (Definition~\ref{definition:direct:descendant}) and order-acceptability of $P$ with respect to $\{Q\}$ 
via $\geq$, it follows that $P$ is order-acceptable with respect
to $\{Q\}$ via $\succeq$. Moreover, it is clear that 
$\succeq\;\subseteq\;\geq$. To complete the
proof we show that $A\not\succ B$. 

For the sake of contradiction assume that $A\succ B$. Then, 
since $\succ$ is defined as a transitive closure, there exists a
sequence of atoms $A_0,\ldots, A_n$ such that $A_i > A_{i+1}$ 
and $A_{i+1}$ is a direct descendant of $A_i$ for all $0\leq i < n$.
Since $A_i\in \mbox{\sl Call}(P,Q)$, $A_i$ is a selected atom of
some query $Q_i$. Thus, we have found a directed derivation $Q_0, \ldots,
Q_n$ as described by the lemma for $A$ and $B$. Therefore, 
our assumption was wrong and $A\not\succ B$. Hence, 
$\succeq\;\subset\;\geq$, that contradicts the minimality of $\geq$.
\end{proof}

Finally, we can prove the main result of this section, namely, that
double extended meta-interpreters
do not violate termination (under certain extra conditions).
Since the proof is long and technical only the general out-line is presented.
Technical details can be found in~\ref{appendix:non-violation}.

\begin{theorem}
\label{theorem:de:1}
Let $P$ be an interpreted program, $D$ a double extended 
meta-interpreter, and $Q\in B^E_{D\cup \mbox{\sl ce}^D(P)}$, 
such that $Q$ is terminating with respect to $P$.
Let $u_1,\ldots,u_n$ be a sequence of terms such that 
$\{A\mid A\in \mbox{\sl Call}(D\cup \mbox{\sl ce}^D(P), \mbox{\sl solve}(Q, u_1,\ldots,u_n)), \mbox{\sl solve}\neq \mbox{\sl rel}(A)\}$ is terminating 
with respect to $D$.  Then $\mbox{\sl solve}(Q, u_1,\ldots,u_n)$ terminates 
with respect to $D\cup \mbox{\sl ce}^D(P)$.
\end{theorem}
\begin{proof}[Proof (sketch)]
Let $M_0$ be the ``vanilla'' meta-interpreter. One can show that 
$M_0\cup \mbox{\sl ce}(P)$ is 
order-acceptable with respect to $S = \{A\eta\mid A\in \mbox{\sl Call}(M_0\cup \mbox{\sl ce}(P),\mbox{\sl solve}(Q)), \eta\;\mbox{\rm is a substitution}\}$. Let $\geq_1$ be a minimal well-founded quasi-ordering, such that 
$M_0\cup \mbox{\sl ce}(P)$ is order-acceptable with respect to $S$ via it. 
Similarly, let $\geq_2$ be a well-founded quasi-ordering such that $D$ is 
order-acceptable with respect to 
$\{A\mid A\in \mbox{\sl Call}(D\cup \mbox{\sl ce}^D(P), \mbox{\sl solve}(Q, u_1,
\ldots, u_n)), \mbox{\sl solve}\neq \mbox{\sl rel}(A)\}$ via $\geq_2$.

We have to show that there exists a well-founded quasi-ordering 
$\succeq$ such that
$D\cup \mbox{\sl ce}^D(P)$ is order-acceptable with respect to 
$\{\mbox{\sl solve}(Q, u_1, \ldots, u_n)\}$ via $\succeq$. 
By Theorem~\ref{taset:term} this will imply termination.

Let $\succeq$ be defined on $B^E_{D\cup\mbox{\sl ce}^D(P)}$ as follows
for any terms $t_1, t_2, t^1_{1},\ldots, t^1_{n}, t^2_{1},\ldots, t^2_{n}$ 
and any atoms $a_1, a_2$:
\begin{enumerate}
\item $\mbox{\sl solve}(t_1, t^1_{1},\ldots, t^1_{n}) \succ \mbox{\sl solve}(t_2, t^2_{1},\ldots, t^2_{n})$, if there is a term $t$, such that 
$\mbox{\sl solve}(t_1) >_1 \mbox{\sl solve}(t)$ and $t_2 = t\theta$ for some 
substitution $\theta$;
\item $a_1 \succ a_2$, if $\mbox{\sl rel}(a_1) \not = \mbox{\sl solve}$,
$\mbox{\sl rel}(a_2) \not = \mbox{\sl solve}$ and $a_1 >_2 a_2$;
\item $\mbox{\sl solve}(t_1, t^1_{1},\ldots, t^1_{n}) \succ a_1$, if 
$\mbox{\sl rel}(a_1) \not = \mbox{\sl solve}$;
\item $a_1 \preceq\succeq a_2$, if $a_1$ and $a_2$ are identical.
\end{enumerate}

First we show that $\succ$ is an ordering and that this ordering
is well-founded. To this end we make use of the minimality of $>_1$ 
and of the Lifting Theorem (Theorem 3.22~\cite{Apt:Book}). 
Next we prove that $D\cup \mbox{\sl ce}^D(P)$ is order-acceptable with 
respect to $\mbox{\sl solve}(Q, u_1, \ldots,$ $u_n)$. 
Let $A_0\in \mbox{\sl Call}(D\cup \mbox{\sl ce}^D(P), \mbox{\sl solve}(Q, 
u_1,\ldots, u_n))$. 
If $\mbox{\sl rel}(A_0) \not = \mbox{\sl solve}$ the desired decrease
follows from the order-acceptability of $D$ with respect to
$\{A\mid A\in \mbox{\sl Call}(D\cup \mbox{\sl ce}^D(P),$ $\mbox{\sl solve}(Q,$ $u_1,\ldots, u_n)),$ 
$\mbox{\sl solve}\neq \mbox{\sl rel}(A)\}$ via $>_2$. If
$\mbox{\sl rel}(A_0) = \mbox{\sl solve}$ each clause defining {\sl solve}
has to be considered separately. In each one of the cases we show that
if all arguments but the first one are dropped from $A_0$ the resulting
atom belongs to $S$ and hence order-acceptability via $\geq_1$ can be used.
Next we make use of Lemma~\ref{lemma:de:call:set} and complete the proof
by reasoning on the substitutions involved.
\end{proof}

This theorem provides an important theoretical result, namely, that
double extended meta-interpreters do not violate termination if the
interpreter terminates with respect to a set of calls to predicates
different from {\sl solve} and generated by the meta-program and the
meta-query. 

\begin{example}
\label{example:nv}
Meta-interpreters presented in Examples~\ref{example:4port:box},
\ref{example:proof:tree} and~\ref{example:instance:needed} are
double extended meta-inter\-pre\-ters. Termination of the calls to 
predicates different from {\sl solve} is immediate because in all
the examples these predicates can be completely unfolded, i.e.,
they do not depend (directly or indirectly) on recursive predicates.
Thus, by Theorem~\ref{theorem:de:1} meta-interpreters of
Examples~\ref{example:4port:box}, \ref{example:proof:tree} 
and~\ref{example:instance:needed} do not violate termination.

The quasi-ordering derived by Theorem~\ref{theorem:de:1} for
Example~\ref{example:proof:tree} has been discussed in 
Example~\ref{example:proof:tree:nv}.
$\hfill\Box$\end{example}

In general, to ensure that given a program $P$ a 
double extended meta-interpreter $D$ terminates with respect to
$\{A\mid A\in \mbox{\sl Call}(D\cup \mbox{\sl ce}^D(P), \mbox{\sl solve}(Q, u_1,\ldots,u_n)), \mbox{\sl solve}\neq \mbox{\sl rel}(A)\}$ one can require
$D$ to terminate for all calls to predicates different from {\sl solve}.
To verify the latter condition one can use the notion of order-acceptability
(Theorem~\ref{taset:term}).

\subsection{Non-improving LD-termination}
In this section we are going to study the opposite direction of the 
implication, namely, given that a meta-program 
$D\cup \mbox{\sl ce}^D(P)$ terminates with respect to
$\mbox{\sl solve}(Q,$ $u_1,\ldots,u_n)$ we would like to prove that 
\begin{itemize}
\item the meta-interpreter $D$ terminates with respect to $\{A\mid A\in \mbox{\sl Call}(D\cup \mbox{\sl ce}^D(P),\\ \mbox{\sl solve}(Q,u_1,\ldots,u_n)), \mbox{\sl solve} \neq \mbox{\sl rel}(A)\}$
\item the interpreted program $P$ terminates with respect to the interpreted
query $Q$
\end{itemize}

\eat{
Our proof of the first statement is based on 
Proposition~\ref{tb:prop:LDterm}. 
\begin{proposition}
\label{lemma:de:D}
Let $P$ be an interpreted program and $D$ a double extended meta-inter\-preter,
such that $D\cup \mbox{\sl ce}^D(P)$ is terminating with respect to
$\mbox{\sl solve}(Q,$ $u_1,\ldots,u_n)$. Then, 
$D$ terminates with respect to
$\{A\mid A\in \mbox{\sl Call}(D\cup \mbox{\sl ce}^D(P), \mbox{\sl solve}(Q,$ $u_1,\ldots,u_n)), \mbox{\sl solve} \neq \mbox{\sl rel}(A)\}$.
\end{proposition}
\begin{proof}
Termination 
with respect to $\mbox{\sl solve}(Q,u_1,\ldots,u_n)$ means termination 
with respect to $\mbox{\sl Call}(D\cup \mbox{\sl ce}^D(P), \mbox{\sl solve}(Q,u_1,\ldots,u_n))$. Next, by Proposition~\ref{tb:prop:LDterm},
termination with respect to a set of queries $S$
always implies termination with respect to a set $S_1\subseteq S$.
Since
\begin{eqnarray*}
&& \{A\mid A\in \mbox{\sl Call}(D\cup \mbox{\sl ce}^D(P), \mbox{\sl solve}(Q,u_1,\ldots,u_n)), 
\mbox{\sl solve}\neq \mbox{\sl rel}(A)\} \\
&&\hspace{1.0cm} \subset\; \mbox{\sl Call}(D\cup \mbox{\sl ce}^D(P), \mbox{\sl solve}(Q,u_1,\ldots,u_n))
\end{eqnarray*}
$D\cup \mbox{\sl ce}(P)$ terminates with respect to
\[\{A\mid A\in \mbox{\sl Call}(D\cup \mbox{\sl ce}^D(P), \mbox{\sl solve}(Q,u_1,\ldots,u_n)), 
\mbox{\sl solve}\neq \mbox{\sl rel}(A)\}.\] 
Moreover, by Proposition~\ref{tb:prop:LDterm},
$D$ terminates with respect to
$\{A\mid A\in \mbox{\sl Call}(D\cup \mbox{\sl ce}^D(P), \mbox{\sl solve}(Q,u_1,
\ldots,u_n)), \mbox{\sl solve}\neq \mbox{\sl rel}(A)\}$,
because of $D\subseteq (D\cup \mbox{\sl ce}^D(P))$.
\end{proof}
}

To prove the first statement, observe that $\{A\mid A\in \mbox{\sl Call}(D\cup \mbox{\sl ce}^D(P), \mbox{\sl solve}(Q,u_1,\ldots,u_n)),$ $\mbox{\sl solve} \neq \mbox{\sl rel}(A)\}$ is a subset of $\mbox{\sl Call}(D\cup \mbox{\sl ce}^D(P), \mbox{\sl solve}(Q, u_1,\ldots,u_n))$. Since $D\cup \mbox{\sl ce}^D(P)$ terminates
with respect to the latter set, it terminates with respect to the former set
as well. This implies that any subset of $D\cup \mbox{\sl ce}^D(P)$ terminates 
with respect to the same set of queries.

For general double extended meta-interpreter the
second statement we would like to prove, i.e., 
{\em ``if the meta-program terminates then
the interpreted program terminates as well''}
does not necessarily hold. Indeed, one can find many
double extended meta-inter\-pre\-ters that are designed to improve termination.
However, we are interested in termination non-improvement and would like
to establish conditions that ensure it.
Given a non-terminating interpreted program, termination of a meta-program 
can result from one of the following problems:
\begin{itemize}
\item failure while unifying a call and a clause head.
\begin{example}
\label{example:meta:ab}
Indeed, consider the following meta-interpreter $D$:
\begin{eqnarray*}
&& \mbox{\sl solve}(\mbox{\sl true}, a)\mbox{.}\\
&& \mbox{\sl solve}((A,B), a) \leftarrow \mbox{\sl solve}(A, a), \mbox{\sl solve}(B, a)\mbox{.}\\
&& \mbox{\sl solve}(A, a) \leftarrow \mbox{\sl clause}(A,B), \mbox{\sl solve}(B, a)\mbox{.}
\end{eqnarray*}
Let $P$ be an interpreted program, such that $Q$ does not terminate with 
respect to it. However, $\mbox{\sl solve}(Q,b)$ terminates with respect to
$D\cup\mbox{\sl ce}^D(P)$.
$\hfill\Box$\end{example}
To eliminate this problem we require that, for every call, the unification
success or failure with the head of the clause depends only on 
their first arguments. In general, predicting success of the unification
during the execution is known to be an undecidable problem. However, sufficient
conditions ensuring unification success can be proposed.
\item failure of intermediate body subgoals. 
\begin{example}
Indeed, consider the following meta-interpreter $D$:
\begin{eqnarray*}
&& \mbox{\sl solve}(\mbox{\sl true})\mbox{.}\\
&& \mbox{\sl solve}((A,B)) \leftarrow \mbox{\sl solve}(A), \mbox{\sl solve}(B)\mbox{.}\\
&& \mbox{\sl solve}(A) \leftarrow \mbox{\sl fail}, \mbox{\sl clause}(A,B), \mbox{\sl solve}(B)\mbox{.}
\end{eqnarray*}
Let $P$ be an interpreted program, such that $Q$ does not terminate with 
respect to it. However, $\mbox{\sl solve}(Q)$ terminates with respect to
$D\cup\mbox{\sl ce}^D(P)$.
$\hfill\Box$\end{example}

To solve this problem, one has to guarantee non-failure of 
the intermediate body subgoals. The general problem of non-failure
analysis is well-known to be 
undecidable~\cite{Debray:LopezGarcia:Hermenegildo}. Fortunately,
the problem is decidable for a restricted class of 
problems~\cite{Debray:LopezGarcia:Hermenegildo}. For the specific
meta-interpreters
we consider, failure of the corresponding intermediate body subgoals
turns out to be decidable.

It should also be noted that failure of the body atoms to the right of 
the last recursive call may influence termination as well. 
\begin{example}
\begin{eqnarray*}
&& \mbox{\sl solve}(\mbox{\sl true}) \leftarrow \mbox{\sl fail}\mbox{.}\\
&& \mbox{\sl solve}((A,B)) \leftarrow \mbox{\sl solve}(A), \mbox{\sl solve}(B)\mbox{.}\\
&& \mbox{\sl solve}(A) \leftarrow \mbox{\sl clause}(A,B), \mbox{\sl solve}(B)\mbox{.}
\end{eqnarray*}
Let $P$ be the following interpreted program:
\begin{eqnarray*}
&& r\leftarrow p, r\mbox{.}\\
&& p\mbox{.}
\end{eqnarray*}
Clearly, the query $r$ does not terminate with 
respect to $P$. However, $\mbox{\sl solve}(r)$ terminates with respect to
$D\cup\mbox{\sl ce}^D(P)$.
$\hfill\Box$\end{example}
Therefore, non-failure is required also for these subgoals.
\item different call sets. In principle, even if no failure occurs 
during the execution, $D$ can
change the call set, as the following example illustrates:
\begin{example}
\label{example:meta:Ap0}
\begin{eqnarray*}
&& \mbox{\sl solve}(\mbox{\sl true})\mbox{.}\\
&& \mbox{\sl solve}((A,B))\leftarrow A = p(0), \mbox{\sl solve}(A), \mbox{\sl solve}(B)\mbox{.}\\
&& \mbox{\sl solve}(A)\leftarrow \mbox{\sl clause}(A,B), \mbox{\sl solve}(B)\mbox{.}
\end{eqnarray*}
Let $P$ be the following program: 
\begin{eqnarray*}
&& q\leftarrow p(X),r\mbox{.} \\
&& p(f(X))\leftarrow p(X)\mbox{.}\\
&& p(0)\mbox{.}\\
&& r \mbox{.} 
\end{eqnarray*}
The termination behaviour of $q$ with respect to $P$ differs from the 
termination behaviour of $\mbox{\sl solve}(q)$ with respect to 
$D\cup \mbox{\sl ce}(P)$. The former computation does not terminate,
while the latter terminates. The reason for this is that $D$ changes
the set of calls, due to the unification
$A = p(0)$ in the body of the second clause. 
Indeed, the call set of $P$ with respect to $\{q\}$ contains 
$p(f(X))$, while the call set of $D\cup \mbox{\sl ce}(P)$ with respect 
to $\mbox{\sl solve}(q)$ does not contain the corresponding atom
$\mbox{\sl solve}(p(f(X)))$. 
$\hfill\Box$\end{example}
Similar problem occurs in Example~\ref{example:instance:needed}.
To solve this problem one has to ensure that meta-variables are not affected
by the intermediate body atoms. 
\eat{
Unfortunately, even this property is not local,i.e., if it holds
for each one of the clauses separately, it does not necessarily hold
for the entire program. Consider the following example:
\begin{example}
\begin{eqnarray*}
&& \mbox{\sl solve}(\mbox{\sl true}, \mbox{\sl true})\mbox{.}\\
&& \mbox{\sl solve}((A,B),X)\leftarrow X = (A,B), Y = A, \mbox{\sl solve}(A,Y), Z = B, \mbox{\sl solve}(B,Z)\mbox{.}\\
&& \mbox{\sl solve}(A, X)\leftarrow X = p(0), \mbox{\sl clause}(A,B), Z = B, \mbox{\sl solve}(B,Z).
\end{eqnarray*}
Let $P$ be the following program: 
\begin{eqnarray*}
&& q\leftarrow p(X),r\mbox{.}\\
&& p(f(X))\leftarrow p(X)\mbox{.}\\
&& p(0)\mbox{.}\\
&& r.
\end{eqnarray*}
Observe that $Y = A$ does not affect $A$, since $Y$ is a fresh variable.
Similarly, when the third clause is considered, $X = p(0)$ also cannot
influence the meta-variables. However, combining the clauses results in
$Y = A$ (the second clause), $Y = X_1, A = A_1$ (unification with the
head of the third clause; for the sake of convenience variables in the third
clause are subscribed), and $X_1 = p(0)$. Namely, $A_1$ is unified with $p(0)$,
i.e., sequence of the unifications above changed this meta-variable.
$\hfill\Box$\end{example}
Therefore, we restrict our attention only to meta-interpreters such that 
their intermediate body atoms either do not mention meta-variables, or
if the meta-variables are mentioned, no extra bindings can be created.
}
\end{itemize}

We summarise the discussion above in the following definition.
\begin{definition}
\label{definition:de:restricted}
Let $D$ be the following double extended meta-interpreter:
\begin{eqnarray*}
&& \mbox{\sl solve}(\mbox{\sl true}, t_{11}, \ldots, t_{1n})\leftarrow C_{11},\ldots, C_{1m_1}\mbox{.}\\
&& \mbox{\sl solve}((A,B), t_{21}, \ldots, t_{2n})\leftarrow \\
&& \hspace{1.0cm}D_{11},\ldots,D_{1k_1}, \mbox{\sl solve}(A, t_{31}, \ldots, t_{3n}), \\
&& \hspace{1.0cm}D_{21},\ldots,D_{2k_2}, \mbox{\sl solve}(B, t_{41}, \ldots, t_{4n}),\\
&& \hspace{1.0cm} C_{21},\ldots, C_{2m_2}\mbox{.}\\
&& \mbox{\sl solve}(A, t_{51}, \ldots, t_{5n})\leftarrow \\
&& \hspace{1.0cm}D_{31},\ldots,D_{3k_3},\mbox{\sl clause}(A,B,s_{1}, \ldots, s_{k}), \\
&& \hspace{1.0cm}D_{41},\ldots,D_{4k_4},\mbox{\sl solve}(B, t_{61}, \ldots, t_{6n}),\\
&& \hspace{1.0cm} C_{31},\ldots, C_{3m_3}\mbox{.}
\end{eqnarray*}
(together with the clauses defining $C_{kl}$ and $D_{pq}$), such that
\begin{itemize}
\item for any computed answer of the preceding atoms,
\begin{itemize}
\item either the corresponding instances of 
$(t_{31}, \ldots, t_{3n})$, $(t_{41}, \ldots, t_{4n})$,
$(t_{61}, \ldots, $ $t_{6n})$, and
$(s_{1}, \ldots, s_{k})$ are linear sequences of free variables,
\item or $(t_{11}, \ldots, t_{1n})$, $(t_{21}, \ldots, t_{2n})$, 
$(t_{51}, \ldots, t_{5n})$, are linear sequences of free variables
and for every program $P$ and for every
$\mbox{\sl clause}(s,t,t_1,\ldots,t_k)\in \mbox{\sl ce}^D(P)$,
$t_1,\ldots,t_k$ is a linear sequence of free variables;
\end{itemize}
\item $D_{11},\ldots,D_{1k_1}$, $D_{21},\ldots,D_{2k_2}$, $D_{31},\ldots,D_{3k_3}$, $D_{41},\ldots,D_{4k_4}$, $C_{11},\ldots, C_{1m_1}$, $C_{21},\ldots, $
 $C_{2m_2}$, and $C_{31},\ldots, C_{3m_3}$ do not fail for the corresponding calls (independently on the values of the meta-variables), i.e.,
for all $P$, for all $Q$ and for all $t_{1},\ldots,t_{n}$, 
calls to $C_{kl}$ and $D_{pq}$ in
$\mbox{\sl Call}(D\cup \mbox{\sl ce}^D(P),\mbox{\sl solve}(Q,t_{1},\ldots,t_{n})$
do not fail;
\item for any instance $(D_{i1},\ldots,D_{ik_i})\theta$ of 
$D_{i1},\ldots,D_{ik_i}$ and for
any computed answer substitution $\sigma$ for 
$(D_{i1},\ldots,D_{ik_i})\theta$, 
$A\theta\sigma$ is identical to $A\theta$ and $B\theta\sigma$ is 
identical to $B\theta$.
\end{itemize}
Then $D$ is called {\em restricted}.
\end{definition}

It should be noted that Definition~\ref{definition:de:restricted} is not 
syntactical. However, syntactical conditions implying it can
be proposed. We postpone discussing them  until after we formulate the 
termination non-violation theorem (Theorem~\ref{theorem:de:2}). 

We also introduce a notion of a {\em restricted query}, corresponding
to Definition~\ref{definition:de:restricted}.
\begin{definition}
\label{definition:de:restricted:query}
Given a restricted double extended meta-interpreter $D$ and a query 
$\mbox{\sl solve}(Q, v_1, \ldots, v_n)$, the query is called {\em
restricted} if:
\begin{itemize}
\item either  $(v_1, \ldots, v_n) \in \mbox{\it Vars}_n$,
\item or $(t_{11}, \ldots, t_{1n}), (t_{21}, \ldots, t_{2n}),
(t_{51}, \ldots, t_{5n})\in \mbox{\it Vars}_n$. 
\end{itemize}
\end{definition}

\begin{example}
\label{example:proof:tree:rq}
Recall the meta-interpreter that constructs proof-trees we considered in
Example~\ref{example:proof:tree}. Note that this meta-interpreter
is restricted. Observe that $t_{11}$ is {\em true}, i.e., 
$t_{11}\not\in \mbox{\it Vars}_n$. Thus, for a query 
$\mbox{\sl solve}(Q, u)$ to be restricted, $u$ should be in 
$\mbox{\it Vars}_1$, i.e., $u$ should be a free variable.
$\hfill\Box$\end{example}

First of all, we are going to
see that the condition imposed on the arguments of the head (or
recursive body subgoals) ensures the requirement stated after 
Example~\ref{example:meta:ab}.

\begin{lemma}
\label{lemma:de:un}
Let $D$ be a restricted double extended meta-interpreter, $P$
be an interpreted program, $Q_0$ be an interpreted query and
$(v_1, \ldots, v_n)$ a sequence of terms such that 
$\mbox{\sl solve}(Q_0, v_1, \ldots, v_n)$ is restricted. Then,
for any call $\mbox{\sl solve}(Q, u_1, \ldots, u_n)\in
\mbox{\sl Call}(D\cup \mbox{\sl ce}^D(P),\mbox{\sl solve}(Q_0, v_1, \ldots, v_n))$ and for any $\mbox{\sl solve}(H, t_1,\ldots, t_n)\leftarrow B_1,\ldots, B_k$ in $D$, $u_1, \ldots, u_n$ is unifiable with
$t_1,\ldots, t_n$.
\end{lemma}
\begin{proof}
We distinguish between two cases. 

If $(t_{11}, \ldots, t_{1n})\not\in \mbox{\it Vars}_n$, then,
$(v_1, \ldots, v_n)\in \mbox{\it Vars}_n$. Hence, the unification step
succeeds and if $\theta$ is the mgu, then $(t_{11}, \ldots, t_{1n})\theta = 
(t_{11}, \ldots, t_{1n})$. The first condition in Definition~\ref{definition:de:restricted}
implies that for any
computed answer of the preceding atoms, the corresponding instances of 
$(t_{31}, \ldots, t_{3n})$, $(t_{41}, \ldots, t_{4n})$,
$(t_{61}, \ldots, t_{6n})$, and
$(s_{1}, \ldots, s_{k})$ are linear sequences of free variables. Thus,
for any call $\mbox{\sl solve}(Q, u_1, \ldots, u_n)\in
\mbox{\sl Call}(D\cup \mbox{\sl ce}^D(P),\mbox{\sl solve}(Q_0, v_1, \ldots, v_n))$, $u_1, \ldots, u_n$ is a linear sequence of free variables.
Therefore, for any sequence of terms $t_1,\ldots, t_n$, the
sequence $u_1, \ldots, u_n$ can be unified with $t_1,\ldots, t_n$.

If $(t_{11}, \ldots, t_{1n})\in \mbox{\it Vars}_n$, then the second subcase of the
first condition in Definition~\ref{definition:de:restricted} holds, i.e., 
$(t_{21}, \ldots, t_{2n}), (t_{51}, \ldots, t_{5n})\in \mbox{\it Vars}_n$
as well. Recall, that $\mbox{\sl solve}(H, $ $t_1,\ldots, t_n)\leftarrow B_1,\ldots, B_k$ a clause in $D$. Thus, $t_1,\ldots, t_n$ is one of the 
$(t_{11}, \ldots, t_{1n}), (t_{21}, $ 
$\ldots, t_{2n}), (t_{51}, \ldots,$ $t_{5n})$.
In other words, $(t_1,\ldots, t_n)$ is a linear sequence of free
variables. Therefore, for any sequence of atoms $(u_1,\ldots, u_n)$,
$(t_1,\ldots, t_n)$ is unifiable with it.
\end{proof}

The remainder of the section is dedicated to the proof that restricted
double extended meta-interpreters do not improve termination. 
In order to
provide some intuition how the actual proof will proceed we recall
Example~\ref{example:proof:tree} and show that it does not improve
termination. The treatment is done on the intuitive level. More precise 
discussion of this example can be found in Example~\ref{example:ni}.

\begin{example}
\label{example:proof:tree:ni}
Example~\ref{example:proof:tree}, continued. 
In order to prove that this meta-interpreter
does not improve LD-termination, we have to show that
for any definite program $P$, and for any query $Q$ 
if a restricted query $\mbox{\sl solve}(Q, u)$ terminates with respect to
$D\cup\mbox{\sl ce}^D(P)$, then $Q$ terminates
with respect to $P$. Observe that the requirement that 
$\mbox{\sl solve}(Q, u)$ should be 
restricted means that $u$ should be a free variable (Example~\ref{example:proof:tree:rq}).

Given that $\mbox{\sl solve}(Q, u)$ terminates with respect to 
$D\cup \mbox{\sl ce}^D(P)$ 
there exists a quasi-ordering
$\geq$ such that $D\cup\mbox{\sl ce}^D(P)$ is order-acceptable with respect to 
$\mbox{\sl solve}(Q, u)$ via $\geq$.
Let $\succeq$ be a quasi-ordering on $B^E_{M_0\cup\mbox{\sl ce}(P)}$ defined
as follows: for any atom $a$, it holds that $a\preceq\succeq a$, and
for any terms $s,t,u_1$ and $u_2$, 
if $\mbox{\sl solve}(s, u_1) > \mbox{\sl solve}(t, u_2)$ then $\mbox{\sl solve}(s) \succ \mbox{\sl solve}(t)$.
Next, one has to show that $\succeq$ is a well-founded quasi-ordering
and that $M_0\cup\mbox{\sl ce}(P)$ is order-acceptable with respect to 
$\mbox{\sl solve}(Q)$. Since $D$ is restricted, termination cannot
be enforced by the information contained in the second argument of
{\sl solve}. Hence, both well-foundedness and order-acceptability 
follow from the corresponding properties of $\geq$.
$\hfill\Box$\end{example}

In order to prove that restricted double extended meta-interpreters do not 
improve termination, we would like to show the completeness result for 
this class of meta-interpreters. However, as the following example illustrates,
completeness does not necessarily hold.
\begin{example}
\label{example:no:completeness}
Consider the following restricted double extended meta-interpreter.
\begin{eqnarray*}
&& \mbox{\sl solve}(\mbox{\sl true}, X)\mbox{.}\\
&& \mbox{\sl solve}((A,B), X)\leftarrow \mbox{\sl solve}(A, Y), \mbox{\sl solve}(B, Z)\mbox{.}\\
&& \mbox{\sl solve}(A, X)\leftarrow p, \mbox{\sl clause}(A,B), \mbox{\sl solve}(B, Y)\mbox{.}\\
&& p\leftarrow p\mbox{.}
\end{eqnarray*}
This meta-interpreter is not complete due to its non-termination.
$\hfill\Box$\end{example}

Therefore, we prove a more restricted result that will be sufficient for
showing termination non-improvement.
\begin{lemma}
\label{lemma:de:complete}
Let $D$ be a restricted double extended meta-interpreter, $P$ be an interpreted
program, $Q$ be an interpreted query and $(v_1,\ldots,v_n)$ be terms, 
such that $\mbox{\sl solve}(Q, v_1,\ldots,v_n)$ is restricted, and
$D\cup\mbox{\sl ce}^D(P)$ terminates with respect to 
$\{A\mid A\in \mbox{\sl Call}(D\cup \mbox{\sl ce}^D(P), \mbox{\sl solve}(Q,v_1,$
 $\ldots,v_n)), \mbox{\sl solve}\neq \mbox{\sl rel}(A)\}$. Then,
for every computed answer $t$ for $Q$ obtained with respect to $P$,
there exists a computed answer
$\mbox{\sl solve}(t',t_1,\ldots,t_n)$ for 
$\mbox{\sl solve}(Q,v_1,\ldots,v_n)$ with respect to $D\cup \mbox{\sl ce}^D(P)$,
such that $t$ is a variant of $t'$.
\end{lemma}
\begin{proof}
The proof, analogous to Lemma~\ref{lemma:de:sound},
can be found in~\ref{appendix:non-improvement}.
\end{proof}

Lemma~\ref{lemma:de:complete} implies that there is a one-to-one 
correspondence between the calls to {\sl solve} obtained with respect 
to $D$ and those obtained with respect to $M_0$. More formally,
there is a call $\mbox{\sl solve}(Q)$
in $\mbox{\sl Call}(M_0\cup\mbox{\sl ce}(P),\mbox{\sl solve}(Q_0))$ 
if and only if, there is a call $\mbox{\sl solve}(Q', u_1, \ldots, $ $u_n)$ 
in the set $\mbox{\sl Call}(D\cup\mbox{\sl ce}^D(P),\mbox{\sl solve}(Q_0,
v_1, \ldots, v_n))$, such that $Q$ and $Q'$ are variants. Indeed, if 
there are no intermediate calls
to $\mbox{\sl solve}$, the claim follows from 
Definition~\ref{definition:de:restricted}. Otherwise, it follows
from the preceding Lemma~\ref{lemma:de:complete}. 

\eat{
Before proving the termination preservation result for restricted 
double extended meta-interpreters, we establish an important property of
orderings that are used to prove termination of the corresponding 
meta-programs. Recall that by a minimal (quasi-)ordering satisfying some
property we understand a (quasi-)ordering such that no it proper subset
satisfies the given property.

\begin{lemma}
\label{lemma:de:ordering:independent}
Let $D$ be a double extended meta-interpreter, $P$ be an interpreted program,
$Q$ be an interpreted query and $(v_1,\ldots,v_n)$ be a sequence of terms,
such that $\mbox{\sl solve}(Q, v_1,\ldots,v_n)$ is restricted, and
$D\cup\mbox{\sl ce}^D(P)$ is order-acceptable with respect to 
$\{\mbox{\sl solve}(Q, v_1,\ldots,v_n)\}$ via a minimal ordering $>$. 

Then, 

for any terms $t, s_1,\ldots, s_n, t_1,\ldots,t_n$,
such that $\mbox{\sl solve}(t,s_1,\ldots,s_n)\in 
\mbox{\sl Call}(D\cup\mbox{\sl ce}^D(P), \mbox{\sl solve}(Q, v_1,\ldots,v_n))$
and $\mbox{\sl solve}(t,t_1,\ldots,t_n)\in 
\mbox{\sl Call}(D\cup\mbox{\sl ce}^D(P), \mbox{\sl solve}(Q, v_1,\ldots,v_n))$
 either
\[
\mbox{\sl solve}(t,s_1,\ldots,s_n)\leq\geq \mbox{\sl solve}(t,t_1,\ldots,t_n)
\]
or
\[
\mbox{\sl solve}(t,s_1,\ldots,s_n)\|_\geq \mbox{\sl solve}(t,t_1,\ldots,t_n)
\]
\end{lemma}
\begin{proof}
Similarly to Lemma~\ref{lemma:de:un} two cases are distinguished.
If $(t_{11},\ldots,t_{1n})\not \in \mbox{\it Vars}_n$, then all 
\end{proof}
} 
Now we are ready
to prove that restricted meta-interpreters preserve termination.
Similarly to Theorem~\ref{theorem:de:1} only a proof sketch is included,
technical details can be found in~\ref{appendix:non-improvement}.

\begin{theorem}
\label{theorem:de:2}
Let $D$ be a restricted double extended meta-interpreter.
Let $P$ be an interpreted program and let $Q$ be an interpreted query,
such that $D\cup\mbox{\sl ce}^D(P)$ LD-terminates for $\mbox{\sl solve}(Q,
v_1,\ldots,$ $v_n)$, where $(v_1,\ldots,v_n)$ are terms such that
$\mbox{\sl solve}(Q, v_1,\ldots,v_n)$ is restricted.
Then, $P$ LD-terminates with respect to $Q$.
\end{theorem}
\begin{proof}[Proof (sketch)]
In order to show that $P$ LD-terminates for $Q$ it is sufficient to 
prove that $M_0\cup\mbox{\sl ce}(P)$ LD-terminates with respect to
$\mbox{\sl solve}(Q)$.
Then, by Theorem~\ref{meta:interpreted} $P$ LD-terminates with respect 
to $Q$. Thus, we aim to establish order-acceptability of 
$M_0\cup\mbox{\sl ce}(P)$ with respect to $\mbox{\sl solve}(Q)$.

First of all, we define a relationship on $B^E_{M_0\cup \mbox{\sl ce}(P)}$.
Since $D\cup\mbox{\sl ce}^D(P)$ LD-terminates for $\mbox{\sl solve}(Q,v_1,\ldots,v_n)$, $D\cup\mbox{\sl ce}^D(P)$ is order-accep\-tab\-le with respect to 
$\mbox{\sl solve}(Q,v_1,\ldots,v_n)$ via a quasi-ordering. Let a minimal
quasi-ordering such that $D\cup\mbox{\sl ce}^D(P)$ is order-accep\-tab\-le with 
respect to $\mbox{\sl solve}(Q,$ $v_1,\ldots,v_n)$ via it,
be denoted $\geq$. Then, we define 
$\mbox{\sl solve}(s) \succ \mbox{\sl solve}(t)$
if there exist $\mbox{\sl solve}(s, s_1,\ldots, s_n), \mbox{\sl solve}(t, t_1,$ $\ldots, t_n)\in$ $\mbox{\sl Call}(D\cup\mbox{\sl ce}^D(P), \mbox{\sl solve}(Q,$ $v_1,\ldots,v_n))$ such that $\mbox{\sl solve}(s, s_1,\ldots, s_n) > 
\mbox{\sl solve}(t, t_1,$ $\ldots, t_n)$ and $\mbox{\sl solve}(s) \preceq\succeq \mbox{\sl solve}(t)$ if $\mbox{\sl solve}(s)$ and $\mbox{\sl solve}(t)$ are identical.

One can show that $\succ$ is indeed a well-founded ordering. In order to
show that $M_0\cup \mbox{\sl ce}(P)$ is order-acceptable with respect to
$\mbox{\sl solve}(Q)$ via $\succeq$ we make use of 
Lemmata~\ref{lemma:de:un} and ~\ref{lemma:de:complete},
and of Definition~\ref{definition:de:restricted}.
\end{proof}

\begin{example}
\label{example:ni}
Example~\ref{example:4port:box}, continued. The meta-inter\-preter is restricted and Theorem~\ref{theorem:de:2} ensures that it preserves termination.
However, if the clause (\ref{example:debugger:non-fail}) would have been removed, the meta-interpreter would no longer
be restricted. Indeed, the call to $\mbox{\sl before}$ would 
fail, violating the second requirement of Definition~\ref{definition:de:restricted}. Thus, Theorem~\ref{theorem:de:2} would not have been applicable.
It should be noted that indeed, this meta-interpreter improves termination. 
\eat{
\item The meta-interpreter that constructs proof trees (Example~\ref{example:proof:tree}). The meta-inter\-preter is restricted. Thus, given a restricted
query of the form $\mbox{\sl solve}(Q,v)$ for a variable $v$, Theorem~\ref{theorem:de:2} guarantees that the termination is preserved. In other words,
for any interpreted program $P$ and any interpreted query $Q$,
$P$ LD-terminates with respect to $Q$ if and only if
$D\cup \mbox{\sl ce}(P)$ LD-terminates with respect 
$\mbox{\sl solve}(Q,v)$ for a variable $v$. This is exactly the 
desired behaviour for a debugger, for instance, as we would like to
be sure that program execution inside the debugger reflects 
exactly its actual execution.
\end{itemize}
}
$\hfill\Box$\end{example}

The following example illustrates that the 
fact that $D$ is restricted is  {\em 
sufficient but not necessary} for LD-termination of a restricted 
query $\mbox{\sl solve}(Q,v_1,\ldots,v_n)$  with respect to 
$D\cup\mbox{\sl ce}^D(P)$ to imply LD-termination of $Q$ with respect to $P$.

\begin{example}
Let $D$ be the following meta-interpreter.
\begin{eqnarray*}
&& \mbox{\sl solve}(\mbox{\sl true})\mbox{.}\\
&& \mbox{\sl solve}((A,B))\leftarrow
\mbox{\sl fail}, \mbox{\sl solve}(A),\mbox{\sl solve}(B)\mbox{.}\\
&& \mbox{\sl solve}(A)\leftarrow \mbox{\sl loop}, \mbox{\sl clause}(A,B),\mbox{\sl solve}(B)\mbox{.}\\
&& \mbox{\sl loop}\leftarrow \mbox{\sl loop}\mbox{.}
\end{eqnarray*}
This meta-interpreter is double extended, but it is not restricted since one
of the intermediate body atoms fails. Observe that
$D$ does not improve termination. Indeed, for any program $P$ and
for any query $Q$ the restricted query $\mbox{\sl solve}(Q)$ does not
terminate with respect to $D\cup\mbox{\sl ce}^D(P)$. Thus, the following
implication is trivially {\sl true}: 
``if $\mbox{\sl solve}(Q)$ LD-terminates with 
respect to $D\cup\mbox{\sl ce}^D(P)$ then $Q$ LD-terminates with respect to 
$P$''. 
$\hfill\Box$\end{example}

To conclude this section we discuss syntactical conditions that can be used
to ensure that a double extended meta-interpreter is restricted. The first
condition of Definition~\ref{definition:de:restricted} requires certain 
sequences of arguments to be linear sequences of variables for any computed 
answer of the preceding atoms. Recall that we made a distinction between two
possibilities:
\begin{itemize}
\item either the corresponding instances of 
$(t_{31}, \ldots, t_{3n})$, $(t_{41}, \ldots, t_{4n})$,
$(t_{61}, \ldots, t_{6n})$, and
$(s_{1}, \ldots, s_{k})$ are linear sequences of free variables.
To ensure this we require that $(t_{31}, \ldots, t_{3n})$, 
$(t_{41}, \ldots, t_{4n})$, $(t_{61}, \ldots, t_{6n})$, and
$(s_{1}, \ldots, s_{k})$ are linear sequences of free variables
{\em and} that none of these variables appear in the preceding subgoals.
\item or $(t_{11}, \ldots, t_{1n})$, $(t_{21}, \ldots, t_{2n})$, 
$(t_{51}, \ldots, t_{5n})$, are linear sequences of free variables
and for every program $P$ and for every
$\mbox{\sl clause}(s,t,t_1,\ldots,t_k)\in \mbox{\sl ce}^D(P)$,
$t_1,\ldots,t_k$ is a linear sequence of free variables.
Observe that the sequences of arguments appear in heads of the clauses
in this case. Thus, there are {\em no preceding atoms}. In other
words, if the condition holds on the syntactic level, then it trivially holds
for any computed answer of the preceding atoms.
\end{itemize}

In order to ensure the remaining conditions we
require every atom $a$ among $C_{kl}$ and $D_{pq}$ to satisfy one of the 
following ($A$ and $B$ denote the meta-variables of the clauses):
\begin{enumerate}
\item $a$ is {\sl true}
\item $a$ is $u = f(u_1,\ldots,u_n)$ and $u$ is a fresh variable
and $A,B\not\in \mbox{\sl Var}(f(u_1,\ldots,u_n))$;
\item $a$ is a call to a built-in predicate $p$ and $p$ is either $\mbox{\sl write}$ or $\mbox{\sl nl}$;
\item $a$ is $p(u_1,\ldots,u_n)$ for a user-defined predicate $p$, $p$ cannot fail.
\end{enumerate}
The latter condition can be safely approximated by compiler by means of determinism analysis
(cf.\ \cite{Henderson:Somogyi:Conway}).

These requirements may seem to be very restrictive. However, 
they are satisfied by the majority of the meta-interpreters considered
including a  meta-interpreter that constructs proof trees presented in 
Example~\ref{example:proof:tree}, as well as a 
reasoner with uncertainty~\cite{Sterling:Shapiro}, and meta-interpreters allowing to reason about theories and provability~\cite{Brogi:Mancarella:Pedreschi:Turini,Martens:DeSchreye:inbook}. However, as the following example illustrates,
not every restricted meta-interpreter satisfies these conditions.
\begin{example}
\label{example:proof:tree:variant}
Consider the following variant of the meta-interpreter that constructs 
proof trees (cf.\ \ref{example:proof:tree}).
\begin{eqnarray*}
&& \mbox{\sl solve}(\mbox{\sl true}, \mbox{\sl true})\mbox{.}\\
&& \mbox{\sl solve}((A,B), (\mbox{\it ProofA},\mbox{\it ProofB}))\leftarrow
\mbox{\sl solve}(A,\mbox{\it ProofA}),\mbox{\sl solve}(B,\mbox{\it ProofB})\mbox{.}\\
&& \mbox{\sl solve}(A,(A\leftarrow \mbox{\it Proof}))\leftarrow \mbox{\sl clause}(A,B),\mbox{\sl foo}(\mbox{\it Proof}),\mbox{\sl solve}(B, \mbox{\it Proof})\mbox{.}\\
&& \mbox{\sl foo}(\_)\mbox{.}
\end{eqnarray*}
This meta-interpreter is a restricted double extended meta-interpreter. 
However,
the syntactic conditions specified above do not hold, since
$\mbox{\it Proof}$ appears in the preceding subgoal of $\mbox{\sl foo}$.
$\hfill\Box$\end{example}

\section[Extending the language]{Extending the language of the interpreted programs}
\label{section:ground:representation}

So far we have considered only definite programs. However, in order
to make our approach practical, the language of the underlying interpreted
programs should be extended to include negation, frequently appearing
in applications of the meta-interpreters. 

As earlier, in order to prove that meta-interpreters with negation 
preserve termination, we use among others a termination analysis framework 
based on order-acceptability. 
By using this result and applying the
same methodology as above one can prove that the following meta-interpreter
$M_4$, being an immediate extension of the ``vanilla'' meta-interpreter to normal programs~\cite{Hill:Gallagher}, preserves LDNF-termination. By 
LDNF-termination we understand finiteness of the LDNF-forest. 
Soundness and 
completeness of $M_4$ are proved in Theorem 2.3.3~\cite{Hill:Lloyd}.
\begin{eqnarray*}
&& \mbox{\sl solve}(true)\mbox{.}\\
&& \mbox{\sl solve}((\mbox{\sl Atom},\mbox{\sl Atoms})) \leftarrow \mbox{\sl solve}(\mbox{\sl Atom}), \mbox{\sl solve}(\mbox{\sl Atoms})\mbox{.}\\
&& \mbox{\sl solve}(\neg \mbox{\sl Atom})\leftarrow \neg\mbox{\sl solve}(\mbox{\sl Atom})\mbox{.}\\
&& \mbox{\sl solve}(\mbox{\sl Head}) \leftarrow \mbox{\sl clause}(\mbox{\sl Head},\mbox{\sl Body}), \mbox{\sl solve}(\mbox{\sl Body})\mbox{.}
\end{eqnarray*}

\begin{theorem}
\label{theorem:normal} 
Let $P$ be a normal program, $S$ be a set of queries.
Then $P$ LDNF-terminates with respect to $S$ if and only if
$M_4\cup \mbox{\sl ce}(P)$ LDNF-terminates with respect to
$\{\mbox{\sl solve}(Q) \mid Q\in S\}$.
\end{theorem}
\begin{proof}
Mimicking the proof of Theorem~\ref{meta:interpreted} and the result of
Pedreschi and Ruggieri~\cite{Pedreschi:Ruggieri}. First, for each one
of the LDNF-trees, calls and semantics are preserved by the corresponding
results for definite programs. Second, given the definition of an
LDNF-forest, if $\neg A$ is discovered, the new tree with $A$ as a root is
started. Observe that $\neg \mbox{\sl solve}(A)$ is ground if and only if
$A$ is ground. Thus, the derivation obtained with respect to 
$M_4\cup \mbox{\sl ce}(P)$ flounders if and only if the derivation obtained 
with respect to $P$ does. 
\end{proof}

Theorem~\ref{theorem:normal} allows us to consider termination of
different kinds of meta-interpreters, namely, those using the {\sl ground} 
representation of interpreted 
programs~\cite{Bowen:Kowalski,Godel,Hill:Gallagher}. 
This idea can be traced back to G\"{o}del, who suggested a one-to-one 
mapping, called {\em G\"{o}del numbering}, of expressions in a first order 
language to natural 
numbers~\cite{Godel:original}. The idea of numbering is a key idea of  
ground representation. Intuitively, we number predicates
$p(0), p(1), p(2),\ldots$, then functors $f(0), f(1), f(2),\ldots$, constants
$c(0), c(1), c(2),\ldots$ and, finally, variables used in the program 
$v(0), v(1), v(2),\ldots$ Each one of the sets is finite since the program
itself is finite.
Atoms are represented as $\mbox{\sl atom}(P, L)$,
where $P$ is the encoding of the predicate and $L$ is the list of the 
encodings of the arguments. Terms are represented in a similar way
as $\mbox{\sl term}(F, L)$, where $F$ is the encoding of the 
main functor and $L$ is the list of the encodings of the arguments.
To represent a clause, we use {\sl and}, denoting a conjunction, and 
{\sl if}, standing for an implication.

For example, instead of representing a clause $${\mbox {\sl permute}}(L,[El|T])\leftarrow {\mbox {\sl delete}}(El,L,L1), {\mbox {\sl permute}}(L1,T),$$ as 
$\mbox{\sl clause}({\mbox {\sl permute}}(L,[El|T]), ({\mbox {\sl delete}}(El,L,L1), {\mbox {\sl permute}}(L1,T)))$
as we used to do, it is represented as 
\begin{eqnarray*}
&& \mbox{\sl if}(\mbox{\sl atom}(p(0), [v(0), \mbox{\sl term}(f(0), [v(1),\mbox{\sl term}(f(0),[v(2), c(0)])])]),\\
&& \hspace{1.0cm} \mbox{\sl and}(\mbox{\sl atom}(p(1), [v(1), v(0), v(3)]), \mbox{\sl atom}(p(0), [v(3), v(2)]))),
\end{eqnarray*}
where the following correspondence holds:
\[
\begin{array}{lll}
\mbox{\rm predicates} & p(0) & \mbox{\sl permute} \\
                      & p(1) & \mbox{\sl delete} \\
\mbox{\rm functors} & f(0) & ./2\;\mbox{\rm also known as}\;\mbox{\sl cons}\\
\mbox{\rm constants} & c(0) & []\\
\mbox{\rm variables} & v(0) & L\\
                     & v(1) & El\\
                     & v(2) & T\\
                     & v(3) & L1\mbox{.}
\end{array}
\]
\eat{
where $p(0)$ corresponds to $\mbox{\sl permute}$, $p(1)$ to 
$\mbox{\sl delete}$, $f(0)$ to $\mbox{.}/2$, $c(0)$ to $[]$,
$v(0)$ to $L$, $v(1)$ to $El$, $v(2)$ to $T$, $v(3)$ to $L1$.
}

Meta-interpreters using the ground representation can be considered
``more pure'' than other meta-interpreters we studied, 
as the meta language and the language of the interpreted program
are strictly separated. However,
a number of primitive operations, such as unification, provided for 
the non-ground case by the underlying Prolog system, have to be defined 
explicitly. Moreover, while the preceding meta-interpreters can be recognised
as such by looking for the built-in predicate {\em clause},
recognising a meta-interpreter based on the ground representation can not
be done easily, unless some extra information, such as type declarations, is 
provided explicitly.
 
The following meta-interpreter {\sl idemo}, inspired 
by~\cite{Kowalski:problems} has been borrowed 
from~\cite{Hill:Gallagher}.
Given the ground representation of a normal program
and the ground representation of a query the meta-interpreter
returns ground representations
of the computed answers corresponding to the query.
\begin{example}
\label{example:idemo}
Given the ground representation of a program
and the ground representation of a query, predicate {\sl idemo}
proceeds in two steps.
First it computes the non-ground version of a query by 
calling predicate {\sl instance\_of}, i.e., replaces $v(i)$'s with variables,
while recording bindings. Secondly, it calls an alternative
version of the meta-interpreter ({\sl idemo1}) to resolve the
non-ground version of a query with the ground representation of a
given program. Observe how a non-ground instance of clause is
computed in the last clause of {\sl idemo1}.
\eat{
\begin{eqnarray*}
&& \mbox{\sl idemo}(P,X,Y)\leftarrow \mbox{\sl instance\_of}(X,Y), \mbox{\sl idemo1}(P,Y)\mbox{.}\\
&& \\
&& \mbox{\sl idemo1}(\_,{\mbox{\sl true}})\mbox{.}\\
&& \mbox{\sl idemo1}(P,\mbox{\sl and}(X,Y)) \leftarrow \mbox{\sl idemo1}(P,X), \mbox{\sl idemo1}(P,Y)\mbox{.}\\
&& \mbox{\sl idemo1}(P,\mbox{\sl not}(X)) \leftarrow \neg \mbox{\sl idemo1}(P,X)\mbox{.}\\
&& \mbox{\sl idemo1}(P,\mbox{\sl atom}(Q,Xs)) \leftarrow \mbox{\sl member}(Z,P), \\
&& \hspace{1.0cm} \mbox{\sl instance\_of}(Z, \mbox{\sl if}(\mbox{\sl atom}(Q,Xs), B)), \mbox{\sl idemo1}(P,B)\mbox{.}\\
&& \\
&& \mbox{\sl instance\_of}(X,Y)\leftarrow \mbox{\sl inst\_formula}(X,Y,[],\_)\mbox{.}\\
&& \\
&& \mbox{\sl inst\_formula}(\mbox{\sl atom}(Q,Xs), \mbox{\sl atom}(Q,Ys), S, S1)\leftarrow \mbox{\sl inst\_args}(Xs, Ys, S, S1)\mbox{.}\\
&& \mbox{\sl inst\_formula}(\mbox{\sl and}(X,Y), \mbox{\sl and}(Z,W), S, S2)\leftarrow \mbox{\sl inst\_formula}(X,Z, S, S1), \\
&& \hspace{1.0cm} \mbox{\sl inst\_formula}(Y, W, S1, S2)\mbox{.}\\
&& \mbox{\sl inst\_formula}(\mbox{\sl if}(X,Y), \mbox{\sl if}(Z,W), S, S2)\leftarrow \mbox{\sl inst\_formula}(X,Z, S, S1), \\
&& \hspace{1.0cm} \mbox{\sl inst\_formula}(Y,W, S1,S2)\mbox{.}\\
&& \mbox{\sl inst\_formula}(\mbox{\sl not}(X), \mbox{\sl not}(Z), S, S1) \leftarrow \mbox{\sl inst\_formula}(X,Z, S, S1)\mbox{.}\\
&& \mbox{\sl inst\_formula}(true, true, S, S)\mbox{.}\\
&& \\
&& \mbox{\sl inst\_args}([], [], S, S)\mbox{.}\\
&& \mbox{\sl inst\_args}([X|Xs], [Y|Ys], S, S2)\leftarrow \mbox{\sl inst\_term}(X,Y,S,S1), \\
&& \hspace{1.0cm} \mbox{\sl inst\_args}(Xs,Ys,S1,S2)\mbox{.}\\
&& \\
&& \mbox{\sl inst\_term}(v(N), X, [], [\mbox{\sl bind}(N,X)])\mbox{.}\\
&& \mbox{\sl inst\_term}(v(N), X, [\mbox{\sl bind}(N,X)|S], [\mbox{\sl bind}(N,X)|S])\mbox{.}\\
&& \mbox{\sl inst\_term}(v(N), X, [\mbox{\sl bind}(M,Y)|S], [\mbox{\sl bind}(M,Y)|S1])\leftarrow \\
&& \hspace{1.0cm} N \neq M, \mbox{\sl inst\_term}(v(N),X,S,S1)\mbox{.}\\
&& \mbox{\sl inst\_term}(\mbox{\sl term}(F,Xs),\mbox{\sl term}(F,Ys),S,S1)\leftarrow \mbox{\sl inst\_args}(Xs,Ys,S,S1).
\end{eqnarray*}
}
\[\begin{array}{ll}
 \mbox{\sl idemo}(P,X,Y)\leftarrow & \mbox{\sl instance\_of}(X,Y)\leftarrow \\
 \hspace{1.0cm}\mbox{\sl instance\_of}(X,Y), & \hspace{1.0cm} \mbox{\sl inst\_formula}(X,Y,[],\_)\mbox{.}\\
 \hspace{1.0cm}\mbox{\sl idemo1}(P,Y)\mbox{.} & \\
 &   \mbox{\sl inst\_formula}(\mbox{\sl atom}(Q,Xs), \\ 
 \mbox{\sl idemo1}(\_,{\mbox{\sl true}})\mbox{.} & \hspace{0.5cm}\mbox{\sl atom}(Q,Ys), S, S1)\leftarrow \\
 \mbox{\sl idemo1}(P,\mbox{\sl and}(X,Y)) \leftarrow & \hspace{1.0cm} \mbox{\sl inst\_args}(Xs, Ys, S, S1)\mbox{.}\\
 \hspace{1.0cm}\mbox{\sl idemo1}(P,X),& \mbox{\sl inst\_formula}(\mbox{\sl and}(X,Y), \\
 \hspace{1.0cm}\mbox{\sl idemo1}(P,Y)\mbox{.} & \hspace{0.5cm}\mbox{\sl and}(Z,W), S, S2)\leftarrow\\
 \mbox{\sl idemo1}(P,\mbox{\sl not}(X)) \leftarrow & \hspace{1.0cm} \mbox{\sl inst\_formula}(X,Z, S, S1),\\ 
 \hspace{1.0cm}\neg \mbox{\sl idemo1}(P,X)\mbox{.} & \hspace{1.0cm} \mbox{\sl inst\_formula}(Y,W, S1,S2)\mbox{.}\\
 \mbox{\sl idemo1}(P,\mbox{\sl atom}(Q,Xs)) \leftarrow & \mbox{\sl inst\_formula}(\mbox{\sl if}(X,Y), \\
 \hspace{1.0cm}\mbox{\sl member}(Z,P), & \hspace{0.5cm}\mbox{\sl if}(Z,W), S, S2)\leftarrow\\
 \hspace{1.0cm}\mbox{\sl instance\_of}(Z, & \hspace{1.0cm} \mbox{\sl inst\_formula}(X,Z, S, S1),\\
\hspace{1.5cm}\mbox{\sl if}(\mbox{\sl atom}(Q,Xs), B)),& \hspace{1.0cm} \mbox{\sl inst\_formula}(Y,W, S1,S2)\mbox{.}\\
 \hspace{1.0cm}\mbox{\sl idemo1}(P,B)\mbox{.} & \mbox{\sl inst\_formula}(\mbox{\sl not}(X), \mbox{\sl not}(Z), S, S1) \leftarrow \\
& \hspace{1.0cm} \mbox{\sl inst\_formula}(X,Z, S, S1)\mbox{.}\\
\mbox{\sl inst\_term}(v(N), X, [], [\mbox{\sl bind}(N,X)])\mbox{.} & \mbox{\sl inst\_formula}(true, true, S, S)\mbox{.}\\
\mbox{\sl inst\_term}(v(N), X, [\mbox{\sl bind}(N,X)|S], & \\
\hspace{0.5cm}[\mbox{\sl bind}(N,X)|S])\mbox{.} & \mbox{\sl inst\_args}([], [], S, S)\mbox{.}\\
\mbox{\sl inst\_term}(v(N), X, [\mbox{\sl bind}(M,Y)|S],& \mbox{\sl inst\_args}([X|Xs], [Y|Ys], S, S2)\leftarrow \\
\hspace{0.5cm}[\mbox{\sl bind}(M,Y)|S1])\leftarrow & \hspace{1.0cm} \mbox{\sl inst\_term}(X,Y,S,S1), \\
\hspace{1.0cm}  N \neq M, & \hspace{1.0cm} \mbox{\sl inst\_args}(Xs,Ys,S1,S2)\mbox{.}\\
\hspace{1.0cm} \mbox{\sl inst\_term}(v(N),X,S,S1)\mbox{.}  & \\
\mbox{\sl inst\_term}(\mbox{\sl term}(F,Xs), & \\
\hspace{0.5cm}\mbox{\sl term}(F,Ys),S,S1)\leftarrow  & \\
\hspace{1.0cm} \mbox{\sl inst\_args}(Xs,Ys,S,S1)\mbox{.} & 
\end{array}\]
We are interested in proving termination of $\mbox{\sl idemo}(p,q,v)$,
where $p$ is the ground representation of a program, $q$ is a ground
representation of a query and $v$ is a free variable
that will be bound to the ground representations
of computed answers corresponding to the query.

Existing termination techniques, such as~\cite{Dershowitz:Lindenstrauss:Sagiv:Serebrenik}
are powerful enough to prove termination of {\sl idemo1}
calls to $\mbox{\em instance\_of}(t, v)$, where $t$ is a term,
being a ground representation of a term, atom or clause, and $v$ is
a variable that will be bounded to the non-ground representation of the
same object. However, they are not powerful enough to analyse correctly
this example, both due to imprecise representation of all possible ground 
terms (in particular all possible ground representations of programs, 
by the same abstraction) and due to the nature of {\em idemo1} as a 
meta-interpreter. 

It should be noted that the ``troublesome'' part of this 
example is a definition of {\em idemo1}. However, {\em idemo1}
is very similar to the meta-interpreter $M_4$ discussed. The only 
differences are that the clauses of the interpreted program are stored
in the first argument and that a {\em non}-ground instance of a 
a clause has to be computed before resolving a query with it. Despite
of these differences a theorem analogous to Theorem~\ref{theorem:normal} 
holds. Hence,
termination of the meta-program is equivalent to termination of the
interpreted program.
$\hfill\Box$\end{example}

\eat{
Moreover, we can extend double extended meta-interpreters to normal
programs and prove results similar to Theorem~\ref{theorem:de:1}. 
More formally, we introduce the following notion.

\begin{definition}
A normal program of the following form 
\begin{eqnarray*}
&& \mbox{\sl solve}(\mbox{\sl true}, t_{11}, \ldots, t_{1n})\leftarrow C_{11},\ldots, C_{1m_1}\mbox{.}\\
&& \mbox{\sl solve}((A,B), t_{21}, \ldots, t_{2n})\leftarrow \\
&& \hspace{1.0cm}D_{11},\ldots,D_{1k_1}, \mbox{\sl solve}(A, t_{31}, \ldots, t_{3n}), \\
&& \hspace{1.0cm}D_{21},\ldots,D_{2k_2}, \mbox{\sl solve}(B, t_{41}, \ldots, t_{4n})\\
&& \hspace{1.0cm} C_{21},\ldots, C_{2m_2}\mbox{.}\\
&& \mbox{\sl solve}(A, t_{51}, \ldots, t_{5n})\leftarrow \\
&& \hspace{1.0cm}D_{31},\ldots,D_{3k_3},\mbox{\sl clause}(A,B,s_{1}, \ldots, s_{k}), \\
&& \hspace{1.0cm}D_{41},\ldots,D_{4k_4},\mbox{\sl solve}(B, t_{61}, \ldots, t_{6n})\\
&& \hspace{1.0cm} C_{31},\ldots, C_{3m_3}\mbox{.}\\
&& \mbox{\sl solve}(\neg A, t_{71}, \ldots, t_{7n})\leftarrow \\
&& \hspace{1.0cm}D_{51},\ldots,D_{5k_5},\neg \mbox{\sl solve}(A, t_{81}, \ldots, t_{8n}),\\
&& \hspace{1.0cm} C_{41},\ldots, C_{4m_4}.
\end{eqnarray*}
together with defining clauses for any other predicates occurring in the 
$C_{kl}$ and $D_{pq}$ (none of which contain $\mbox{\sl solve}$ or 
$\mbox{\sl clause}$) is called a {\em normal double extended meta-interpreter}.
\end{definition}

Then, the following results can be shown to hold similarly to Theorem~\ref{theorem:de:1} and Lemma~\ref{lemma:de:D}.
\begin{theorem}
\label{theorem:nde:1}
Let $P$ be an interpreted program, $N$ a normal double extended 
meta-interpreter, and $Q\in B^E_{N\cup \mbox{\sl ce}(P)}$, 
such that $Q$ is terminating with respect to $P$ and $\{A\mid A\in \mbox{\sl Call}(N\cup \mbox{\sl ce}(P), \mbox{\sl solve}(Q)), \mbox{\sl solve}\neq \mbox{\sl rel}(A)\}$ is terminating with respect to  $N$
then $\mbox{\sl solve}(Q)$ terminates with respect to $N\cup \mbox{\sl ce}(P)$.
\end{theorem}

\begin{lemma}
\label{lemma:nde:D}
Let $P$ be an interpreted program and $N$ a normal 
double extended meta-interpreter,
such that $N\cup \mbox{\sl ce}^D(P)$ is terminating with respect to
$\mbox{\sl solve}(Q,s_1,\ldots,s_n)$. Then, 
$N$ terminates with respect to
$\{A\mid A\in \mbox{\sl Call}(N\cup \mbox{\sl ce}^D(P), \mbox{\sl solve}(Q,s_1,\ldots,s_n)), \mbox{\sl solve} \neq \mbox{\sl rel}(A)\}$.
\end{lemma}
}
\eat{
\subsection{An alternative meta-interpreter}
Some authors~\cite{Bruynooghe:DeSchreye:Martens,Cheng:van:Emden:Strooper,%
van:Harmelen} considered the following 
meta-interpreter that uses list of goals instead 
of traditional conjunction of goals. 

\begin{example}
\label{example:alt:mi}
Let $M_{\mbox{\sl alt}}$ be the following meta-interpreter:
\begin{eqnarray*}
&&\mbox{\sl solve}([])\mbox{.}\\
&&\mbox{\sl solve}([Q|Qs])\leftarrow \mbox{\sl clause}(Q,B),
\mbox{\sl append}(B,Qs,NewQs), \mbox{\sl solve}(NewQs)\mbox{.}\\
&&\\
&&\mbox{\sl append}([],L,L)\mbox{.}\\
&&\mbox{\sl append}([H|X],Y,[H|Z])\leftarrow \mbox{\sl append}(X,Y,Z).
\end{eqnarray*}
Recall, that while executing a ``vanilla''-based meta-program,
an interpreted query is represented as a sequence of meta-atoms, each
one of them corresponding to (a remaining part of) a query, introduced
by SLD-one resolution step. Unlike this, during the execution
of an $M_{\mbox{\sl alt}}$-based meta-program the entire interpreted
query is represented as one meta atom. 
This representation is closer to Prolog computation, since
Prolog does not ``remember'' from where did a query atom come.
$\hfill\Box$\end{example}

Similarly to ``vanilla'', meta-interpreter $M_{\mbox{\sl alt}}$ can also 
serve as a basis for developing applications, such as proof tree constructors 
and debuggers. By following general principles of 
Theorems~\ref{theorem:de:1} and \ref{theorem:de:2}
one can reduce reasoning on termination of the extended meta-interpreters to 
reasoning on termination of $M_{\mbox{\sl alt}}$. Therefore, we
restrict our attention to $M_{\mbox{\sl alt}}$.

In order to use $M_{\mbox{\sl alt}}$ slightly different clause encoding is
used. Instead of representing a clause $H\leftarrow B_1,\ldots,B_n$ as
$\mbox{\sl clause}(H,(B_1,\ldots,B_n))$, it is represented as
$\mbox{\sl clause}(H,[B_1,\ldots,B_n])$. The clause encoding used in this
case is denoted $\gamma_{cea}$.

Before discussing termination behaviour,
we show that $M_{\mbox{\sl alt}}$ preserves the queries of the LD-tree.
\begin{lemma}
\label{lemma:alt:calls:preserved}
Let $P$ be an interpreted program, $M_{\mbox{\sl alt}}$ be an alternative
meta-interpreter of Example~\ref{example:alt:mi} and $Q\in B^E_{P}$, then
a query $Q_1,\ldots,Q_n$ appears in the LD-tree of $P$ and $Q$ if, and only
if, a query $\mbox{\sl solve}([Q'_1,\dots,Q'_n])$ appears in the 
LD-tree of $M_{\mbox{\sl alt}}\cup \gamma_{cea}(P)$ and 
$\mbox{\sl solve}([Q])$, where for all $i$, $Q'_i$ is a variant of $Q_i$.
\end{lemma}
\begin{proof}
Proof is done by induction on the LD-derivation. Induction base is
straightforward, since the lemma holds for $Q$ and $\mbox{\sl solve}([Q])$.

Let $Q_1,\ldots,Q_n$ be a non-empty query in the LD-tree of $P$ and $Q$. Assume
that $\mbox{\sl solve}([Q'_1,\dots,Q'_n])$ appears in the 
LD-tree of $M_{\mbox{\sl alt}}\cup \gamma_{cea}(P)$ and 
$\mbox{\sl solve}([Q])$, where for all $i$, $Q'_i$ is a variant of $Q_i$.
Let $H\leftarrow B_1,\ldots,B_m$ be a renamed apart version of a
clause $c$ in $P$ such that $H$ is unifiable
with $Q_1$ via $\theta$. Then, the next query in the LD-tree of $P$ and $Q$
is $(B_1,\ldots,B_m,Q_2,\ldots,Q_n)\theta$.

Since $Q_1,\ldots,Q_n$ is assumed to be non-empty, only the second
clause of $M_{\mbox{\sl alt}}$ can be used to resolve
$\mbox{\sl solve}([Q'_1,\dots,Q'_n])$. After the resolution step, the 
following query is obtained:
\[
\mbox{\sl clause}(Q'_1,B),\mbox{\sl append}(B,[Q'_2,\dots,Q'_n],NewQs), \mbox{\sl solve}(NewQs)
\]
Let $clause(H',[B'_1,\ldots,B'_m])$ be a renamed apart version of
a clause encoding of $c$. Since $H$ is unifiable with $Q_1$ via $\theta$,
$H'$ is unifiable with $Q'_1$ via $\theta'$, which is identical to $\theta$
up to variable renaming. Thus, $\mbox{\sl clause}(Q'_1,B)$
can be unified with $clause(H',[B'_1,\ldots,B'_m])$, resulting in
\[
\mbox{\sl append}([B'_1\theta',\ldots,B'_m\theta'],[Q'_2\theta,\dots,Q'_n\theta],NewQs), \mbox{\sl solve}(NewQs)
\]
which leads to $\mbox{\sl solve}([B'_1\theta',\ldots,B'_m\theta',Q'_2\theta',\dots,Q'_n\theta'])$, satisfying conditions of the lemma. Similar reasoning can 
be performed to show the opposite direction of the implication.
\end{proof}

Now we are ready to prove that $M_{\mbox{\sl alt}}$ preserves termination 
behaviour.
\begin{theorem}
\label{alt:meta:interpreted}
Let $P$ be a definite program, $S$ a set of atomic queries, 
such that $P$  LD-terminates for all queries in $S$, 
and  let $M_{\mbox{\sl alt}}$ be the meta-interpreter as above. Then, 
$M_{\mbox{\sl alt}}\cup \gamma_{cea}(P)$ is LD-terminating for all queries in
$\{\mbox{\sl solve}([Q]) \mid Q\in S\}$. 
\end{theorem}
\begin{proof} 
By Theorem~\ref{taset:term} 
$P$ is order-acceptable with respect to $S$.
Let $>$ be a minimal ordering 
such that $P$ is order-acceptable 
with respect to $S$ via it, i.e., there exists no ordering $>_1$
such that $P$ is order-acceptable 
with respect to $S$ via $>_1$ and $>_1\subset >$.
We are going to prove  
order-acceptability of $M_{\mbox{\sl alt}}\cup \gamma_{cea}(P)$ 
with respect to $\{\mbox{\sl solve}([Q]) \mid Q\in S\}$. 
By Theorem~\ref{taset:term} termination is implied. 

Let $>^{*}$ be defined on $B^E_P$ as following:
\[
A >^{*} B\;\mbox{\rm if}\;
\left\{
\begin{array}{l}
\mbox{\rm if}\;\mbox{\sl rel}(A)\simeq \mbox{\sl rel}(B),\;\mbox{\rm and there exists $\theta$, such that $A > B\theta$}\; \\
\mbox{\rm if}\;\mbox{\sl rel}(A)\neq \mbox{\sl rel}(B)
\end{array}
\right.
\]

We have to show that $>^{*}$ is irreflexive, antisymmetric, transitive
and well-founded.
\begin{itemize}
\item Let $A$ be such that $A >^* A$ holds. Since $\mbox{\sl rel}(A)\simeq \mbox{\sl rel}(A)$ is always true, there exists a substitution $\theta$, such that
$A > A\theta$. By Lemma~\ref{lemma:minimal:ds}, there exists a sequence of 
queries $A = Q, \ldots, Q_n = A\theta$ such that for all $0 \leq i < n$, 
$Q_{i+1}$ is an instantiated body subgoal of a clause $c_i$ whose head was 
unified with $Q_i$. 

\end{itemize}

Let $\succ$ be defined as following:
\begin{eqnarray*}
&& \mbox{\sl solve}([A_1,\ldots,A_m])\succ \mbox{\sl solve}([B_1,\ldots,B_n])\\
&&\hspace{1.0cm}\mbox{\rm if}\;\{A_1,\ldots,A_m\}\gg \{B_1,\ldots,B_n\},\\
&& \mbox{\sl append}(X_1,X_2,X_3)\succ \mbox{\sl append}(Y_1,Y_2,Y_3)\\
&&\hspace{1.0cm}\mbox{\rm if}\;\mid~X_1\mid_l\;>_{\mathbb N}\;\mid~Y_1\mid_l
\end{eqnarray*}
where $\gg$ denotes the extension of $>^{\sqsupset}$ to multisets,
$\mid\cdot\mid_l$ is a list-length norm and $>_{\mathbb N}$ is the usual
ordering on the natural numbers. 

To show that
$M_{\mbox{\sl alt}}\cup \gamma_{cea}(P)$ 
with respect to $\{\mbox{\sl solve}([Q]) \mid Q\in S\}$ via $\succ$
we distinguish three cases.
\begin{itemize}
\item For atoms of the predicate $\mbox{\sl clause}$ the claim is 
trivial, since there are no recursive clauses defining this predicate.
\item For atoms of the predicate $\mbox{\sl append}$ the claim is 
immediate from the definition of $\succ$ and observation that for
all calls of $\mbox{\sl append}$ their first argument will be a list of
a finite length.
\item Let $\mbox{\sl solve}([A_1,\ldots,A_m])$ be an element in 
the call set of $M_{\mbox{\sl alt}}\cup \gamma_{cea}(P)$ and one of the
queries of $\{\mbox{\sl solve}([Q]) \mid Q\in S\}$. The only clause
that contains a recursive body subgoal is the second clause of
$\mbox{\sl solve}$, unifying $\mbox{\sl solve}([A_1,\ldots,A_m])$ with its
head does not affect $A_i$'s and the only recursive subgoal is its last atom.
Then, for any computed answer substitution $\sigma$ for a query
\[\mbox{\sl clause}(A_1,B),\mbox{\sl append}(B,[A_2,\ldots,A_m],NewQs),\]
$\mbox{\sl solve}([A_1,\ldots,A_m])\succ \mbox{\sl solve}(NewQs)$ should hold.

To obtain the desired result, it is sufficient to show that 
for any atom $\mbox{\sl clause}(A_1,[B_1,\ldots,B_n])$ and
for any $1\leq i\leq n$ holds $A_1 >^{\sqsupset} B_i$. Indeed, if
$\mbox{\sl rel}(A_1)\sqsupset \mbox{\sl rel}(B_i)$ this follows from
the definition of $>^{\sqsupset}$. 
\end{itemize}
\end{proof}
}

\section{Conclusion}
\label{section:meta:conclusion}
We have presented a methodology for proving termination properties of 
meta-programs.  It is well-known that termination verification plays a 
crucial role in meta-programming \cite{Pfenning:Schuermann}.
Our main contribution is in providing a technique linking 
termination behaviour of an interpreted program with a termination behaviour
of the meta-program. We have shown that for a wide variety of meta-interpreters,
a relatively simple relation can be defined between the ordering that 
satisfies the requirements of order-acceptability for an interpreted program 
and the ordering that 
satisfies the requirements of order-acceptability 
for the meta-interpreter extended by this interpreted program and a 
corresponding set of queries. This category of meta-interpreters
includes many important ones, such as extended meta-interpreters
studied by~\cite{Martens:DeSchreye},  
meta-interpreter, that constructs proof trees~\cite{Sterling:Shapiro}, 
reasoners about theories and 
provability~\cite{Brogi:Mancarella:Pedreschi:Turini,Martens:DeSchreye:inbook},
and reasoners with uncertainty~\cite{Sterling:Shapiro}. Moreover, it
also describes a depth tracking tracer for Prolog, a reasoner with threshold 
cutoff~\cite{Sterling:Shapiro}, a pure four port box execution model 
tracer~\cite{Bowles:Wilk} and the {\em idemo} meta-interpreter
of~\cite{Hill:Gallagher}. 
The relationship established between the orderings allows
termination proofs to be reused, i.e., a termination proof
obtained for an interpreted program can be used for showing termination of 
the meta-program and vice versa.
Example~\ref{meta:example:nogood:linear:lm} demonstrated 
such a simple relation cannot be  established if linear level mappings 
were considered instead of general orderings.

Ease of meta-programming is often considered to be one of
the advantages of logic programming. From the early days 
meta-interpreters were developed to implement different control strategies for 
Prolog~\cite{Gallaire:Lasserre,Beckstein:Stolle:Tobermann:generalised}. 
Furthermore, meta-programming finds a
wide variety of applications in such areas as artificial intelligence,
compilation, constraints solving, debugging, and
program analysis~\cite{Codish:Taboch,Hill:Gallagher,Lamma:Milano:Mello,%
Martens:DeSchreye,Sterling:Shapiro}. Meta-interpreters 
have also been successfully applied to aspect-oriented 
programming~\cite{DeVolder:DHondt,Brichau:Mens:DeVolder}. 
Recently, Sheard presented a number of challenges
in meta-programming~\cite{Sheard}.

Despite the intensive research on meta-programming inside the 
logic programming community~\cite{Apt:Ben-Eliyahu,Apt:Turini,Levi:Ramundo,%
Martens:DeSchreye},
termination behaviour of meta-programs has attracted relatively little
attention, with Pedreschi and Ruggieri being the only known notable exception.
In their work~\cite{Pedreschi:Ruggieri}, a generic
verification method is used, 
based on specifying preconditions and postconditions.
Unfortunately, their termination results are restricted only to the ``vanilla''
meta-interpreter $M_0$. It is not immediately obvious 
how their results can be extended
to alternative meta-inter\-preters, nor if a relationship between termination
characterisation of the interpreted program and the meta-program can be established.

Researchers working on modular termination aim to discover how level mappings
required to prove termination of separate modules can be combined to obtain
a termination proof for the entire program~\cite{Apt:Pedreschi,Bossi:Cocco:Etalle:Rossi:modular,Pedreschi:Ruggieri:modular,Verbaeten:Sagonas:DeSchreye}. Since
meta-program can be viewed as a union of a meta-interpreter and of the
clause-encoding of an interpreted program, these results might seem applicable.
However, clause-encoding represents a program as a set of facts. Therefore,
for any program $P$, termination of $\mbox{\sl clause}(H,B)$ with respect to 
$\mbox{\sl ce}(P)$ is trivial and any level-mapping is sufficient to show 
termination. Hence, no useful information on termination of $P$ is provided by
the level-mapping and termination of the meta-program cannot be established.
  
Our methodology gains its power from the use of the 
integrated approach presented in~\cite{DeSchreye:Serebrenik:Kowalski}, 
which extends the traditional notion of 
acceptability by adding a 
wide class of orderings that have been studied in the 
context of the term-rewriting systems. Theoretically, this approach has exactly
the same power as the classical level mappings based results, but in practice,
quite often a simple ordering is sufficient to prove termination in
an example that would otherwise require the application of 
a complex level mapping. 
Meta-programs provide typical examples of this kind. 

The study of 
termination preservation for general meta-interpreters is an extremely
difficult task. We do not believe that termination preservation conditions
can be formulated without assuming any additional information on the 
meta-interpreter or on the interpreted programs. Therefore, we have identified
a number of important classes of meta-interpreters and proposed conditions 
implying termination preservation for each one of the classes. 
Some authors~\cite{Bruynooghe:DeSchreye:Martens,Cheng:van:Emden:Strooper,%
van:Harmelen} have 
studied a meta-interpreter that uses a list of goals instead 
of a traditional conjunction of goals. Study of termination preservation
properties of this meta-interpreter is considered as a future work.

The paper by Pedreschi and Ruggieri~\cite{Pedreschi:Ruggieri} is, to the
best of our knowledge, the only one to study additional verification properties
of the meta-interpreters such as absence of errors and partial correctness.
Their results hint at further research directions in the 
context of verification
of meta-interpreters. 

\bibliographystyle{acmtrans}
\bibliography{/home/alexande/M.Sc.Thesis/main}

\appendix

\section{``Vanilla'' preserves the calls set}
\label{appendix:vanilla:preserves:calls}

In this section we present a formal proof that the ``vanilla'' meta-interpreter
$M_0$ preserves the set of calls (Lemma~\ref{lemma:vanilla:calls:preserved}). 

Before presenting the actual proof, we introduce an auxiliary notion of {\sl partition}.
Intuitively, a sequence of sequences $(x_{1,1},\ldots,x_{1,n_1}),\ldots,
(x_{m,1},\ldots,x_{m,n_m})$ forms a partition of a sequence $(x_1,\ldots,
x_{\Sigma n_i})$ if they are exactly the same, except for additional 
division into subsequences. More formally,
\begin{definition}
\label{definition:partition}
Let $S$ be a set, and let $(x_1,\ldots, x_k)$ be a sequence of elements
of this set. Let $(x_{11},\ldots, x_{1n_1}),$ 
$\ldots, (x_{m1},\ldots, x_{mn_m})$
be sequences of elements of $S$. 
We say that $(x_{11},\ldots, x_{1n_1}),\ldots,
(x_{m1},$ $\ldots, x_{mn_m})$ forms a {\em partition} of
$(x_1,\ldots, x_k)$ if the following holds:
\begin{itemize}
\item $x_1 = x_{11}$
\item $x_k = x_{mn_m}$
\item If $x_l = x_{ij}$ then 
\[
x_{l+1} = 
\left\{
\begin{array}{ll}
x_{i(j+1)} & \mbox{\rm if $j < n_i$} \\
x_{(i+1)1} & \mbox{\rm otherwise}
\end{array}
\right.\]
\end{itemize}
\end{definition}

\begin{proposition}
\label{partition:prop}
The following properties hold:
\begin{itemize}
\item Let $S$ be a set, $(x_1,\ldots, x_k)$ be a sequence of elements
of $S$ and let \[(x_{11},\ldots, x_{1n_1}),\ldots,
(x_{m1},\ldots, x_{mn_m})\] form a partition of
$(x_1,\ldots, x_k)$. Then, for any substitution $\theta$,
\[(x_{11},\ldots, x_{1n_1})\theta,\ldots,
(x_{m1},\ldots, x_{mn_m})\theta\] forms a partition of
$(x_1,\ldots, x_k)\theta$.
\item Let $S$ be a set, $x_1,\ldots, x_k$ and $y_1,\ldots,y_l$ be sequences 
of elements of $S$ and let 
$(x_{11},\ldots,$ $x_{1n_1}),\ldots, (x_{m1},\ldots, x_{mn_m})$ and
$(y_{11},\ldots, y_{1p_1}),\ldots, (y_{q1},\ldots, y_{qp_q})$
 form a partition of $x_1,\ldots, x_k$ and 
$y_1,\ldots,y_l$, respectively. Then,
\[(x_{11},\ldots, x_{1n_1}),\ldots, (x_{m1},\ldots, x_{mn_m}),
  (y_{11},\ldots, y_{1p_1}),\ldots, (y_{q1},\ldots, y_{qp_q})\]
 forms a partition of $(x_1,\ldots, x_k,y_1,\ldots,y_l)$.
\end{itemize}
\end{proposition}
\begin{proof}
Immediately from Definition~\ref{definition:partition}.
\end{proof}

\begin{proof}[Proof of Lemma~\ref{lemma:vanilla:calls:preserved}]
In order to prove that the ``vanilla'' meta-interpreter $M_0$ preserves 
the calls set we have to show 
$$\{\mbox{\sl solve}(A)\mid A\in \mbox{\sl Call}(P, Q)\} \equiv \mbox{\sl Call}(M_0\cup \mbox{\sl ce}(P), 
\mbox{\sl solve}(Q))\;\;\cap\;\;\{\mbox{\sl solve}(A)\mid A\in B^E_{P}\},$$
where $\equiv$ means equality up to variable renaming.

We prove the set-equality by proving containment in both directions. We start by 
proving that the left-hand side set is contained in the right-hand side set 
and then prove the other direction.

$(\subseteq)$ Clearly, $\mbox{\sl Call}(P, Q)\subseteq B^E_{P}$.
Thus, $\{\mbox{\sl solve}(A)\mid A\in \mbox{\sl Call}(P, Q)\} \subseteq
\{\mbox{\sl solve}(A)\mid A\in B^E_{P}\}$. To prove the inclusion we
need, therefore, to prove that 
$\{\mbox{\sl solve}(A)\mid A\in \mbox{\sl Call}(P, Q)\}$ $\subseteq 
\mbox{\sl Call}(M_0\cup \mbox{\sl ce}(P), \mbox{\sl solve}(Q))$.
We show that every element of  
$\{\mbox{\sl solve}(A)\mid A\in \mbox{\sl Call}(P,$ $Q)\}$ is also
an element of $\mbox{\sl Call}(M_0\cup \mbox{\sl ce}(P), \mbox{\sl solve}(Q))$.

Let $K\in \{\mbox{\sl solve}(A)\mid A\in \mbox{\sl Call}(P, Q)\}$. That is,
$K =  \mbox{\sl solve}(K')$ for some $K' \in \mbox{\sl Call}(P,$ $Q)$.
The proof is inductive and based on the derivation of $K'$.

Note that we are going to prove a stronger claim than we actually
need: we prove that for every query $A_1,A_2,\ldots,A_k$
in the LD-tree of $P$ and $Q$, there is a query 
$\mbox{\sl solve}(A_0),$ $\mbox{\sl solve}((A_{21},$ $\ldots,A_{2n_2})),
\ldots,\mbox{\sl solve}((A_{m1},\ldots,A_{mn_m}))$
in $M_0\cup \mbox{\sl ce}(P)$ and $\mbox{\sl solve}(Q)$ such that 
$A_0,(A_{21},$ $\ldots,A_{2n_2}),\ldots,(A_{m1},\ldots,A_{mn_m})$ 
forms a partition of $A_1,\ldots,A_k$ up to variable renaming. Observe that
this also means that $A_0$ is a variant of $A_1$.

\begin{itemize}
\item {\sl Induction base} 
$K' = Q$ and $K = \mbox{\sl solve}(Q)$, implying
$K\in \mbox{\sl Call}(M_0\cup \mbox{\sl ce}(P),$ $\mbox{\sl solve}(Q))$.
Since $Q$ is an atomic query, the proof is completed.
\item {\sl Inductive assumption}
Assume that for some query $A_1,\ldots, A_k$ in the LD-tree of $P$ and $Q$, there exists a query 
\[\mbox{\sl solve}(A_0), \mbox{\sl solve}((A_{21},\ldots,A_{2n_2})),\ldots,\mbox{\sl solve}((A_{m1},\ldots,A_{mn_m}))\] 
in the LD-tree of $M_0\cup \mbox{\sl ce}(P)$ and 
$\mbox{\sl solve}(Q)$, such that 
\[A_0,(A_{21},\ldots,A_{2n_2}),\ldots,(A_{m1},\ldots,A_{mn_m})\] 
forms a partition of $A_1,\ldots,A_k$ up to variable renaming. This also 
implies that $A_0$ is a variant of $A_1$.
\item {\sl Inductive step}
Let $H\leftarrow H_1,\ldots,H_l$ be a renamed apart version of a clause in $P$.
Then, the query $(H_1,\ldots,H_l,A_2,\ldots,A_k)\theta$,
is the 
LD-resolvent of  $A_1,\ldots, A_k$
and $H\leftarrow H_1,\ldots,H_l$, where 
$\theta = \mbox{\sl mgu}(A_1, H)$. 

We are going to construct the required query. The only clause in 
$M_0\cup \mbox{\sl ce}(P)$
that is applicable for 
$\mbox{\sl solve}(A_0), \mbox{\sl solve}((A_{21},\ldots,A_{2n_2})), \ldots,
\mbox{\sl solve}((A_{m1},\ldots,A_{mn_m}))$ is 
$\mbox{\sl solve}(\mbox{\sl Head}) \leftarrow \mbox{\sl clause}(\mbox{\sl Head}
,\mbox{\sl Body}), \mbox{\sl solve}(\mbox{\sl Body})$.
Thus, the LD-resolvent is
\begin{eqnarray*}
&& \mbox{\sl clause}(\mbox{\sl Head}\tau,\mbox{\sl Body}\tau), 
\mbox{\sl solve}(\mbox{\sl Body}\tau), \\
&& \hspace{1.0cm}\mbox{\sl solve}((A_{21},\ldots,A_{2n_2}))\tau, \ldots, \mbox{\sl solve}((A_{m1},\ldots,A_{mn_m}))\tau,
\end{eqnarray*} 
where
$\tau = \mbox{\sl mgu}(\mbox{\sl solve}(A_0), \mbox{\sl solve}(\mbox{\sl Head}))$. We denote this resolvent $R$.

Recall that $\mbox{\sl Head}$ is a variable. 
Thus, $\mbox{\sl Head}\tau = A_0\tau = A_0$.
Moreover, $\mbox{\sl Head}$ is the only variable affected by $\tau$.
Therefore, $\mbox{\sl Body} = \mbox{\sl Body}\tau$ and
$A_{ij}\tau = A_{ij}$ for all $i$ and $j$.
Thus, $R$ is
$$\mbox{\sl clause}(A_0,\mbox{\sl Body}), 
\mbox{\sl solve}(\mbox{\sl Body}), \mbox{\sl solve}((A_{21},\ldots,A_{2n_2})),\ldots, \mbox{\sl solve}((A_{m1},\ldots, A_{mn_m}))\mbox{.}$$
Let ${H'}\!\leftarrow\! {H'}_1,\ldots,{H'}_l$ be a renamed apart
variant of the same clause as $H\!\leftarrow\! H_1,\ldots,H_l$. By definition 
of $\mbox{\sl ce}$, the atom 
$\mbox{\sl clause}({H'}, ({H'}_1,\ldots,{H'}_l))$ is a variant of
an element in $\mbox{\sl ce}(P)$. 
Thus, $R$ can be resolved with 
$\mbox{\sl clause}({H'}, ({H'}_1,\ldots,$ ${H'}_l))$. 
Let $\sigma$ be the most general unifier of
$\mbox{\sl clause}(A_0,\mbox{\sl Body})$ and of
$\mbox{\sl clause}({H'}, ({H'}_1,\ldots,{H'}_l)))$. The resolvent of $R$ and
$\mbox{\sl clause}({H'},$ $({H'}_1,\ldots,{H'}_l))$ is, therefore,
$$\mbox{\sl solve}(({H'}_1,\ldots,{H'}_l)\sigma), \mbox{\sl solve}((A_{21},\ldots,A_{2n_2}))\sigma,\ldots,\mbox{\sl solve}((A_{m1},\ldots, A_{mn_m}))\sigma\mbox{.}$$
Since $\mbox{\sl Body}$ is a variable,
$\sigma$ is the most general unifier of $A_0$ and $H'$, i.e.,
$\sigma = \mbox{\sl mgu}(A_0, H')\cup \{\mbox{\sl Body}/({H'}_1,\ldots,{H'}_l)\sigma\}$.
Thus, the following holds:
\begin{eqnarray*}
&& \theta = \mbox{\sl mgu}(A_1, H)\\
&& A_0\;\mbox{\rm is a variant of}\; A_1\\
&& H'\;\mbox{\rm is a variant of}\;H\\
&& A_{mn_m}\;\mbox{\rm is a variant of}\;A_k
\end{eqnarray*}
Thus,  
$A_0\sigma, {H'}_1\sigma,\ldots, {H'}_l\sigma$ are variants of
$A_1\theta, H_1\theta,\ldots, H_l\theta$. 
Therefore, by Inductive assumption
${H'}_1\sigma,\ldots,{H'}_l\sigma,(A_{21},\ldots,A_{2n_2})\sigma,\ldots,(A_{m1},\ldots,A_{mn_m})\sigma$
forms a partition of $(H_1,\ldots,H_l,A_2,\ldots,A_k)\theta$
up to variable renaming. 

If $l = 1$ the proof is completed. Otherwise, the main functor of 
$({H'}_1,\ldots,{H'}_l)\sigma$ is comma and the first rule of $M_0$
should be applied. Reasoning as above one can show that 
the unification with the head of the rule binds the meta-variables only.
The resolvent is thus,
\begin{eqnarray*}
&&\mbox{\sl solve}({H'}_1\sigma),\mbox{\sl solve}(({H'}_2,\ldots,{H'}_l)\sigma),\\
&&\hspace{1.0cm} \mbox{\sl solve}((A_{21},\ldots,A_{2n_2}))\sigma,\ldots,\mbox{\sl solve}((A_{m1},\ldots,A_{mn_m}))\sigma
\end{eqnarray*}
Recalling our observation on the relation between $\sigma$ and $\theta$ and 
applying Inductive assumption completes the proof.
\end{itemize}

($\supseteq$) 
Now we are going to prove the second part of the equality. That is
\begin{eqnarray*}
&& \{\mbox{\sl solve}(A)\mid A\in \mbox{\sl Call}(P, Q)\} \supseteq \\
&& \hspace{1.0cm}\mbox{\sl Call}(M_0\cup \mbox{\sl ce}(P), 
\mbox{\sl solve}(Q))\;\;\cap\;\;\{\mbox{\sl solve}(A)\mid A\in B^E_{P}\}
\end{eqnarray*}

Let $K\in \mbox{\sl Call}(M_0\cup \mbox{\sl ce}(P), 
\mbox{\sl solve}(Q))\;\;\cap\;\;\{\mbox{\sl solve}(A)\mid A\in B^E_{P}\}$.
Then, $K = \mbox{\sl solve}(K')$ for some $K'\in B^E_{P}$.
We need to show that $K'\in \mbox{\sl Call}(P, Q)$.

As earlier, we are going to prove this claim inductively and, similarly,
we prove a stronger claim than we need. We show that for every query 
$\mbox{\sl solve}(A_0), \mbox{\sl solve}((A_{21},$ $\ldots,A_{2n_2})),$ 
$\ldots, \mbox{\sl solve}((A_{m1},\ldots,A_{mn_m}))$
in the LD-tree of $M_0\cup \mbox{\sl ce}(P)$ and $\mbox{\sl solve}(Q)$,
such that $A_0$ differs from {\sl true},
there is a query $A_1,A_2,\ldots,A_k$ in the LD-tree of $P$ and
$Q$, such that $A_0,(A_{21},\ldots,A_{2n_2}),\ldots,$ $(A_{m1},\ldots,A_{mn_m})$ forms a partition of $A_1,\ldots,A_k$. Proving this will imply the statement 
we would like to prove, since $\mbox{\sl true}\not\in B^E_{P}$.

\begin{itemize}
\item {\sl Induction base}
If $K = \mbox{\sl solve}(Q)$, then $K' = Q$, and
 $K'\in \mbox{\sl Call}(P, Q)$. $Q$ is an atomic query, 
and it obviously partitions itself.
\item {\sl Inductive step}
As above, assume that given a query \[\mbox{\sl solve}(A_0), 
\mbox{\sl solve}((A_{21},\ldots,A_{2n_2})),\ldots, \mbox{\sl solve}((A_{m1},\ldots,A_{mn_m}))\]
in the LD-tree of $M_0\cup \mbox{\sl ce}(P)$ and $\mbox{\sl solve}(Q)$,
such that $A_0\neq \mbox{\sl true}$,
there exists a query $A_1,A_2,\ldots,A_k$ in the LD-tree of $P$ and
$Q$, such that 
$A_0,(A_{21},\ldots,A_{2n_2}),\ldots,(A_{m1},$ $\ldots,A_{mn_m})$ 
forms a partition of $A_1,\ldots,A_k$. In particular, this means that $A_1$
is a variant of $A_0$.

Since $A_0\neq \mbox{\sl true}$, there is only one clause in $M_0$ that
can be used to resolve $\mbox{\sl solve}(A_0),$ 
$\mbox{\sl solve}((A_{21},\ldots,A_{2n_2})),\ldots, \mbox{\sl solve}((A_{m1},\ldots,A_{mn_m}))$. This clause is 
$\mbox{\sl solve}(\mbox{\sl Head}) \leftarrow \mbox{\sl clause}(\mbox{\sl Head}
,\mbox{\sl Body}), \mbox{\sl solve}(\mbox{\sl Body})$.
The resolvent obtained is
\begin{eqnarray*}
&& \mbox{\sl clause}(\mbox{\sl Head}\tau,\mbox{\sl Body}\tau), 
\mbox{\sl solve}(\mbox{\sl Body}\tau),\\
&&\hspace{1.0cm} \mbox{\sl solve}((A_{21},\ldots,A_{2n_2})\tau),\ldots, \mbox{\sl solve}((A_{m1},\ldots,A_{mn_m})\tau),
\end{eqnarray*}
where $\tau =\mbox{\sl mgu}(\mbox{\sl solve}(A_0),
\mbox{\sl solve}(\mbox{\sl Head}))$. Recall that $\mbox{\sl Head}$ is a 
variable, i.e., $\mbox{\sl Head}\tau = A_0\tau = A_0$ and that 
neither $\mbox{\sl Body}$ nor any of $A_{ij}$'s can be instantiated by
$\tau$.

Next, the $\mbox{\sl clause}$-atom of the resolvent has to be unified with 
a renamed apart variant of one of the facts in the clause encoding of the 
interpreted program. Let it be 
$\mbox{\sl clause}(H, (H_1,\ldots,H_l))$ and let $\theta$ be the most
general unifier of $\mbox{\sl clause}(A_0,\mbox{\sl Body})$
and $\mbox{\sl clause}(H, (H_1,\ldots,H_l))$. In fact, since 
$\mbox{\sl Body}$ is a variable, $\theta$ is $\mbox{\sl mgu}(A_0,H)\cup
\{\mbox{\sl Body}/(H_1,\ldots,H_l)\theta\}$. Then, the next resolvent is
\begin{eqnarray*}
&& \leftarrow \mbox{\sl solve}((H_1,\ldots,H_l)\theta),
\mbox{\sl solve}((A_{21},\ldots,A_{2n_2})\theta),\ldots, \\
&&\hspace{1.0cm} \mbox{\sl solve}((A_{m1},\ldots,A_{mn_m})\theta).
\end{eqnarray*}
If $l > 1$ we can apply another rule of $M_0$ and in one step obtain 
\begin{eqnarray*}
&& \leftarrow 
\mbox{\sl solve}(H_1\theta), \mbox{\sl solve}(H_2\theta,\ldots,H_l)\theta),\\
&&\hspace{1.0cm} 
\mbox{\sl solve}((A_{21},\ldots,A_{2n_2})\theta),\ldots, \mbox{\sl solve}((A_{m1},\ldots,A_{mn_m})\theta).
\end{eqnarray*}

If $H_1\theta$ is {\sl true}, the proof is completed. Otherwise,
we need to show that there exists a query  
in the LD-tree of $P$ and $Q$ satisfying the conditions above.
We are going to construct this query. Let $H'\leftarrow {H'}_1,\ldots, {H'}_l$ be a renamed apart variant of the same clause as represented by an a variant 
of $\mbox{\sl clause}(H, (H_1,\ldots,H_l))$. Such a clause exists by 
definition of $\mbox{\sl ce}(P)$. Since $A_1$ is a variant of $A_0$ and $A_0$ 
is unifiable with $H$ (via $\theta$), $A_1$ is unifiable with ${H'}$. 
Moreover, if $\sigma = \mbox{\sl mgu}(A_1, {H'})$,
then ${H'}_1\theta$ is a variant of $H_1\sigma$ and 
$(H_2\theta,\ldots,H_l\theta)$ forms a partition of
${H'}_2\sigma, \ldots, {H'}_l\sigma$.  

The inductive assumption implies
that 
$(A_{21},\ldots,A_{2n_2}),\ldots,(A_{m1},\ldots,A_{mn_m})$ forms a 
partition of $A_2,\ldots, A_k$. Thus, by Proposition~\ref{partition:prop}
$(A_{21},\ldots,A_{2n_2})\theta,\ldots,(A_{m1},\ldots,$ $A_{mn_m})\theta$
forms a partition of $A_2\sigma,\ldots, A_k\sigma$.

Therefore, by Proposition~\ref{partition:prop}
${H'}_1\sigma, \ldots, {H'}_l\sigma, A_2\sigma,\ldots, A_k\sigma$
satisfies
the condition, completing the proof.
\end{itemize}
\end{proof}

\section{Double extended meta-interpreters do not violate LD-termination}
\label{appendix:non-violation}

In this section we present the formal proofs of the statements in Section~\ref{section:non-violation}. The first result we need to prove is soundness of the
double extended meta-interpreters. As mentioned above, 
in order to prove the soundness result, we use the $s$-semantics
approach presented in~\cite{Bossi:Gabbrielli:Levi:Martelli}. For the sake of 
clarity, we present first the relevant results of their work (notation has 
been adapted).

\begin{definition}{\bf (Definition 3.2~\cite{Bossi:Gabbrielli:Levi:Martelli})}
(Computed answer substitutions semantics, $s$-semantics)
Let $P$ be a definite program. 
\[\begin{array}{lll}
{\cal O}(P) = \{A\mid &
\exists x_1,\ldots,x_n\in\mbox{\it Vars}_n, \exists \sigma, & \\
 & p(x_1,\ldots,x_n)\stackrel{\sigma}{\mapsto}_P\Box & \\
 & A = p(x_1,\ldots,x_n)\sigma & \},\\
\end{array}
\]
where $Q\stackrel{\sigma}{\mapsto}_P\Box$ denotes the LD-refutation of $Q$
in the program $P$ with computed answer substitution $\sigma$.
\end{definition}

Instead of considering Herbrand interpretations, we are going to study
$\pi$-interpretations, defined as subsets of $B^E_P$. 
Next, an immediate consequence operator $T^{\pi}_P$ on $\pi$-interpretations
is defined:
\begin{definition}{\bf (Definition 3.13~\cite{Bossi:Gabbrielli:Levi:Martelli})}
Let $P$ be a definite program and $I$ be a $\pi$-interpretation. 
\[\begin{array}{lll}
T^{\pi}_P(I) = \{A\in B^E_P\mid &
\exists A'\leftarrow B_1,\ldots,B_n\in P,& \\
 & \exists\;B'_1,\ldots,B'_n\;\mbox{\rm variants of atoms in $I$ and renamed apart,}& \\
 & \exists\;\theta = \mbox{\it mgu}((B_1,\ldots,B_n),(B'_1,\ldots,B'_n))\;\mbox{\rm and}\;A = A'\theta & \}
\end{array}
\]
\end{definition}

As usual, we abbreviate $T^{\pi}_D(\emptyset)$ to $T^{\pi}_D$,
$T^{\pi}_D(T^{\pi}_D(\emptyset))$ to $(T^{\pi}_D)^2$, etc.
\begin{example}
\label{example:tpw}
Let $D$ be the meta-interpreter presented in Example~\ref{example:proof:tree}.
Then, the following holds:
\begin{eqnarray*}
T^{\pi}_D &=& \{\mbox{\sl solve}(\mbox{\sl true}, \mbox{\sl true})\}\\
(T^{\pi}_D)^2 &=& \{\mbox{\sl solve}((\mbox{\sl true}, \mbox{\sl true}),(\mbox{\sl true}, \mbox{\sl true})),\mbox{\sl solve}(\mbox{\sl true}, \mbox{\sl true})\}\\
&\vdots&\\
(T^{\pi}_D)^{\omega} &=& \{\mbox{\sl solve}(t,t)\mid t\;\mbox{\rm is a finite sequence of {\sl true}}\}
\end{eqnarray*}
$\hfill\Box$\end{example}

The existence of $(T^{\pi}_P)^{\omega}$ as observed in Example~\ref{example:tpw}
is not a coincidence: one can show that $(T^{\pi}_P)^{\omega}$ exists
and that it is a fixpoint of the computation. Moreover,
${\cal O}(P) = (T^{\pi}_P)^{\omega}$. Formally, 
this relationship is given by the 
following theorem (cf.\ Theorems 3.14 and 3.21~\cite{Bossi:Gabbrielli:Levi:Martelli} and~\cite{Falaschi:Levi:Palamidessi:Martelli}).
\begin{theorem}
\label{theorem:s:semantics}
Let $P$ be a positive program. Then $(T^{\pi}_P)^{\omega}$ exists,
it is a fixpoint of the computation and ${\cal O}(P) = (T^{\pi}_P)^{\omega}$.
\end{theorem}

Using this result, we can show formally 
that double extended meta-interpreters are sound. 

\begin{proof}[Proof of Lemma~\ref{lemma:de:sound}]
Let $D$ be a double extended meta-interpreter. Let $P$ be an interpreted 
program, $Q_0$ be an interpreted query and let $u_1,\ldots, u_n$ be a 
sequence of terms. Then we need to show that for every call 
$\mbox{\sl solve}(Q,t_1,\ldots,t_n)$ in $\mbox{\sl Call}(D\cup\mbox{\sl ce}^D(P),\mbox{\sl solve}(Q_0,u_1,\ldots, u_n))$ there exists 
a call $\mbox{\sl solve}(G)$ in 
$\mbox{\sl Call}(M_0\cup\mbox{\sl ce}(P),Q_0)$ such that $Q$ is an instance 
of $G$.

Since $M_0$ is known to be sound and complete, 
we are going to compare computed answers
obtained with respect to $D$ with computed answers obtained with respect
to $M_0$.

Let $P$ be an interpreted program, 
$Q$ be an interpreted query and let $u_1,\ldots,
u_n$ be a sequence of terms. We have to show that for any computed answer
$\mbox{\sl solve}(t^D,t_1,\ldots,$ $t_n)$ for 
$\mbox{\sl solve}(Q,u_1,\ldots,u_n)$ 
(with respect to $D\cup \mbox{\sl ce}^D(P)$)
there exists a computed answer $\mbox{\sl solve}(t^{M_0})$ 
for $\mbox{\sl solve}(Q)$ 
(with respect to $M_0\cup \mbox{\sl ce}(P)$), such that $t^D$ is an instance of
$t^{M_0}$. 
By Theorem~\ref{theorem:s:semantics}, instead of reasoning on the
computed answers we can apply the $T^{\pi}$ operator. Formally,
we have to show that the following holds:
\begin{eqnarray*}
&& \forall\;\mbox{\sl solve}(t^D,t_1,\ldots,t_n)\in (T^{\pi}_{D\cup \mbox{\sl ce}^D(P)})^{\omega}\\
&& \hspace{1.0cm}\exists\;\mbox{\sl solve}(t^{M_0})\in (T^{\pi}_{M_0\cup \mbox{\sl ce}(P)})^{\omega}\;\mbox{\rm such that}\;t^D\;\mbox{\rm is an instance of}\;t^{M_0}
\end{eqnarray*}
We prove the claim by induction on the power $\alpha$ of $T^{\pi}_{D\cup \mbox{\sl ce}^D(P)}$.

\begin{itemize}
\item If $\alpha = 0$, then $(T^{\pi}_{D\cup \mbox{\sl ce}^D(P)})^\alpha = \emptyset$ and the claim holds vacuously.
\item If $\alpha$ is a successor ordinal, then $(T^{\pi}_{D\cup \mbox{\sl ce}^D(P)})^\alpha = T^{\pi}_{D\cup \mbox{\sl ce}^D(P)}((T^{\pi}_{D\cup \mbox{\sl ce}^D(P)})^{\alpha-1})$. Let $\mbox{\sl solve}(t^D,t_1,\ldots,t_n)\in (T^{\pi}_{D\cup \mbox{\sl ce}^D(P)})^\alpha$.
By definition of the immediate consequence operator, there exist a
clause $(A'\leftarrow B_1,\ldots,B_k)\in D$
and atoms $B'_1,\ldots,B'_k$ (variants of atoms in $(T^{\pi}_{D\cup \mbox{\sl ce}^D(P)})^{\alpha-1}$) such that there exists $\theta\! =\! \mbox{\it mgu}((B_1,\ldots,B_k), (B'_1,\ldots,B'_k))$ and $A'\theta\! =\! \mbox{\sl solve}(t^D,t_1,\ldots,t_n)$. Then, one of the following holds:
\begin{itemize}
\item $A'$ is $\mbox{\sl solve}(\mbox{\sl true},t_{11},\ldots,t_{1n})$.
Then, $t^D = \mbox{\sl true}$. By definition of $M_0$, $\mbox{\sl solve}(\mbox{\sl true})\in (T^{\pi}_{M_0\cup \mbox{\sl ce}(P)})^1$. Thus, the claim
holds for $t^{M_0} = \mbox{\sl true}$.
\item $A'$ is $\mbox{\sl solve}((A,B), t_{21},\ldots,t_{2n})$. Then,
\begin{eqnarray*}
&& \theta = \mbox{\it mgu}((B'_1,\ldots,B'_k),\\
&& \hspace{1.0cm}(D_{11},\ldots,D_{1k_1}, \mbox{\sl solve}(A, t_{31}, \ldots, t_{3n}), D_{21},\ldots,D_{2k_2}, \\
&& \hspace{1.0cm} \mbox{\sl solve}(B, t_{41}, \ldots, t_{4n}), C_{21},\ldots, C_{2m_2}))
\end{eqnarray*}
Let $B'_i = \mbox{\sl solve}(t^{D,i}, t^i_{1}, \ldots, t^i_{n})$ and 
$B'_j = \mbox{\sl solve}(t^{D,j}, t^j_{1}, \ldots, t^j_{n})$ be atoms that 
are unified with $\mbox{\sl solve}(A, t_{31}, \ldots, t_{3n})$ and 
$\mbox{\sl solve}(B, t_{41}, \ldots, t_{4n})$, respectively.
By definition of $\theta$, $\theta$ unifies $A$ with $t^{D,i}$,
$B$ with $t^{D,j}$, and maps $(A,B)\theta$ to $t^D$. In other words,
$t^D = (t^{D,i},t^{D,j})\theta$, i.e., $t^D$ is an instance of
$(t^{D,i},t^{D,j})$.

Both $B'_i$ and $B'_j$ are variants of atoms in 
$(T^{\pi}_{D\cup \mbox{\sl ce}^D(P)})^{\alpha-1}$. Therefore, 
the inductive assumption is applicable, and there exist atoms
$\mbox{\sl solve}(t^{M_0,i})$ and $\mbox{\sl solve}(t^{M_0,j})$
in $(T^{\pi}_{M_0\cup \mbox{\sl ce}(P)})^{\omega}$, such that
$t^{D,i} = t^{M_0,i}\delta^i$ and $t^{D,j} = t^{M_0,j}\delta^j$, 
for some substitutions $\delta^i$ and $\delta^j$. 
Let $s^{M_0,i}$ and $s^{M_0,j}$ be renamed apart variants of
$t^{M_0,i}$ and $t^{M_0,j}$ respectively.

Then, given the clause $\mbox{\sl solve}((A,B)) 
\leftarrow \mbox{\sl solve}(A), \mbox{\sl solve}(B)$,
$s^{M_0,i}$ can be unified with $A$ and $s^{M_0,j}$ with $B$. 
Since $A$ and $B$ are free variables in  the body, 
there exists an mgu $\theta^{M_0}$
of $(\mbox{\sl solve}(A), \mbox{\sl solve}(B))$ and
$(\mbox{\sl solve}(s^{M_0,i}),$ $\mbox{\sl solve}(s^{M_0,j}))$
and $\mbox{\sl solve}((A,B))\theta^{M_0} = 
\mbox{\sl solve}((s^{M_0,i}, s^{M_0,j}))$ holds.
Since $(T^{\pi}_{M_0\cup \mbox{\sl ce}(P)})^{\omega}$ is a fixpoint for
$T^{\pi}_{M_0\cup \mbox{\sl ce}(P)}$ (see~\cite{Falaschi:Levi:Palamidessi:Martelli,Bossi:Gabbrielli:Levi:Martelli}), 
$\mbox{\sl solve}((s^{M_0,i},$ $s^{M_0,j}))\in (T^{\pi}_{M_0\cup \mbox{\sl ce}(P)})^{\omega}$. Moreover, since $s^{M_0,i}$ and $s^{M_0,j}$ are variable disjoint,
$t^{D}$ is an instance of $(s^{M_0,i},s^{M_0,j})$, completing 
the proof in this case.
\item $A'$ is $\mbox{\sl solve}(A, t_{51},\ldots, t_{5n})$, where $A$
is an atom. Then,
\begin{eqnarray*}
&& \theta = \mbox{\it mgu}((B'_1,\ldots,B'_k),\\
&& \hspace{1.0cm}(D_{31},\ldots,D_{3k_3},\mbox{\sl clause}(A,B,s_{1}, \ldots, s_{l}), \\
&& \hspace{1.0cm}D_{41},\ldots,D_{4k_4},\mbox{\sl solve}(B, t_{61}, \ldots, t_{6n})\\
&& \hspace{1.0cm}C_{31},\ldots, C_{3m_3}))
\end{eqnarray*}
Then, let $\mbox{\sl clause}(t^{D}_1,t^{D}_2,t^1_1,\ldots,t^1_l)$ 
and $\mbox{\sl solve}(t^{D}_3,t^2_1,\ldots,t^2_n)$ be elements of
$B'_1,\ldots,$ $B'_k$ that can be unified with 
$\mbox{\sl clause}(A,B,s_{1}, \ldots, s_{l})$ and
$\mbox{\sl solve}(B, t_{61},$ $ \ldots, t_{6n})$, respectively. 
By definition of $\theta$, $t^D = t^{D}_1\theta$ and 
$t^{D}_2\theta = t^{D}_3\theta$.

Since $\mbox{\sl ce}^D$ and $\mbox{\sl ce}$ encode the same program,
there exists an atom $\mbox{\sl clause}(t^{D}_1,$ 
$t^{D}_2)$ in $\mbox{\sl ce}(P)$.
By the inductive assumption, there exists 
$\mbox{\sl solve}(t^{M_0}_3)\in (T^{\pi}_{M_0\cup \mbox{\sl ce}(P)})^{\omega}$,
such that $t^{D}_3$ is an instance of $t^{M_0}_3$. Let $s^{M_0}_3$ be
a variant of $t^{M_0}_3$, such that $s^{M_0}_3$ is variable disjoint from
$\mbox{\sl clause}(t^{D}_1,t^{D}_2)$. 
Let $\rho$ be a variable renaming such that 
$t^{M_0}_3 = s^{M_0}_3\rho$. The only variables that can be affected
by $\rho$ are the variables of $s^{M_0}_3$. By choice of $s^{M_0}_3$,
this implies that $t^{D}_1\rho =  t^{D}_1$. It should also
be observed that $t^{D}_2$ and $s^{M_0}_3$ are unifiable. 

The composition $\rho\theta$ is a unifier of
$(\mbox{\sl clause}(A,B),\mbox{\sl solve}(B))$ and
$(\mbox{\sl clause}(t^{D}_1,$ $t^D_2),\mbox{\sl solve}(s^{M_0}_3))$.
Let $\theta^{M_0}$ be the most general unifier of these expressions.
Thus, there exists a substitution $\delta$ such that 
$\rho\theta = \theta^{M_0}\delta$.
The head instance $t^{M_0}$ that will be inferred is $A\theta^{M_0}$,
that is $t^{D}_1\theta^{M_0}$. Therefore, the following holds:
$t^D = t^{D}_1\theta = t^{D}_1(\rho\theta) = t^{D}_1(\theta^{M_0}\delta) =
(t^{D}_1\theta^{M_0})\delta = t^{M_0}\delta$, completing the proof.
\end{itemize}
\item If $\alpha$ is a limit ordinal,
$(T^{\pi}_{D\cup \mbox{\sl ce}^D(P)})^\alpha = \bigcup_{\beta < \alpha} (T^{\pi}_{D\cup \mbox{\sl ce}^D(P)})^{\beta}$.
In our case, the only limit ordinal is $\omega$. In other words,
$(T^{\pi}_{D\cup \mbox{\sl ce}^D(P)})^\omega = \bigcup_{n < \omega} (T^{\pi}_{D\cup \mbox{\sl ce}^D(P)})^{n}$.
Thus, if an atom belongs to $(T^{\pi}_{D\cup \mbox{\sl ce}^D(P)})^\omega$,
there exists a natural number $n$, such that it belongs
to $(T^{\pi}_{D\cup \mbox{\sl ce}^D(P)})^n$. For such atoms the claim
follows inductively from the previous cases.
\end{itemize}
\end{proof}

As we have seen already in Example~\ref{example:instance:needed}, 
unlike the ``vanilla'' interpreter
$M_0$, double extended meta-interpreters do not necessarily preserve
the set of calls. However, Lemma~\ref{lemma:de:call:set} allowed us to
establish a correspondence between the sets 
of calls obtained with respect to $D$ and those obtained 
with respect to $M_0$. We present the formal proof of the lemma. 

\begin{proof}[Proof of Lemma~\ref{lemma:de:call:set}]
The proof is similar to the proof of Lemma~\ref{lemma:vanilla:calls:preserved}.
It is inductive and based on the derivation of 
$\mbox{\sl solve}(Q,t_1,\ldots,t_n)$. 
We are going to prove a stronger claim 
than we need: we prove that for every query 
\begin{eqnarray*}
&& \mbox{\sl solve}(A_0,t^0_1,\ldots,t^0_n), E_{01},\ldots, E_{0p_0},\\
&& \hspace{1.0cm}\mbox{\sl solve}((A_{21},\ldots,A_{2n_2}),t^2_1,\ldots,t^2_n), E_{21},\ldots, E_{2p_2},\\
&& \hspace{1.0cm}\ldots,\mbox{\sl solve}((A_{m1},\ldots,A_{mn_m}),t^m_1,\ldots,t^m_n), E_{m1},\ldots, E_{mp_m}
\end{eqnarray*}
in the LD-tree of $D\cup\mbox{\sl ce}^D(P)$ and
$\mbox{\sl solve}(Q_0,u_1,\ldots, u_n)$, such that
$A_0$ differs from {\sl true} and $\mbox{\sl rel}(E_{ij})$
differs from $\mbox{\sl solve}$ for all $i$ and $j$, there is a query
\[
\mbox{\sl solve}(B_0), 
\mbox{\sl solve}((B_{21},\ldots,B_{2n_2})), 
\ldots,
\mbox{\sl solve}((B_{m1},\ldots,B_{mn_m})),
\]
such that $A_0$ is an instance of $B_0$ and for all $i,j$, $A_{ij}$
is an instance of $B_{ij}$.
\begin{itemize}
\item {\sl Induction base}
If $\mbox{\sl solve}(Q, t_1,\ldots,t_n) = \mbox{\sl solve}(Q_0, u_1, \ldots, u_n)$, then $Q = Q_0$, and the statement of the lemma holds.
\item {\sl Inductive step} As above, assume that given a query
\begin{eqnarray*}
&& \mbox{\sl solve}(A_0,t^0_1,\ldots,t^0_n), E_{01},\ldots, E_{0p_0},\\
&& \hspace{1.0cm}\mbox{\sl solve}((A_{21},\ldots,A_{2n_2}),t^2_1,\ldots,t^2_n), E_{21},\ldots, E_{2p_2},\\
&& \hspace{1.0cm}\ldots,\mbox{\sl solve}((A_{m1},\ldots,A_{mn_m}),t^m_1,\ldots,t^m_n), E_{m1},\ldots, E_{mp_m}
\end{eqnarray*}
in the LD-tree of $D\cup\mbox{\sl ce}^D(P)$ and
$\mbox{\sl solve}(Q_0,u_1,\ldots, u_n)$, such that
$A_0$ differs from {\sl true} and for all $i,j$,, $\mbox{\sl rel}(E_{ij})$
differs from $\mbox{\sl solve}$. 

Since $A_0$ differs from {\sl true}, there is only one clause in $D$
that can be applied to resolve the query. This clause is (variables
are renamed for the clarity of presentation):
\begin{eqnarray*}
&& \mbox{\sl solve}(\mbox{\sl Head}, t_{51}, \ldots, t_{5n})\leftarrow \\
&& \hspace{1.0cm}D_{31},\ldots,D_{3k_3},\mbox{\sl clause}(\mbox{\sl Head},
\mbox{\sl Body},s_{1}, \ldots, s_{k}), \\
&& \hspace{1.0cm}D_{41},\ldots,D_{4k_4},\mbox{\sl solve}(\mbox{\sl Body}, 
t_{61}, \ldots, t_{6n})\\
&& \hspace{1.0cm} C_{31},\ldots, C_{3m_3}.
\end{eqnarray*}
If the unification fails, the next query is not produced.
Thus, let $\tau$ be the most general unifier of 
$\mbox{\sl solve}(A_0,t^0_1,\ldots,t^0_n)$ and the head of the clause. In particular, $A_0\tau = \mbox{\sl Head}\tau$.
The resolvent obtained is thus,
\begin{eqnarray*}
&& D_{31}\tau,\ldots,D_{3k_3}\tau,
\mbox{\sl clause}(\mbox{\sl Head},\mbox{\sl Body},s_{1}, \ldots, s_{k})\tau,\\
&& \hspace{1.0cm}D_{41}\tau,\ldots,D_{4k_4}\tau,\mbox{\sl solve}(\mbox{\sl Body}, t_{61}, \ldots, t_{6n})\tau,C_{31}\tau,\ldots, \\
&& \hspace{1.0cm}C_{3m_3}\tau,E_{01}\tau, \ldots, E_{0p_0}\tau,\mbox{\sl solve}((A_{21},\ldots,A_{2n_2}),t^2_1,\ldots,t^2_n)\tau, \\
&& \hspace{1.0cm}E_{21}\tau, \ldots,E_{2p_2}\tau,\ldots,
\mbox{\sl solve}((A_{m1},\ldots,A_{mn_m}),t^m_1,\ldots,t^m_n)\tau,\\
&& \hspace{1.0cm}E_{m1}\tau, \ldots, E_{mp_m}\tau
\end{eqnarray*}

Let $\sigma$ be a computed answer substitution for
$D_{31}\tau,\ldots,D_{3k_3}\tau$. If no such a substitution exists
the next query is not produced. Next, the appropriately instantiated
atom of clause has to be unified with one of the facts in the clause
encoding of the interpreted program. Let it be 
$\mbox{\sl clause}(H, (H_1,\ldots,H_l), s^1, \ldots, $ $s^k)$ and let $\theta$
be the corresponding most general unifier. The unifier $\theta$
should exist, otherwise, the computation would have failed.
Let $\delta$ be a computed answer substitution for
$D_{41}\tau,\ldots,D_{4k_4}\tau\sigma\theta$. Then, the next resolvent is
\begin{eqnarray*}
&& \mbox{\sl solve}((H_1,\ldots,H_l), t_{61}, \ldots, t_{6n})\tau\sigma\theta\delta, \\
&&\hspace{1.0cm}C_{31}\tau\sigma\theta\delta,\ldots, C_{3m_3}\tau\sigma\theta\delta,\\
&& \hspace{1.0cm}E_{01}\tau\sigma\theta\delta, \ldots, E_{0p_0}\tau\sigma\theta\delta,\\
&&\hspace{1.0cm}\mbox{\sl solve}((A_{21},\ldots,A_{2n_2}),t^2_1,\ldots,t^2_n)\tau\sigma\theta\delta,
\\
&& \hspace{1.0cm}E_{21}\tau\sigma\theta\delta, \ldots, E_{2p_2}\tau\sigma\theta\delta,\ldots,\\
&& \hspace{1.0cm}\mbox{\sl solve}((A_{m1},\ldots,A_{mn_m}),t^m_1,\ldots,t^m_n)\tau\sigma\theta\delta,\\
&&\hspace{1.0cm}E_{m1}\tau\sigma\theta\delta, \ldots, E_{mp_m}\tau\sigma\theta\delta 
\end{eqnarray*}
If $l > 1$ another clause of $D$ can be applied and the following resultant is
obtained after solving also appropriately instantiated calls to
$D_{11},\ldots,D_{1k_1}$ (where $\rho$ denotes a substitution obtained during this application):
\begin{eqnarray*}
&& \mbox{\sl solve}(H_1, t_{31},\ldots, t_{3n})\tau\sigma\theta\delta\rho, \\
&&\hspace{1.0cm}D_{21}\tau\sigma\theta\delta\rho,\ldots,D_{2k_2}\tau\sigma\theta\delta\rho,\\
&& \hspace{1.0cm}\mbox{\sl solve}(H_2,\ldots,H_l), t_{41}, \ldots, t_{4n})\tau\sigma\theta\delta\rho, \\
&&\hspace{1.0cm}C_{21}\tau\sigma\theta\delta,\ldots, C_{2m_2}\tau\sigma\theta\delta\rho,\\
&& \hspace{1.0cm}C_{31}\tau\sigma\theta\delta,\ldots, C_{3m_3}\tau\sigma\theta\delta,\\
&&\hspace{1.0cm}E_{01}\tau\sigma\theta\delta\rho, \ldots, E_{0p_0}\tau\sigma\theta\delta\rho,\\
&& \hspace{1.0cm}\mbox{\sl solve}((A_{21},\ldots,A_{2n_2}),t^2_1,\ldots,t^2_n)\tau\sigma\theta\delta,\\
&&\hspace{1.0cm}E_{21}\tau\sigma\theta\delta\rho, \ldots, E_{2p_2}\tau\sigma\theta\delta\rho,\ldots,\\
&& \hspace{1.0cm}\mbox{\sl solve}((A_{m1},\ldots,A_{mn_m}),t^m_1,\ldots,t^m_n)\tau\sigma\theta\delta\rho, \\
&&\hspace{1.0cm}E_{m1}\tau\sigma\theta\delta\rho, \ldots, E_{mp_m}\tau\sigma\theta\delta\rho .
\end{eqnarray*}

We need to show that there exists a query in the LD-tree of $M_0\cup\mbox{\sl ce}(P)$ and $\mbox{\sl solve}(Q_0)$ that satisfies our inductive statement.
We are going to construct such a query.

By the Inductive Assumption, there exists a query 
\[
\mbox{\sl solve}(B_0), 
\mbox{\sl solve}((B_{21},\ldots,B_{2n_2})), 
\ldots,
\mbox{\sl solve}((B_{m1},\ldots,B_{mn_m})),
\]
such that $A_0$ is an instance of $B_0$ and for all $i,j$, $A_{ij}$
is an instance of $B_{ij}$. The only clause that can be applied to
resolve this query is the clause corresponding to the clause applied
to resolve the corresponding query above. Let $\mbox{\sl clause}(H',({H'}_1,
\ldots, {H'}_l))$ be a clause encoding of the same clause as represented
by $\mbox{\sl clause}(H, (H_1,\ldots,H_l), s^1,$ $\ldots, s^k)$. In order to show that
the unification of $B_0$ and $H'$ succeeds observe that the following holds:
\begin{eqnarray*}
&& A_0\;\mbox{\rm is an instance of}\;B_0\\
&& A_0\tau\sigma\;\mbox{\rm is unifiable with}\;H\\
&& H'\;\mbox{\rm is a variant of}\;H
\end{eqnarray*}
Let $\mu$ be the most general unifier of $B_0$ and $H'$. Then,
$(H_1,\ldots,H_l)\tau\sigma\theta$ is an instance of $({H'}_1,
\ldots, {H'}_l))\mu$. The following resolvent is obtained
\begin{eqnarray*}
&& \mbox{\sl solve}(({H'}_1,\ldots, {H'}_l)\mu),
\mbox{\sl solve}((B_{21},\ldots,B_{2n_2})\mu), \\
&& \hspace{1.0cm} \ldots,\mbox{\sl solve}((B_{m1},\ldots,B_{mn_m})\mu)
\end{eqnarray*}

If $l > 1$ another rule of $M_0$ can be applied and the following resolvent
will be obtained:
\begin{eqnarray*}
&& \mbox{\sl solve}({H'}_1\mu),\mbox{\sl solve}(({H'}_2,\ldots, {H'}_l)\mu),
\mbox{\sl solve}((B_{21},\ldots,B_{2n_2})\mu), \\
&& \hspace{1.0cm} \ldots, \mbox{\sl solve}((B_{m1},\ldots,B_{mn_m})\mu),
\end{eqnarray*}

We claim that the latter query satisfies the conditions of the lemma.
Indeed, since $(H_1,\ldots,H_l)\tau\sigma\theta$ is an instance of $({H'}_1,
\ldots, {H'}_l))\mu$, the query 
$(H_1,\ldots,$ $H_l)\tau\sigma\theta\delta\rho$ is also 
an instance of $({H'}_1,\ldots, {H'}_l))\mu$. Thus,
$H_1\tau\sigma\theta\delta\rho$ is an instance of ${H'}_1\mu$
and $(H_2,\ldots,H_l)\tau\sigma\theta\delta\rho$ is 
an instance of $({H'}_2,\ldots, {H'}_l))\mu$. Recalling the inductive
assumption allows us to complete the proof.
\end{itemize}
\end{proof}

Finally we present a formal proof of Theorem~\ref{theorem:de:1}.
\\$\;$\\
{\it Theorem~\ref{theorem:de:1}}\ \\
Let $P$ be an interpreted program, $D$ a double extended 
meta-interpreter, and $Q\in B^E_{D\cup \mbox{\sl ce}^D(P)}$, 
such that $Q$ is terminating with respect to $P$.
Let $u_1,\ldots,u_n$ be a sequence of terms such that 
$\{A\mid A\in \mbox{\sl Call}(D\cup \mbox{\sl ce}^D(P), \mbox{\sl solve}(Q, u_1,\ldots,u_n)), \mbox{\sl solve}\neq \mbox{\sl rel}(A)\}$ is terminating 
with respect to $D$.  Then $\mbox{\sl solve}(Q, u_1,\ldots,u_n)$ terminates 
with respect to $D\cup \mbox{\sl ce}^D(P)$.
\begin{proof}
Let $M_0$ be the ``vanilla'' meta-interpreter. By Corollary~\ref{corollary:pres:term} $M_0\cup \mbox{\sl ce}(P)$ terminates with respect to $\mbox{\sl solve}(Q)$.
This is equivalent to saying that $M_0\cup \mbox{\sl ce}(P)$ terminates with 
respect to $\mbox{\sl Call}(M_0\cup \mbox{\sl ce}(P),\mbox{\sl solve}(Q))$.
Moreover, this also means that $M_0\cup \mbox{\sl ce}(P)$ terminates with 
respect to $S = \{A\eta\mid A\in \mbox{\sl Call}(M_0\cup \mbox{\sl ce}(P),\mbox{\sl solve}(Q)), \eta\;\mbox{\rm is a substitution}\}$.
By Theorem~\ref{taset:term} $M_0\cup \mbox{\sl ce}(P)$ is order-acceptable
with respect to $S$. Let $\geq_1$ be a minimal well-founded 
quasi-ordering, such that $M_0\cup \mbox{\sl ce}(P)$ is order-acceptable with 
respect to $S$ via it. 

Similarly, let $\geq_2$ be a well-founded quasi-ordering such that $D$ is 
order-acceptable with respect to 
$\{A\mid A\in \mbox{\sl Call}(D\cup \mbox{\sl ce}^D(P), \mbox{\sl solve}(Q, u_1,
\ldots, u_n)), \mbox{\sl solve}\neq \mbox{\sl rel}(A)\}$ via $\geq_2$.

We have to show that there exists a well-founded quasi-ordering 
$\succeq$ such that
$D\cup \mbox{\sl ce}^D(P)$ is order-acceptable with respect to 
$\{\mbox{\sl solve}(Q, u_1, \ldots, u_n)\}$ via $\succeq$. 
By Theorem~\ref{taset:term} this 
will imply termination.

Let $\succeq$ be defined on $B^E_{D\cup\mbox{\sl ce}^D(P)}$ as follows
for any terms $t_1, t_2, t^1_{1},\ldots, t^1_{n}, t^2_{1},\ldots, t^2_{n}$ 
and any atoms $a_1, a_2$:
\begin{enumerate}
\item $\mbox{\sl solve}(t_1, t^1_{1},\ldots, t^1_{n}) \succ \mbox{\sl solve}(t_2, t^2_{1},\ldots, t^2_{n})$, if there is a term $t$, such that 
$\mbox{\sl solve}(t_1) >_1 \mbox{\sl solve}(t)$ and $t_2 = t\theta$ for some 
substitution $\theta$;
\item $a_1 \succ a_2$, if $\mbox{\sl rel}(a_1) \not = \mbox{\sl solve}$,
$\mbox{\sl rel}(a_2) \not = \mbox{\sl solve}$ and $a_1 >_2 a_2$;
\item $\mbox{\sl solve}(t_1, t^1_{1},\ldots, t^1_{n}) \succ a_1$, if 
$\mbox{\sl rel}(a_1) \not = \mbox{\sl solve}$;
\item $a_1 \preceq\succeq a_2$, if $a_1$ and $a_2$ are identical.
\end{enumerate}

In order to prove that $\succ$ is an ordering and that this ordering
is well-founded we make use of the minimality of $>$ and of the Lifting 
Theorem (Theorem 3.22~\cite{Apt:Book}). 
We prove irreflexivity only as
antisymmetry, transitivity and well-foundedness can be proved
in a similar way.
Let $A$ be an atom such that $A\succ A$. If $A$ is of the form
$\mbox{\sl solve}(t_1, t^1_{1},\ldots, t^1_{n})$, then
$\mbox{\sl solve}(t_1) >_1 \mbox{\sl solve}(t)$ and $t_1 = t\theta$ 
should hold for some substitution $\theta$. By Lemma~\ref{lemma:minimal:ds}, 
there exists a directed derivation $Q_0,\ldots, Q_n$, such that
$Q_0 = \mbox{\sl solve}(t_1) ( = \mbox{\sl solve}(t\theta))$, 
$Q_n = \mbox{\sl solve}(t)$, for all
$0\leq i < n$, $Q_i >_1 Q_{i+1}$. In other words,
$Q_0 = Q_n\theta$. Then, by the Lifting Theorem (Theorem 3.22~\cite{Apt:Book})
there exists a derivation starting with $Q_n$, selecting the same atoms
as in $Q_0,\ldots, Q_n$, and resulting in $Q_{2n}$, such that
$Q_n$ is an instance of $Q_{2n}$. 
Proceeding in this way we can construct an infinite
directed derivation, contradicting the well-foundedness of $>_1$.
Alternatively, if the predicate of $A$ differs from
$\mbox{\sl solve}$, $A >_2 A$ should hold, contradicting the irreflexivity
of $>_2$. 
\eat{
\item  Let $A$ and $B$ be atoms such that $A\succ B$ and $B\succ A$.
If $A = \mbox{\sl solve}(t_1, t^1_{1},\ldots, t^1_{n})$ and $B = \mbox{\sl solve}(t_2, t^2_{1},\ldots, t^2_{n})$, the first case is applicable, i.e., there
exist terms $t_3$ and $t_4$ and substitutions $\theta$ and $\sigma$, 
such that $\mbox{\sl solve}(t_1) >_1 \mbox{\sl solve}(t_3)$,
$\mbox{\sl solve}(t_2) >_1 \mbox{\sl solve}(t_4)$, $t_2 = t_3\theta$
and $t_1 = t_4\sigma$. Reasoning in the same way as above one can
construct an infinite directed derivation, contradicting the well-foundedness 
of $>_1$.

If $\mbox{\sl rel}(A) \not = \mbox{\sl solve}$ and $\mbox{\sl rel}(B) \not = \mbox{\sl solve}$, the second case is applicable providing a contradiction with the asymmetry of $>_2$. The remaining cases are eliminated by definition
of $\succ$---indeed, there exist no atoms $A$ and $B$ such that
$\mbox{\sl rel}(A) \not = \mbox{\sl solve}$, 
$\mbox{\sl rel}(B) = \mbox{\sl solve}$, and $A\succ B$.

\item Let $A$, $B$ and $C$ be atoms such that $A\succ B$ and $B\succ C$. If
either all of $A$, $B$ and $C$ are atoms of {\sl solve}, or none of them is an
atom of {\sl solve}, $A\succ C$ follows from the transitivity of 
$>_2$ or from a directed derivation construction similar to above. 
Assuming $A$ to be an atom of {\sl solve} and $C$ not to 
be an atom of {\sl solve} leads to $A\succ C$ by definition of $\succ$.
In the remaining cases, either $\mbox{\sl rel}(A) \not = \mbox{\sl solve}$ and $\mbox{\sl rel}(B) = \mbox{\sl solve}$ or $\mbox{\sl rel}(B) \not = \mbox{\sl solve}$ and $\mbox{\sl rel}(C) = \mbox{\sl solve}$, which 
are inconsistent with the definition of $\succ$.

\item Let $A_1,A_2,\ldots$ be an infinite sequence of atoms such that
$A_n \succ A_{n+1}$ for all $n$. Then, since there are only finitely
many different predicates, there exists a predicate $p$, such that 
its atoms occur infinitely often
in this sequence. If $p$ is not $\mbox{\sl solve}$, contradiction with the
well-foundedness of $>_2$ is obtained. Otherwise, an infinite directed 
derivation can be constructed as above, contradicting the well-foundedness
of $>_1$.
}

Next we prove that $D\cup \mbox{\sl ce}^D(P)$ is order-acceptable with respect to
$\mbox{\sl solve}(Q, u_1, \ldots,$ $u_n)$. 
Let $A_0\in \mbox{\sl Call}(D\cup \mbox{\sl ce}^D(P), \mbox{\sl solve}(Q, 
u_1,\ldots, u_n))$. We distinguish between the following two cases.

First, assume that $\mbox{\sl rel}(A_0) \not = \mbox{\sl solve}$.
If $\mbox{\sl rel}(A_0) = \mbox{\sl clause}$ the order-acceptability 
condition holds immediately, since there are no recursive clauses 
defining this predicate.
Otherwise, order-acceptability of $D$ with respect to
$\{A\mid A\in \mbox{\sl Call}(D\cup \mbox{\sl ce}^D(P),$ $\mbox{\sl solve}(Q, u_1,\ldots, u_n)),$ 
$\mbox{\sl solve}\neq \mbox{\sl rel}(A)\}$ via $>_2$ implies that 
for any clause $A'\leftarrow B_1,\ldots, B_s$, such that 
$\mbox{\sl mgu}(A_0,$ $A') = \theta$ exists, for any atom $B_i$, such that 
$\mbox{\sl rel}(A_0) \simeq \mbox{\sl rel}(B_i)$ and for any computed
answer substitution $\sigma$ for $\leftarrow (B_1,\ldots, B_{i-1})\theta$
holds $A_0 >_2 B_i\theta\sigma$. Therefore, $A_0 > B_i\theta\sigma$ holds.

Hence, in the remainder of the proof we assume that
$\mbox{\sl rel}(A_0) = \mbox{\sl solve}$. In this case,
there are three different kinds of clauses $A'\leftarrow B_1,\ldots, B_s$, 
such that $\mbox{\sl mgu}(A_0,A') = \theta$ exists.
\begin{itemize}
\item  $A'\leftarrow B_1,\ldots, B_s$ is $\mbox{\sl solve}(\mbox{\sl true}, t_{
11}, \ldots, t_{1n})\leftarrow C_{11},\ldots, C_{1m_1}$. In this
case there are no recursive body subgoals and order-ac\-cept\-abil\-i\-ty condition is trivially satisfied.
\item $A'\leftarrow B_1,\ldots, B_s$ is 
\begin{eqnarray*}
&& \mbox{\sl solve}((A,B), t_{21}, \ldots, t_{2n})\leftarrow D_{11},\ldots,D_{1k_1}, \mbox{\sl solve}(A, t_{31}, \ldots, t_{3n}), \\
&& \hspace{1.0cm}D_{21},\ldots,D_{2k_2}, \mbox{\sl solve}(B, t_{41}, \ldots, t_{4n}), C_{21},\ldots, C_{2m_2}\mbox{.}
\end{eqnarray*}
By our assumption for all $k,l,p,q$, neither $\mbox{\sl rel}(C_{kl})$ nor
$\mbox{\sl rel}(D_{pq})$ depend on {\sl solve}. Thus,
the only recursive subgoals are an instance of
$\mbox{\sl solve}(A, t_{31}, \ldots, t_{3n})$, and an 
instance of $\mbox{\sl solve}(B, t_{41}, \ldots, t_{4n})$. We have to show
that $A_0 \succ \mbox{\sl solve}(A, t_{31}, \ldots, t_{3n})\theta\sigma$
and $A_0 \succ \mbox{\sl solve}(B, t_{41}, \ldots, t_{4n})\theta\sigma\delta\rho$,
where $\theta = \mbox{\sl mgu}(A_0,\mbox{\sl solve}((A,B), t_{21}, \ldots, t_{2n}))$, $\sigma$ is a computed answer substitution for $(D_{11},\ldots,D_{1k_1})\theta$, $\delta$ is a computed answer substitution for $\mbox{\sl solve}(A, t_{31},$ $\ldots, t_{3n})\theta\sigma$, 
and $\rho$ is a computed answer substitution for
$(D_{21},\ldots,D_{2k_2})\theta\sigma\delta$.

Let $A'_0$ be obtained from $A_0$ by dropping all the arguments except for the
first one. By definition of $\theta$, $\theta$ is a unifier of
$A'_0$ and $\mbox{\sl solve}((A,B))$. Let $\theta'$ be a most general 
unifier of $A'_0$ and $\mbox{\sl solve}((A,B))$. 
Thus, there exists a substitution
$\theta''$ such that $\theta = \theta'\theta''$. Therefore,
$A\theta\sigma = A(\theta'\theta'')\sigma = A\theta'(\theta''\sigma)$,
i.e., $A\theta\sigma$ is an instance of $A\theta'$. 

Moreover, by Lemma~\ref{lemma:de:call:set}, 
$A'_0$ is an instance of some $A''_0\!\in\! \mbox{\sl Call}(M_0\cup\mbox{\sl ce}(P), \mbox{\sl solve}(Q))$. Thus, $A'_0\in S$. Thus,
by definition of $>_1$, $A'_0 >_1 \mbox{\sl solve}(A\theta')$ holds, and
by definition of
$\succeq$, $A_0 \succ \mbox{\sl solve}(A, t_{31}, \ldots, t_{3n})\theta\sigma$.

Since $\delta$ is a computed answer substitution for $\mbox{\sl solve}(A, t_{31}, \ldots,$ $ t_{3n})\theta\sigma$, by Corollary 3.23~\cite{Apt:Book}, there
exists a computed answer substitution $\delta'$ for $\mbox{\sl solve}(A,$ $t_{31}, \ldots, t_{3n})\theta'$, such that $\mbox{\sl solve}(A, t_{31}, \ldots,$ $t_{3n})\theta'\delta'$ is more general than $\mbox{\sl solve}(A,$ $t_{31}, \ldots,$ $ t_{3n})\theta\sigma\delta$. 
Moreover, by Lemma~\ref{lemma:de:sound} there exists a computed answer
substitution $\delta''$ for $\mbox{\sl solve}(A\theta')$
such that $A\theta'\delta'$ is an instance of $A\theta'\delta''$.
Transitivity implies that $\theta'\delta''$ is more general than
$\theta\sigma\delta\rho$. Moreover, since all the unifiers are
relevant, i.e., the only variables affected are the variables of 
the terms to be unified, $\mbox{\sl solve}(B\theta\sigma\delta\rho)$ 
is an instance of $\mbox{\sl solve}(B\theta'\delta'')$. Furthermore, 
since $M_0\cup\mbox{\sl ce}(P)$ is order-acceptable with respect to $\mbox{\sl solve}(Q)$ via $\geq_1$, and $\delta''$ is a computed answer 
substitution for $\mbox{\sl solve}(A)\theta'$,
$A'_0 >_1 \mbox{\sl solve}(B\theta'\delta'')$. Thus, 
by definition of
$\succeq$, $A_0 \succ \mbox{\sl solve}(B, t_{41}, \ldots, t_{4n})\theta\sigma\delta\rho$.
\item $A'\leftarrow B_1,\ldots, B_s$ is 
\begin{eqnarray*}
&& \mbox{\sl solve}(A, t_{51}, \ldots, t_{5n})\leftarrow D_{31},\ldots,D_{3k_3},\mbox{\sl clause}(A,B,s_{1}, \ldots, s_{k}), \\
&& \hspace{1.0cm}D_{41},\ldots,D_{4k_4},\mbox{\sl solve}(B, t_{61}, \ldots, t_{6n}), C_{31},\ldots, C_{3m_3}\mbox{.}
\end{eqnarray*}
Similarly to the previous case, let $\theta$ be a most general unifier of
$A_0$ and $\mbox{\sl solve}(A,$ $t_{51},\ldots, t_{5n})$. Then, let
$A'_0$ be an atom obtained from $A_0$ by dropping all the arguments except
for the first one. Observe that $\theta$ is a unifier of $A'$ and $A$,
and, therefore, if $\theta'$ is a most general unifier of these atoms, 
$\theta = \theta'\theta''$ for some substitution $\theta''$. Clearly,
if an encoding of a clause in $P$ via $\mbox{\sl ce}^D$ can be unified with
$\mbox{\sl clause}(A,B,s_{1}, \ldots, s_{k})\theta\sigma$ for
a computed answer $\sigma$ for $\leftarrow (D_{31},\ldots,D_{3k_3})\theta$,
the encoding of the same clause by $\mbox{\sl ce}$ can be unified
with $\mbox{\sl clause}(A,B)\theta'$. Thus, if $\delta$ denotes the
computed answer substitution for $\mbox{\sl clause}(A,B,s_{1}, \ldots, s_{k})\theta\sigma$ and $\delta'$ denotes the computed answer substitution for 
$\mbox{\sl clause}(A,B)\theta'$, $B\theta\sigma\delta$ is an instance
of $B\theta'\delta'$. Furthermore, for any computed answer substitution
$\rho$ for $\leftarrow (D_{41},\ldots,D_{4k_4})\theta\sigma\delta$,
$B\theta\sigma\delta\rho$ is an instance of $B\theta'\delta'$. 
Reasoning as above and applying order-acceptability via $\geq_1$,
proves $A'_0 >_1 \mbox{\sl solve}(B\theta'\delta')$ and, thus, by definition 
of $\succeq$,
$A_0 \succ \mbox{\sl solve}(B, t_{61},$ $\ldots, t_{6n})\theta\sigma\delta\rho$.
\end{itemize}
Thus, $D\cup \mbox{\sl ce}^D(P)$ is order-acceptable with respect to
$\mbox{\sl solve}(Q)$ and, by Theorem~\ref{taset:term} terminates with
respect to it.
\end{proof}

\section{Restricted double extended meta-interpreters do not improve LD-termination}
\label{appendix:non-improvement}

Example~\ref{example:no:completeness} illustrated that restricted double 
extended meta-interpreters are not necessarily complete. However,
a weaker result, stated in Lemma~\ref{lemma:de:complete} can be shown. Here
we prove this result formally.

\begin{proof}[Proof of Lemma~\ref{lemma:de:complete}]
By the completeness result of Levi and Ramundo~\cite{Levi:Ramundo} every 
LD-derivation of $P$ and $Q$ can be mimicked by an LD-derivation of
$M_0\cup\mbox{\sl ce}(P)$ and $\mbox{\sl solve}(Q)$. 

Similarly to Lemma~\ref{lemma:de:sound} the proof is done by induction
on powers of the immediate consequence operator. In this case, however,
the immediate consequence operator for $M_0\cup\mbox{\sl ce}(P)$ is considered.
More formally, we have to show that 
\begin{eqnarray*}
&& \forall\;\alpha\leq \omega\;\forall\;\mbox{\sl solve}(t^{M_0})\in (T^{\pi}_{M_0\cup \mbox{\sl ce}(P)})^{\alpha}\\
&& \hspace{1.0cm}\exists\;\mbox{\sl solve}(t^{D}, t_1,\ldots,t_n)\in (T^{\pi}_{D\cup \mbox{\sl ce}^D(P)})^{\omega}\;\mbox{\rm such that}\;t^{M_0}\;\mbox{\rm is a variant of}\;t^{D}
\end{eqnarray*}
\begin{itemize}
\item If $\alpha = 1$, the only value for $t^{M_0}$ is $\mbox{\sl true}$.
Assumptions of the lemma and the fact that $D$ is restricted imply that
$(C_{11},\ldots, C_{1m_1})$ finitely succeeds. Thus, there
exists a finite power $l$ of the immediate consequence operator 
$T^{\pi}_{D\cup\mbox{\sl ce}^D(P)}$
for 
$D\cup\mbox{\sl ce}^D(P)$ such that $\mbox{\sl solve}(\mbox{\sl true}, t_{11},\ldots, t_{1n})$ is contained in
$(T^{\pi}_{D\cup \mbox{\sl ce}^D(P)})^l$.
\item If $\alpha$ is a successor ordinal, $\mbox{\sl solve}(t^{M_0})$
has been produced by applying one of the clauses. We distinguish
between the following cases:
\begin{itemize}
\item Let $\mbox{\sl solve}((A,B))\leftarrow \mbox{\sl solve}(A),
\mbox{\sl solve}(B)$ be the applied clause. Then, there exist
$\mbox{\sl solve}(t^{M_0}_1)$ and $\mbox{\sl solve}(t^{M_0}_2)$, 
variants of atoms in 
$(T^{\pi}_{M_0\cup \mbox{\sl ce}(P)})^{\alpha-1}$, such that there
exists a most general unifier $\theta$ of $(\mbox{\sl solve}(A),
\mbox{\sl solve}(B))$ and $(\mbox{\sl solve}(t^{M_0}_1),
\mbox{\sl solve}(t^{M_0}_2))$. Then, $t^{M_0}$ is 
$(t^{M_0}_1,t^{M_0}_2)\theta$, i.e., $t^{M_0}_1,t^{M_0}_2$, since
$A$ and $B$ are free variables.

By our inductive assumption there exist atoms $\mbox{\sl solve}(t^{D}_1,
t^1_1,\ldots,t^1_n)$ and $\mbox{\sl solve}(t^{D}_2,$ $t^2_1,\ldots,t^2_n)$
in $(T^{\pi}_{D\cup \mbox{\sl ce}^D(P)})^\omega$, such that
$t^{M_0}_1$ is a variant of $t^{D}_1$ and $t^{M_0}_2$ is a variant of 
$t^{D}_2$. Observe that 
$(\mbox{\sl solve}(t^{D}_1,t^1_1,\ldots,t^1_n), \mbox{\sl solve}(t^{D}_2,t^2_1,\ldots,t^2_n))$
is unifiable with $(\mbox{\sl solve}(A, t_{31},\ldots,t_{3n}),
\mbox{\sl solve}(B, t_{41},\ldots, t_{4n}))$, since $A$ and $B$
are free variables and by Lemma~\ref{lemma:de:un}.

Assumptions of the lemma and Definition~\ref{definition:de:restricted}
imply that there exists a finite power $l$, such that all computed answers
of $\mbox{\sl rel}(C_{pq})$ and $\mbox{\sl rel}(D_{rs})$ are contained in
$(T^{\pi}_{D\cup \mbox{\sl ce}^D(P)})^{l}$. Then, there exist atoms $d_{11},
\ldots, d_{1k_1}, d_{21},$ $\ldots, d_{2k_2}, c_{21}, \ldots, c_{2m_2}$
such that there exists 
\begin{eqnarray*}
\theta^D &=& \mbox{\sl mgu}((d_{11},\ldots, d_{1k_1},\mbox{\sl solve}(t^{D}_1,t^1_1,\ldots,t^1_n), d_{21}, \ldots, d_{2k_2}, \\
&& \hspace{1.0cm} \mbox{\sl solve}(t^{D}_2,t^2_1,\ldots,t^2_n),
c_{21}, \ldots, c_{2m_2}), (D_{11},\ldots, D_{1k_1},\\
&& \hspace{1.0cm} \mbox{\sl solve}(A, t_{31},\ldots,t_{3n}),
D_{21},\ldots, D_{2k_2},\mbox{\sl solve}(B, t_{41},\\
&& \hspace{1.0cm} 
\ldots, t_{4n}), C_{21},\ldots,C_{2m_2}))
\end{eqnarray*}
By the third requirement of Definition~\ref{definition:de:restricted},
for every computed answer substitution $\sigma$ for $(D_{11},\ldots,
D_{1k_1})$, $\sigma$ does not affect $A$. Thus, the same holds for
any correct answer substitution as well. In other words, for any
$d_{11},\ldots, d_{1k_1}$, the sequence
\[(d_{11},\ldots, d_{1k_1},\mbox{\sl solve}(t^{D}_1,t^1_1,\ldots,t^1_n))\]
is unifiable with $(D_{11},\ldots, D_{1k_1}, \mbox{\sl solve}(A, t_{31},\ldots,t_{3n}))$ and, in particular, $t^D_1\theta^D = t^D_1$.
Reasoning in a similar way allows us to conclude that $t^D_2\theta^D = t^D_2$.
Thus, an element $t^D$ inferred at this step is $(t^D_1\theta^D, t^D_2\theta^D) = (t^D_1, t^D_2)$. By choice of $t^{D}_1$ and $t^{D}_2$,
$t^{M_0} = (t^{M_0}_1, t^{M_0}_2)$ is a variant of $(t^D_1, t^D_2) = t^D$,
completing the proof.
\item Let $\mbox{\sl solve}(A)\leftarrow \mbox{\sl clause}(A,B),\mbox{\sl solve}(B)$ be the clause applied. Then, there exist atoms $\mbox{\sl clause}(t^{M_0}_1, t^{M_0}_2)$ and $\mbox{\sl solve}(t^{M_0}_3)$, variants of some atoms in 
$(T^{\pi}_{M_0\cup \mbox{\sl ce}(P)})^{\alpha-1}$, such that there
exists an mgu $\theta$ of $(\mbox{\sl clause}(A,B),
\mbox{\sl solve}(B))$ and $(\mbox{\sl clause}(t^{M_0}_1, t^{M_0}_2),$
$\mbox{\sl solve}(t^{M_0}_3))$. Then, $t^{M_0}$ is 
$t^{M_0}_1\theta$, and $t^{M_0}_2\theta =t^{M_0}_3\theta$. Moreover,
the restriction of $\theta$ to the variables of $t^{M_0}_2$ and
of $t^{M_0}_3$ is an mgu of these terms.

Since $\mbox{\sl ce}$ and $\mbox{\sl ce}^D$ encode the same interpreted program 
$P$, there are some $s^1,\ldots,s^k$, such that 
$\mbox{\sl clause}(t^{M_0}_1, t^{M_0}_2, s^1,\ldots,s^k)$ belongs to
$\mbox{\sl ce}^D(P)$. Moreover, by our inductive assumption, there exists
an atom $\mbox{\sl solve}(t^D_3,t^3_1,$ $\ldots,t^3_n)\in (T^{\pi}_{D\cup \mbox{\sl ce}^D(P)})^{\omega}$, such that $t^{M_0}_3$ is a variant of $t^D_3$.
Reasoning as above, observe that there exist atoms $d_{31},\ldots,d_{3k_3},
d_{41},\ldots,d_{4k_4}$ and $c_{31},$ $\ldots,c_{3m_3}$ such that
there exists 
\begin{eqnarray*}
\theta^D &=& \mbox{\sl mgu}((d_{31},\ldots, d_{3k_3},\mbox{\sl clause}(t^{M_0}_1,t^{M_0}_2,s^1,\ldots,s^k), d_{41}, \ldots, \\
&& \hspace{1.0cm}  d_{4k_4}, \mbox{\sl solve}(t^{D}_3,t^3_1,\ldots,t^3_n),
c_{31}, \ldots, c_{3m_3}), (D_{31},\ldots, \\
&& \hspace{1.0cm} D_{3k_3},\mbox{\sl clause}(A, B, s_{1},\ldots,s_{k}),
D_{41},\ldots, D_{4k_4},\\
&& \hspace{1.0cm} \mbox{\sl solve}(B, t_{61}, \ldots, t_{6n}), C_{31},\ldots,C_{3m_3}))
\end{eqnarray*}
and, similarly to the previous case, $t^D = t^{M_0}_1\theta^D$
and $t^{M_0}_2\theta^D = t^D_3\theta^D$. The third condition of
Definition~\ref{definition:de:restricted} implies that $\theta^D$
restricted to variables of $t^{M_0}_2$ and $t^D_3$ is an mgu.
Thus, \[\theta^D\mid_{\mbox{\it Var}(t^{M_0}_2)\cup \mbox{\it Var}(t^{D}_3)}
= \theta\mid_{\mbox{\it Var}(t^{M_0}_2)\cup \mbox{\it Var}(t^{M_0}_3)}\rho\]
for some variable renaming $\rho$. In other words, $t^{M_0}_1\theta 
(= t^{M_0})$ and $t^{M_0}_1\theta^D (= t^D)$ are variants, completing 
the proof.
\end{itemize}
\item Finally, similarly to Lemma~\ref{lemma:de:sound} the only case of 
a limit ordinal is $\omega$, and, since 
\[T^\pi_{M_0\cup \mbox{\sl ce}(P)})^{\omega}
= \bigcup_{n < \omega} T^\pi_{M_0\cup \mbox{\sl ce}(P)})^{n}
\]
the claim follows from the established result for the 
finite powers of the operator. 
\end{itemize}
\end{proof}

Next we present a formal proof of Theorem~\ref{theorem:de:2}.
\\$\;$\\
{\it Theorem~\ref{theorem:de:2}}\ \\
Let $D$ be a restricted double extended meta-interpreter.
Let $P$ be an interpreted program and let $Q$ be an interpreted query,
such that $D\cup\mbox{\sl ce}^D(P)$ LD-terminates for $\mbox{\sl solve}(Q,
v_1,\ldots,$ $v_n)$, where $(v_1,\ldots,v_n)$ are terms such that
$\mbox{\sl solve}(Q, v_1,\ldots,v_n)$ is restricted.
Then, $P$ LD-terminates with respect to $Q$.

\begin{proof}
In order to show that $P$ LD-terminates for $Q$ it is sufficient to 
prove that $M_0\cup\mbox{\sl ce}(P)$ LD-terminates with respect to
$\mbox{\sl solve}(Q)$.
Then, by Theorem~\ref{meta:interpreted} $P$ LD-terminates with respect 
to $Q$. Thus, we aim to establish order-acceptability of 
$M_0\cup\mbox{\sl ce}(P)$ with respect to $\mbox{\sl solve}(Q)$.

First of all, we define a relationship on $B^E_{M_0\cup \mbox{\sl ce}(P)}$.
Then we show that the relationship is a quasi-ordering, that it is
well-founded and that $M_0\cup\mbox{\sl ce}(P)$ is order-acceptable
with respect to $\mbox{\sl solve}(Q)$ via the relationship defined.

Since $D\cup\mbox{\sl ce}^D(P)$ LD-terminates for $\mbox{\sl solve}(Q,v_1,\ldots,v_n)$, $D\cup\mbox{\sl ce}^D(P)$ is order-accep\-tab\-le with respect to 
$\mbox{\sl solve}(Q,v_1,\ldots,v_n)$ via a quasi-ordering. Let a minimal
quasi-ordering such that $D\cup\mbox{\sl ce}^D(P)$ is order-accep\-tab\-le with 
respect to $\mbox{\sl solve}(Q,v_1,\ldots,v_n)$ via it,
be denoted $\geq$. Then, we define 
$\mbox{\sl solve}(s) \succ \mbox{\sl solve}(t)$
if there exist $\mbox{\sl solve}(s, s_1,\ldots, s_n), \mbox{\sl solve}(t, t_1,$ $\ldots, t_n)\in$ $\mbox{\sl Call}(D\cup\mbox{\sl ce}^D(P), \mbox{\sl solve}(Q,v_1,\ldots,v_n))$ such that $\mbox{\sl solve}(s, s_1,\ldots, s_n) > 
\mbox{\sl solve}(t, t_1,$ $\ldots, t_n)$ and $\mbox{\sl solve}(s) \preceq\succeq \mbox{\sl solve}(t)$ if $\mbox{\sl solve}(s)$ and $\mbox{\sl solve}(t)$ are identical.

Next we have to show that $\succ$ is indeed an ordering.
We prove irreflexivity only. 
Antisymmetry and transitivity can shown in a similar fashion, and
well-foundedness of $\succ$ follows immediately from the well-foundedness
of $>$.

Let $\mbox{\sl solve}(t)\succ \mbox{\sl solve}(t)$ for some $t$. Then,
there exist atoms $\mbox{\sl solve}(t, s_1,\ldots, s_n)$ and 
$\mbox{\sl solve}(t, t_1,\ldots, t_n)$ in $\mbox{\sl Call}(D\cup\mbox{\sl ce}^D(P), \mbox{\sl solve}(Q,v_1,\ldots,v_n))$ such that
$\mbox{\sl solve}(t, $ $s_1,\ldots,$ $s_n) > \mbox{\sl solve}(t, t_1,\ldots, t_n)$.

If $(t_{11},\ldots,t_{1n})\not\in \mbox{\it Vars}_n$ then,
$s_1,\ldots, s_n$ and $t_1,\ldots, t_n$ are linear sequences of fresh 
variables. Recall that $\succ$ is defined on the extended Herbrand base, i.e.,
in the factor set obtained with respect to the variance relationship. 
Thus, $\mbox{\sl solve}(t, s_1,\ldots, s_n)$ is, in fact,
identical to $\mbox{\sl solve}(t, t_1,\ldots, t_n)$ up to variable
renaming. Hence, the inequality $\mbox{\sl solve}(t, s_1,\ldots, s_n) > \mbox{\sl solve}(t, t_1,\ldots, t_n)$ contradicts the irreflexivity of $>$.

Alternatively, if $(t_{11},\ldots,t_{1n})\in \mbox{\it Vars}_n$ we have to
use the choice of $\geq$ as a minimal quasi-ordering. By 
Lemma~\ref{lemma:minimal:ds},
there exists a directed derivation $Q_0 = \mbox{\sl solve}(t, s_1,\ldots,$ $s_n),$ $Q_1,\ldots, Q_k = \mbox{\sl solve}(t, t_1,\ldots, t_n)$.
Let $c_i$ be a clause used to resolve $Q_i$ and produce $Q_{i+1}$. 
By Lemma~\ref{lemma:de:un}, $Q_k$ is unifiable with the head of $c_0$.
By the second condition of Definition~\ref{definition:de:restricted} 
and by Lemma~\ref{lemma:de:complete} the atom 
that produced $Q_1$ can be selected to obtain $Q_{k+1}$. By the third
condition, the intermediate body atoms do not affect the first position
of {\sl solve}, i.e., the first argument of $Q_{k+1}$ should coincide with
the first argument of $Q_1$. Moreover, order-acceptability implies that
$Q_{k} > Q_{k+1}$. Proceeding in this way one can construct an infinitely 
decreasing sequence of atoms, contradicting the well-foundedness of $>$
and completing the proof of irreflexivity of $\succ$. As stated above the 
remaining properties of $\succ$ can be established analogously.
\eat{
\item Let $\mbox{\sl solve}(s)\succ \mbox{\sl solve}(t)$ and
$\mbox{\sl solve}(t)\succ \mbox{\sl solve}(s)$ hold
for some $s$ and $t$. Then, there exist terms 
$s^1_1,\ldots, s^1_n$, $s^2_1,\ldots, s^2_n$, 
$t^1_1,\ldots, t^1_n$, and $t^2_1,\ldots, t^2_n$
such that 
$\mbox{\sl solve}(s,$ $s^1_1,\ldots, s^1_n)\! >\! \mbox{\sl solve}(t, t^1_1,\ldots, 
t^1_n)$ and
$\mbox{\sl solve}(t, t^2_1,\ldots, t^2_n)\! >\! \mbox{\sl solve}(s, s^2_1,\ldots, s^2_n)$. 

If $(t_{11},\ldots,t_{1n})\not\in \mbox{\it Vars}_n$ one can
reason as above, and observe that
$\mbox{\sl solve}(t, t^1_1,$ $\ldots, t^1_n)$ and $\mbox{\sl solve}(t, t^2_1,\ldots, t^2_n)$ are variants. Thus, $\mbox{\sl solve}(t, t^1_1,\ldots, t^1_n)\;\leq\geq\;\mbox{\sl solve}(t, t^2_1,\ldots, t^2_n)$ and
$\mbox{\sl solve}(s, s^1_1,\ldots, s^1_n) > \mbox{\sl solve}(s, s^2_1,\ldots, s^2_n)$ contradicting that these atoms are variants as well.

If $(t_{11},\ldots,t_{1n})\in \mbox{\it Vars}_n$ by reasoning as above one can 
construct an infinite decreasing chain contradicting the well-foundedness of
$>$.
\item Let  $\mbox{\sl solve}(s)\succ \mbox{\sl solve}(t)$ and
$\mbox{\sl solve}(t)\succ \mbox{\sl solve}(u)$ for some $s$, $t$ and $u$.
Then, there exist terms 
$s_1,\ldots, s_n$,  
$t^1_1,\ldots, t^1_n$, $t^2_1,\ldots, t^2_n$,
and $u_1,\ldots, u_n$,
such that 
$\mbox{\sl solve}(s,$ 
$s_1,\ldots, s_n)\! >\! \mbox{\sl solve}(t, t^1_1,\ldots, 
t^1_n)$ and
$\mbox{\sl solve}(t, t^2_1,\ldots, t^2_n)\! >\! \mbox{\sl solve}(u, u_1,\ldots, u_n)$. 

If $(t_{11},\ldots,t_{1n})\not\in \mbox{\it Vars}_n$, atoms
$\mbox{\sl solve}(t, t^1_1,\ldots, t^1_n)$ and $\mbox{\sl solve}(t, t^2_1,\ldots, t^2_n)$ are variants. Thus, $\mbox{\sl solve}(t, t^1_1,\ldots, t^1_n)\;\leq\geq\;\mbox{\sl solve}(t, t^2_1,\ldots, t^2_n)$ and
$\mbox{\sl solve}(s, s_1,\ldots, s_n) >$ $\mbox{\sl solve}(u, u_1,\ldots, u_n)$. 
In other words, $\mbox{\sl solve}(s)\!\succ\! \mbox{\sl solve}(u)$ by definition
of $\succ$.

Otherwise, if $(t_{11},\ldots,t_{1n})\in \mbox{\it Vars}_n$, there exist
directed derivations $Q^1_0,\ldots,Q^1_{k_1}$ and $Q^2_0,\ldots,Q^2_{k_2}$
corresponding to each one of the inequalities. Then, by Lemma~\ref{lemma:de:un},
the head of the clause that has been used to resolve $Q^2_0$
can be unified with $Q^1_{k_1}$. Proceeding in this way, one constructs 
a directed derivation $Q^1_0,\ldots,Q^1_{k_1}, \ldots,Q^1_{k_1+k_2}$,
such that $Q^1_0 = \mbox{\sl solve}(s, s_1,\ldots, s_n)$ and
$Q^1_{k_1+k_2} = \mbox{\sl solve}(u, u_1,\ldots, u_n)$, completing the proof of
transitivity.
}

Next we are going to prove order-acceptability.
Let $\mbox{\sl solve}(t_0)$ be a call in $\mbox{\sl Call}(M_0\cup\mbox{\sl ce}(P),\mbox{\sl solve}(Q))$. Distinguish between the following cases.
\begin{itemize}
\item $t_0$ is unifiable with $\mbox{\sl true}$. By definition of
double extended meta-interpreters none of the predicates of $C_{11},\ldots,
C_{1m_1}$ is mutually recursive with {\sl solve}. Thus, the order-acceptability
condition holds trivially. 
\item $t_0$ is unifiable with $(A,B)$ via an mgu $\theta^{M_0}$. 
Then, we have to prove that 
$\mbox{\sl solve}(t_0) \succ \mbox{\sl solve}(A\theta^{M_0})$
and $\mbox{\sl solve}(t_0) \succ \mbox{\sl solve}(B\theta^{M_0}\rho^{M_0})$,
where $\rho^{M_0}$ is a computed answer substitution for 
$\mbox{\sl solve}(A\theta^{M_0})$.

By the observation preceding the theorem, there exists a call
$A^D_0 = \mbox{\sl solve}(t'_0, u_1,$ $\ldots, u_n)$ in 
$\mbox{\sl Call}(D\cup\mbox{\sl ce}^D(P),\mbox{\sl solve}(Q, v_1, 
\ldots, v_n))$, where $t'_0$ is a variant of $t_0$. 
Since  $t_0$ is unifiable with $(A,B)$ and Lemma~\ref{lemma:de:un} 
holds, $A^D_0$ can be unified with the head of the second clause in $D$.
\eat{
\begin{eqnarray*}
&& \mbox{\sl solve}((A,B), t_{21}, \ldots, t_{2n})\leftarrow D_{11},\ldots,D_{1k_1}, \mbox{\sl solve}(A, t_{31}, \ldots, t_{3n}), \\
&& \hspace{1.0cm}D_{21},\ldots,D_{2k_2}, \mbox{\sl solve}(B, t_{41}, \ldots, t_{4n}), C_{21},\ldots, C_{2m_2}\mbox{.}
\end{eqnarray*}
}
Moreover, if $\theta^D$ is the mgu, then $\theta^D = \theta^{M_0}\{u_1/t_{21},\ldots,u_n/t_{2n}\}$. Let $\sigma^D$ be a computed answer substitution for
$(D_{11},\ldots,D_{1k_1})\theta^D$. By the third condition of Definition~\ref{definition:de:restricted}, $A\theta^D\sigma^D$ coincides with $A\theta^D$. Thus,
$A\theta^D\sigma^D = A\theta^D = A\theta^{M_0}$. 

Since $D\cup\mbox{\sl ce}^D(P)$ is order-acceptable, 
$A^D_0 > \mbox{\sl solve}(A\theta^D\sigma^D, t_{31}\theta^D\sigma^D, \ldots, 
t_{3n}\theta^D\sigma^D)$.
Then, it holds that
$\mbox{\sl solve}(t'_0, u_1,\ldots,u_n) > \mbox{\sl solve}(A\theta^{M_0}, t_{31}\theta^D\sigma^D, \ldots, t_{3n}\theta^D\sigma^D)$.
Recalling the definition of $\succ$ and that if $a_1$ and $a_2$ are 
variants, then $a_1\;\preceq\succeq\;a_2$, we conclude
$\mbox{\sl solve}(t_0)\succ \mbox{\sl solve}(A\theta^{M_0})$, 
proving one of the order-acceptabi\-li\-ty decreases.

Next, we are going to see that 
$\mbox{\sl solve}(t_0)\succ \mbox{\sl solve}(B\theta^{M_0}\rho^{M_0})$.
Indeed, order-acceptabi\-li\-ty of $D\cup\mbox{\sl ce}^D(P)$ with respect to
$\mbox{\sl solve}(Q, v_1, \ldots, v_n)$ implies that $A^D_0 >
\mbox{\sl solve}(B,t_{41},$ $\ldots,$ $t_{4n})\theta^D\sigma^D\rho^D\delta^D$,
where $\theta^D$ and $\sigma^D$ as above, $\rho^D$ is a computed answer
substitution for $\mbox{\sl solve}(A,t_{31},\ldots,$ $t_{3n})\theta^D\sigma^D$
and $\delta^D$ is a computed answer substitution for
$(D_{21},\ldots,$ $D_{2k_2})\theta^D\sigma^D\rho^D$.

The third condition of Definition~\ref{definition:de:restricted}
implies that $\sigma^D$ cannot affect instances of $B$. 
In other words, $B\theta^D\sigma^D = B\theta^D$. Thus,
$B\theta^D\sigma^D\rho^D\delta^D = B\theta^D\rho^D\delta^D$. The same 
condition applied to $\delta^D$ implies
$B\theta^D\rho^D\delta^D = B\theta^D\rho^D$. Lemma~\ref{lemma:de:complete}
implies that for every computed answer $A\theta^D\rho^D$ obtained with 
respect to $D\cup\mbox{\sl ce}^D(P)$ there exists a computed answer
$A\theta^D\rho^{M_0}$ obtained with respect to $M_0\cup\mbox{\sl ce}(P)$,
such that $A\theta^D\rho^{M_0}$ is a variant of $A\theta^D\rho^D$. Recall,
that $A\theta^D$ coincides with $A\theta^{M_0}$. This implies as well
that $B\theta^D\rho^D$ is a variant of $B\theta^{M_0}\rho^{M_0}$. Hence,
\eat{
\begin{eqnarray*}
&& \mbox{\sl solve}(t'_0, u_1,\ldots,u_n) > \mbox{\sl solve}(B\theta^{M_0}\rho^{M_0},\\
&& \hspace{1.0cm}t_{41}\theta^D\sigma^D\rho^D\delta^D, \ldots, t_{44}\theta^D\sigma^D\rho^D\delta^D)\mbox{.}
\end{eqnarray*}
}
$$\mbox{\sl solve}(t'_0, u_1,\ldots,u_n) > \mbox{\sl solve}(B\theta^{M_0}\rho^{M_0}, t_{41}\theta^D\sigma^D\rho^D\delta^D, \ldots, t_{44}\theta^D\sigma^D\rho^D\delta^D)\mbox{.}$$
Thus, 
$\mbox{\sl solve}(t_0)\succ \mbox{\sl solve}(B\theta^{M_0}\rho^{M_0})$,
completing the proof in this case.
\item $t_0$ is unifiable with $A$ via $\theta^{M_0}$, i.e., the last clause of 
$M_0$ is applied. In this case we have to show that $\mbox{\sl solve}(t_0)\succ \mbox{\sl solve}(B\theta^{M_0}\rho^{M_0})$, where $\rho^{M_0}$ is a 
computed answer substitution for $\mbox{\sl clause}(A,B)\theta^{M_0}$.

By the same observation on the call set, there exist 
$\mbox{\sl solve}(t'_0,t_1,\ldots,t_n)\in 
\mbox{\sl Call}(D\cup\mbox{\sl ce}^D(P),\mbox{\sl solve}(Q,v_1,\ldots,v_n))$, 
such that $t'_0$ is a variant of $t_0$. Lemma~\ref{lemma:de:un} implies that
$\mbox{\sl solve}(t'_0,t_1,\ldots,t_n)$ can be unified with the head of the
corresponding clause in $D$. Moreover, the restriction of an mgu $\theta^D$
to $A$ coincides with $\theta^{M_0}$ up to a variable renaming. Let
$\sigma^D$ be a computed answer substitution for 
$(D_{31},\ldots,D_{3k_3})\theta^D$. By the third condition of Definition~\ref{definition:de:restricted}, $A\theta^D\sigma^D = A\theta^D = A\theta^{M_0}$.
Thus, the call to $\mbox{\sl clause}$ can be unified with the atom
corresponding to the one used to compute $\rho^{M_0}$. In other words,
there exists a computed answer substitution $\rho^D$ for 
$\mbox{\sl clause}(A\theta^D\sigma^D,B,s_1\theta^D\sigma^D,$
$\ldots,s_k\theta^D\sigma^D)$, such that $\rho^D$ restricted to the variables of $A\theta^D\sigma^D$
and $B$ is $\rho^{M_0}$ (up to a variable renaming). Finally,
the third condition of Definition~\ref{definition:de:restricted} implies that
the computed answer substitution $\delta^D$ for an instance of 
$(D_{41},\ldots,D_{4k_4})$
cannot affect the corresponding instance of $B$. Taking this discussion
in consideration, the order-acceptability decrease implies
\eat{
\begin{eqnarray*}
&& \mbox{\sl solve}(t'_0,t_1,\ldots,t_n) > 
\mbox{\sl solve}(B\theta^{M_0}\rho^{M_0},t_{61}\theta^D\sigma^D\rho^D\delta^D,\\
&&\hspace{1.0cm} \ldots,t_{6n}\theta^D\sigma^D\rho^D\delta^D)
\end{eqnarray*}
}
$$\mbox{\sl solve}(t'_0,t_1,\ldots,t_n) > 
\mbox{\sl solve}(B\theta^{M_0}\rho^{M_0},t_{61}\theta^D\sigma^D\rho^D\delta^D,\ldots,t_{6n}\theta^D\sigma^D\rho^D\delta^D)$$
and, by definition of $\succ$, $\mbox{\sl solve}(t_0) > \mbox{\sl solve}(B\theta^{M_0}\rho^{M_0})$, completing the proof.
\end{itemize}
\end{proof}

\end{document}